\documentclass[12pt]{article}
\usepackage{cite}
\usepackage{axodraw}
\usepackage{epsfig}
\usepackage{amssymb}
\makeatletter

\def\paragraph{\@startsection{paragraph}{4}{\z@}{+2.00ex plus
 +1ex minus +.2ex}{1.5ex plus .2ex}{\it\normalsize}}

\def\section{\@startsection {section}{1}{\z@}{+3.0ex plus +1ex minus
  +.2ex}{2.3ex plus .2ex}{\normalsize\bf\boldmath}}
\def\subsection{\@startsection{subsection}{2}{\z@}{+2.5ex plus +1ex
minus +.2ex}{1.5ex plus .2ex}{\normalsize\bf\boldmath}}
\def\subsubsection{\@startsection{subsubsection}{3}{\z@}{+3.25ex plus
 +1ex minus +.2ex}{1.5ex plus .2ex}{\normalsize\it}}

 \expandafter\ifx\csname mathrm\endcsname\relax\def\mathrm#1{{\rm #1}}\fi

\newcounter{saveeqn}

\@addtoreset{equation}{section}

\def\nl{\nonumber\\}

\newcommand{\gsim}
{\mathrel{\raisebox{-.3em}{$\stackrel{\displaystyle >}{\sim}$}}}
\def\asymp#1%
{\mathrel{\raisebox{-.4em}{$\widetilde{\scriptstyle #1}$}}}

\def\Nequal#1%
{\mathrel{\raisebox{-.5em}{$\stackrel{=}{\scriptstyle\rm#1}$}}}
\newcommand{\dsl}[1]{\not \hspace{-0.7mm}#1}
\def\dsl{\mathpalette\make@slash}
\def\make@slash#1#2{\setbox\z@\hbox{$#1#2$}%
  \hbox to 0pt{\hss$#1/$\hss\kern-\wd0}\box0}

\def\beq{\begin{equation}}
\def\eeq{\end{equation}}
\def\bit{\begin{itemize}}
\def\eit{\end{itemize}}
\def\beqar{\begin{eqnarray}}
\def\eeqar{\end{eqnarray}}
\def\barr#1{\begin{array}{#1}}
\def\earr{\end{array}}
\def\bfi{\begin{figure}}
\def\efi{\end{figure}}
\def\btab{\begin{table}}
\def\etab{\end{table}}
\def\bce{\begin{center}}
\def\ece{\end{center}}
\def\nn{\nonumber}

\def\text{\textstyle}

\newcommand{\eqintext}{}

\def\Ga{\Gamma}
\def\ga{\gamma}
\def\de{\delta}
\def\De{\Delta}

\def\si{\sigma}

\def\refeq#1{\mbox{(\ref{#1})}}
\def\refeqs#1{\mbox{(\ref{#1})}}
\def\refeqf#1{\mbox{(\ref{#1})}}
\def\reffi#1{\mbox{Figure~\ref{#1}}}
\def\reffis#1{\mbox{Figures~\ref{#1}}}
\def\refta#1{\mbox{Table~\ref{#1}}}

\def\refse#1{\mbox{Section~\ref{#1}}}
\def\refses#1{\mbox{Sections~\ref{#1}}}
\def\refapp#1{\mbox{App.~\ref{#1}}}
\def\citere#1{\mbox{Ref.~\cite{#1}}}
\def\citeres#1{\mbox{Refs.~\cite{#1}}}

\newcommand{\TeV}{\unskip\,\mathrm{TeV}}
\newcommand{\GeV}{\unskip\,\mathrm{GeV}}

\newcommand{\ri}{{\mathrm{i}}}
\newcommand{\rI}{{\mathrm{I}}}
\newcommand{\rd}{{\mathrm{d}}}

\newcommand{\rL}{{\mathrm{L}}}
\newcommand{\rT}{{\mathrm{T}}}
\newcommand{\rR}{\mathrm R}
\newcommand{\rF}{\mathrm F}
\newcommand{\rS}{\mathrm S}

\newcommand{\A}{{\cal{A}}}

\newcommand{\M}{{\cal{M}}}

\def\mathswitchr#1{\relax\ifmmode{\mathrm{#1}}\else$\mathrm{#1}$\fi}

\newcommand{\PW}{\mathswitchr W}

\newcommand{\PZ}{\mathswitchr Z}

\newcommand{\Pg}{\mathswitchr g}

\newcommand{\PH}{\mathswitchr H}
\newcommand{\Pe}{\mathswitchr e}
\newcommand{\Pne}{\mathswitch \nu_{\mathrm{e}}}

\newcommand{\Pnmubar}{\mathswitch \bar\nu_{\mu}}
\newcommand{\Pd}{\mathswitchr d}
\newcommand{\Pdbar}{\bar{\mathswitchr d}}
\newcommand{\Pu}{\mathswitchr u}
\newcommand{\Pubar}{\bar{\mathswitchr u}}
\newcommand{\Ps}{\mathswitchr s}

\newcommand{\Pc}{\mathswitchr c}

\newcommand{\Pb}{\mathswitchr b}
\newcommand{\Pbbar}{\mathswitchr{\bar b}}
\newcommand{\Pp}{\mathswitchr p}
\newcommand{\Pt}{\mathswitchr t}
\newcommand{\Ptbar}{\mathswitchr{\bar t}}
\newcommand{\Pep}{\mathswitchr {e^+}}
\newcommand{\Pem}{\mathswitchr {e^-}}

\newcommand{\Pmum}{\mathswitchr {\mu^-}}

\newcommand{\PWp}{\mathswitchr {W^+}}
\newcommand{\PWm}{\mathswitchr {W^-}}

\newcommand{\jet}{\mathswitchr {j}}

\newcommand{\Pq}{q}
\newcommand{\Pqbar}{\bar{q}}
\newcommand{\Pl}{l}

\def\mathswitch#1{\relax\ifmmode#1\else$#1$\fi}

\newcommand{\MW}{\mathswitch {M_\PW}}

\newcommand{\MZ}{\mathswitch {M_\PZ}}
\newcommand{\MH}{\mathswitch {M_\PH}}

\newcommand{\Mt}{\mathswitch {m_\Pt}}

\newcommand{\GW}{\Gamma_{\PW}}

\newcommand{\Gt}{\Gamma_{\Pt}}

\newcommand{\rw}{\mathswitchr w}
\newcommand{\sw}{\mathswitch {s_\rw}}
\newcommand{\cw}{\mathswitch {c_\rw}}

\newcommand{\GF}{\mathswitch {G_\mu}}

\def\solid{\raise.9mm\hbox{\protect\rule{1.1cm}{.2mm}}}
\def\dash{\raise.9mm\hbox{\protect\rule{2mm}{.2mm}}\hspace*{1mm}}

\def\ie{i.e.\ }

\newcommand{\LO}{{\mathrm{LO}}}
\newcommand{\NLO}{{\mathrm{NLO}}}

\newcommand{\DPA}{{\mathrm{DPA}}}
\newcommand{\full}{{\mathrm{full}}}

\newcommand{\lab}{{\mathrm{LAB}}}

\newcommand{\CM}{{\mathrm{CM}}}

\newcommand{\FB}{{\mathrm{FB}}}
\newcommand{\CE}{{\mathrm{CE}}}
\newcommand{\virt}{{\mathrm{virt}}}
\newcommand{\real}{{\mathrm{real}}}
\newcommand{\fact}{{\mathrm{fact}}}

\newcommand{\sub}{{\mathrm{sub}}}

\newcommand{\fin}{{\mathrm{fin}}}
\renewcommand{\min}{{\mathrm{min}}}
\renewcommand{\max}{{\mathrm{max}}}
\newcommand{\miss}{{\mathrm{miss}}}

\def\Li{\mathop{\mathrm{Li}_2}\nolimits}

\def\Re{\mathop{\mathrm{Re}}\nolimits}

\def\lra{\mathop{\mathrm{\leftrightarrow}}\nolimits}

\newcommand{\dynscale}{{E}_\rT}
\newcommand{\mudyn}{\mu_{\mathrm{dyn}}}
\newcommand{\mufix}{\mu_{\mathrm{fix}}}
\newcommand{\pz}{\phantom{-}}

\newcommand{\ord}{\mathcal O}

\newcommand{\muF}{\mu_{\mathrm{F}}}
\newcommand{\muR}{\mu_{\mathrm{R}}}

\newcommand{\kT}{\mathswitch {k_{\mathrm{T}}}}
\newcommand{\kt}{\kT}
\newcommand{\pt}{\mathswitch {p_{\mathrm{T}}}}
\newcommand{\alphas}{\alpha_\mathrm{s}}
\newcommand{\pon}{\hat{p}}

\newcommand{\ttb}{{\Pt\bar\Pt}}
\newcommand{\wwbb}{{\PW\PW\Pb\bar\Pb}}
\newcommand{\WWbb}{{\PW^+\PW^-\Pb\bar\Pb}}
\newcommand{\lvlvbb}{{\nu_\Pe\Pe^+\mu^-{\bar\nu}_\mu\Pb\bar\Pb}}
\newcommand{\wwfs}{\PW^+\PW^-\Pb\bar\Pb}
\newcommand{\leptfs}{{\Pne\Pep\Pmum\Pnmubar\Pb\bar\Pb}}

\newcommand{\NtWA}{\mathrm{NtWA}}
\newcommand{\NwWA}{\mathrm{NwWA}}
\newcommand{\FtW}{\mathrm{FtW}}
\newcommand{\FwW}{\mathrm{FwW}}

\newcommand{\trunc}{\mathrm{trunc}}
\newcommand{\MS}{\mathrm{exp}}

\newcommand{\NLOm}{{\mathrm{NLO}^{+}}}
\newcommand{\cuts}{\mathrm{cuts}}

\newcommand{\GtLO}{\Gamma_{\Pt}^{\LO}}
\newcommand{\GtNLO}{\Gamma_{\Pt}^{\NLO}}
\newcommand{\pw}[2]{\rd\Gamma_{\Pt\to #1}^{#2}}
\newcommand{\tw}[2]{\Gamma_{\Pt\to #1}^{#2}}
\newcommand{\pwb}[2]{\rd\Gamma_{\bar\Pt\to #1}^{#2}}
\newcommand{\twb}[2]{\Gamma_{\bar\Pt\to #1}^{#2}}
\newcommand{\BR}[2]{\mathrm{BR}_{\Pt\to #1}^{#2}}
\newcommand{\BRw}[2]{\mathrm{BR}_{\PW\to #1}^{#2}}
\newcommand{\BRb}[2]{\mathrm{BR}_{\bar\Pt\to #1}^{#2}}

\newcommand{\qparbar}{\raisebox{.6em}{\tiny $(-)$}\hspace{-.83em}q}

\def\draftdate{\relax}
\def\mda{\relax}
\def\mua{\relax}
\def\mla{\relax}
\def\Mda{\relax}
\def\Mua{\relax}
\def\Mla{\relax}
\def\draft{
\def\thtystars{******************************}
\def\sixtystars{\thtystars\thtystars}
\typeout{}
\typeout{\sixtystars**}
\typeout{* Draft mode!
         For final version remove \protect\draft\space in source file *}
\typeout{\sixtystars**}
\typeout{}
\def\draftdate{\today}
\def\mua{\marginpar[\boldmath\hfil$\uparrow$]%
                   {\boldmath$\uparrow$\hfil}%
                    \typeout{marginpar: $\uparrow$}\ignorespaces}
\def\mda{\marginpar[\boldmath\hfil$\downarrow$]%
                   {\boldmath$\downarrow$\hfil}%
                    \typeout{marginpar: $\downarrow$}\ignorespaces}
\def\mla{\marginpar[\boldmath\hfil$\rightarrow$]%
                   {\boldmath$\leftarrow $\hfil}%
                    \typeout{marginpar: $\lra$}\ignorespaces}
\def\Mua{\marginpar[\boldmath\hfil$\Uparrow$]%
                   {\boldmath$\Uparrow$\hfil}%
                    \typeout{marginpar: $\uparrow$}\ignorespaces}
\def\Mda{\marginpar[\boldmath\hfil$\Downarrow$]%
                   {\boldmath$\Downarrow$\hfil}%
                    \typeout{marginpar: $\downarrow$}\ignorespaces}
\def\Mla{\marginpar[\boldmath\hfil$\Rightarrow$]%
                   {\boldmath$\Leftarrow $\hfil}%
                    \typeout{marginpar: $\lra$}\ignorespaces}
\overfullrule 5pt
\oddsidemargin -15mm
\oddsidemargin -10mm
\marginparwidth 29mm
}

\def\stars{\strut\leaders\hbox{*}\hfill\strut}
\def\starline{\hfil\strut\hfil\hbox to \textwidth {\stars}\hfil}

\SetScale{0.85}
\newcommand{\DRtchanneltopggveepmumvmxbbx}{
\begin{picture}(160,135)(-20,0)
\Text (-5.,0)[r]{$\mathrm{\scriptstyle g}$}
\Text (-5.,75)[r]{$\mathrm{\scriptstyle g}$}
\Text (110,0)[l]{$\mathrm{\scriptstyle \bar{b}}$}
\Text (110,15)[l]{$\mathrm{\scriptstyle \mu^-}$}
\Text (110,30)[l]{$\mathrm{\scriptstyle \bar{\nu}_\mu}$}
\Text (110,75)[l]{$\mathrm{\scriptstyle b}$}
\Text (110,45)[l]{$\mathrm{\scriptstyle \nu_e}$}
\Text (110,60)[l]{$\mathrm{\scriptstyle e^+}$}
\Vertex(79,52.5){2.0}
\ArrowLine(79,52.5)(105,45)
\ArrowLine(105,60)(79,52.5)
\Vertex(53,60){2.0}
\ArrowLine(53,60)(105,75)
\Photon(79,52.5)(53,60){2}{3}
\Text (75.5,53.)[rt]{$\mathrm{\scriptstyle W^+}$}
\Vertex(79,22.5){2.0}
\ArrowLine(79,22.5)(105,15)
\ArrowLine(105,30)(79,22.5)
\Vertex(53,15){2.0}
\ArrowLine(105,0)(53,15)
\Photon(79,22.5)(53,15){2}{3}
\Text (75.5,22.)[rb]{$\mathrm{\scriptstyle W^-}$}
\Vertex(26,15){2.0}
\Gluon(0,0)(26,15){3}{4}
\ArrowLine(53,15)(26,15)
\Text (39.5,11)[ct]{$\mathrm{\scriptstyle \bar{t}}$}
\Vertex(26,60){2.0}
\ArrowLine(26,15)(26,60)
\Text (22,37.5)[r]{$\mathrm{\scriptstyle t}$}
\Gluon(0,75)(26,60){3}{4}
\ArrowLine(26,60)(53,60)
\Text (39.5,64)[cb]{$\mathrm{\scriptstyle t}$}
\end{picture}
}
\newcommand{\DRschannelgluonggveepmumvmxbbx}{
\begin{picture}(160,135)(-20,0)
\Text (-5.,0)[r]{$\mathrm{\scriptstyle g}$}
\Text (-5.,75)[r]{$\mathrm{\scriptstyle g}$}
\Text (110,0)[l]{$\mathrm{\scriptstyle b}$}
\Text (110,15)[l]{$\mathrm{\scriptstyle \nu_e}$}
\Text (110,30)[l]{$\mathrm{\scriptstyle e^+}$}
\Text (110,75)[l]{$\mathrm{\scriptstyle \bar{b}}$}
\Text (110,45)[l]{$\mathrm{\scriptstyle \mu^-}$}
\Text (110,60)[l]{$\mathrm{\scriptstyle \bar{\nu}_\mu}$}
\Vertex(88,52.5){2.0}
\ArrowLine(88,52.5)(105,45)
\ArrowLine(105,60)(88,52.5)
\Vertex(88,22.5){2.0}
\ArrowLine(88,22.5)(105,15)
\ArrowLine(105,30)(88,22.5)
\Vertex(71,60){2.0}
\ArrowLine(105,75)(71,60)
\Photon(88,52.5)(71,60){2}{2}
\Text (84.5,53.)[rt]{$\mathrm{\scriptstyle W^-}$}
\Vertex(71,15){2.0}
\ArrowLine(71,15)(105,0)
\Photon(88,22.5)(71,15){2}{2}
\Text (84.5,22.)[rb]{$\mathrm{\scriptstyle W^+}$}
\Vertex(54,37.5){2.0}
\ArrowLine(54,37.5)(71,15)
\Text (59.3085,23.8387)[rt]{$\mathrm{\scriptstyle t}$}
\ArrowLine(71,60)(54,37.5)
\Text (59.3085,51.1613)[rb]{$\mathrm{\scriptstyle \bar{t}}$}
\Vertex(17,37){2.0}
\Gluon(0,0)(17,37){3}{5}
\Gluon(0,75)(17,37){3}{5}
\Gluon(54,37.5)(17,37){3}{5}
\Text (35.446,43.2496)[rb]{$\mathrm{\scriptstyle g}$}
\end{picture}
}
\newcommand{\DRschannelgluonqqxveepmumvmxbbx}{
\begin{picture}(160,135)(-20,0)
\Text (-5.,0)[r]{${\scriptstyle q}$}
\Text (-5.,75)[r]{${\scriptstyle \bar{q}}$}
\Text (110,0)[l]{$\mathrm{\scriptstyle b}$}
\Text (110,15)[l]{$\mathrm{\scriptstyle \nu_e}$}
\Text (110,30)[l]{$\mathrm{\scriptstyle e^+}$}
\Text (110,75)[l]{$\mathrm{\scriptstyle \bar{b}}$}
\Text (110,45)[l]{$\mathrm{\scriptstyle \mu^-}$}
\Text (110,60)[l]{$\mathrm{\scriptstyle \bar{\nu}_\mu}$}
\Vertex(88,52.5){2.0}
\ArrowLine(88,52.5)(105,45)
\ArrowLine(105,60)(88,52.5)
\Vertex(88,22.5){2.0}
\ArrowLine(88,22.5)(105,15)
\ArrowLine(105,30)(88,22.5)
\Vertex(71,60){2.0}
\ArrowLine(105,75)(71,60)
\Photon(88,52.5)(71,60){2}{2}
\Text (84.5,53.)[rt]{$\mathrm{\scriptstyle W^-}$}
\Vertex(71,15){2.0}
\ArrowLine(71,15)(105,0)
\Photon(88,22.5)(71,15){2}{2}
\Text (84.5,22.)[rb]{$\mathrm{\scriptstyle W^+}$}
\Vertex(54,37.5){2.0}
\ArrowLine(54,37.5)(71,15)
\Text (59.3085,23.8387)[rt]{$\mathrm{\scriptstyle t}$}
\ArrowLine(71,60)(54,37.5)
\Text (59.3085,51.1613)[rb]{$\mathrm{\scriptstyle \bar{t}}$}
\Vertex(17,37){2.0}
\ArrowLine(0,0)(17,37)
\ArrowLine(17,37)(0,75)
\Gluon(54,37.5)(17,37){3}{5}
\Text (35.446,43.2496)[rb]{$\mathrm{\scriptstyle g}$}
\end{picture}
}
\newcommand{\diagramsggveepmumvmxbbxSRa}{
\begin{picture}(160,110)(-20,0)
\Text (-5.,0)[r]{$\mathrm{\scriptstyle g}$}
\Text (-5.,75)[r]{$\mathrm{\scriptstyle g}$}
\Text (110,0)[l]{$\mathrm{\scriptstyle \bar{b}}$}
\Text (110,15)[l]{$\mathrm{\scriptstyle \mu^-}$}
\Text (110,30)[l]{$\mathrm{\scriptstyle \bar{\nu}_\mu}$}
\Text (110,75)[l]{$\mathrm{\scriptstyle b}$}
\Text (110,45)[l]{$\mathrm{\scriptstyle \nu_e}$}
\Text (110,60)[l]{$\mathrm{\scriptstyle e^+}$}
\Vertex(88,52.5){2.0}
\ArrowLine(88,52.5)(105,45)
\ArrowLine(105,60)(88,52.5)
\Vertex(71,60){2.0}
\ArrowLine(71,60)(105,75)
\Photon(88,52.5)(71,60){2}{2}
\Text (83.8854,54.4097)[rt]{$\mathrm{\scriptstyle W^+}$}
\Vertex(88,22.5){2.0}
\ArrowLine(88,22.5)(105,15)
\ArrowLine(105,30)(88,22.5)
\Vertex(54,47.5){2.0}
\Photon(88,22.5)(54,47.5){2}{5}
\Text (70.6304,36.7774)[rt]{$\mathrm{\scriptstyle W^-}$}
\ArrowLine(54,47.5)(71,60)
\Text (60.1304,56.9726)[rb]{$\mathrm{\scriptstyle t}$}
\Vertex(37,38){2.0}
\ArrowLine(105,0)(37,38)
\ArrowLine(37,38)(54,47.5)
\Text (43.5487,46.2418)[rb]{$\mathrm{\scriptstyle b}$}
\Vertex(17,37.5){2.0}
\Gluon(0,0)(17,37.5){3}{5}
\Gluon(0,75)(17,37.5){3}{5}
\Gluon(37,38)(17,37.5){3}{2}
\Text (28.9,43.7488)[rb]{$\mathrm{\scriptstyle g}$}
\end{picture}
}

\newcommand{\diagramsuuxveepmumvmxbbxSR}{
\begin{picture}(160,110)(-20,0)
\Text (-5.,0)[r]{${\scriptstyle q}$}
\Text (-5.,75)[r]{${\scriptstyle \bar{q}}$}
\Text (110,0)[l]{$\mathrm{\scriptstyle \bar{b}}$}
\Text (110,15)[l]{$\mathrm{\scriptstyle \mu^-}$}
\Text (110,30)[l]{$\mathrm{\scriptstyle \bar{\nu}_\mu}$}
\Text (110,75)[l]{$\mathrm{\scriptstyle b}$}
\Text (110,45)[l]{$\mathrm{\scriptstyle \nu_e}$}
\Text (110,60)[l]{$\mathrm{\scriptstyle e^+}$}
\Vertex(88,52.5){2.0}
\ArrowLine(88,52.5)(105,45)
\ArrowLine(105,60)(88,52.5)
\Vertex(71,60){2.0}
\ArrowLine(71,60)(105,75)
\Photon(88,52.5)(71,60){2}{2}
\Text (83.8854,54.4097)[rt]{$\mathrm{\scriptstyle W^+}$}
\Vertex(88,22.5){2.0}
\ArrowLine(88,22.5)(105,15)
\ArrowLine(105,30)(88,22.5)
\Vertex(54,47.5){2.0}
\Photon(88,22.5)(54,47.5){2}{5}
\Text (70.6304,36.7774)[rt]{$\mathrm{\scriptstyle W^-}$}
\ArrowLine(54,47.5)(71,60)
\Text (60.1304,56.9726)[rb]{$\mathrm{\scriptstyle t}$}
\Vertex(37,38){2.0}
\ArrowLine(105,0)(37,38)
\ArrowLine(37,38)(54,47.5)
\Text (43.5487,46.2418)[rb]{$\mathrm{\scriptstyle b}$}
\Vertex(17,37.5){2.0}
\ArrowLine(0,0)(17,37.5)
\ArrowLine(17,37.5)(0,75)
\Gluon(37,38)(17,37.5){3}{2}
\Text (28.9,43.7488)[rb]{$\mathrm{\scriptstyle g}$}
\end{picture}
}
\newcommand{\diagramsggveepmumvmxbbxSRbmod}{
\begin{picture}(160,110)(-20,0)
\Text (-5.,0)[r]{$\mathrm{\scriptstyle g}$}
\Text (-5.,75)[r]{$\mathrm{\scriptstyle g}$}
\Text (110,0)[l]{$\mathrm{\scriptstyle \bar{b}}$}
\Text (110,15)[l]{$\mathrm{\scriptstyle \mu^-}$}
\Text (110,30)[l]{$\mathrm{\scriptstyle \bar{\nu}_\mu}$}
\Text (110,75)[l]{$\mathrm{\scriptstyle b}$}
\Text (110,45)[l]{$\mathrm{\scriptstyle \nu_e}$}
\Text (110,60)[l]{$\mathrm{\scriptstyle e^+}$}
\Vertex(84,52.5){2.0}
\ArrowLine(84,52.5)(105,45)
\ArrowLine(105,60)(84,52.5)
\Vertex(63,60){2.0}
\ArrowLine(63,60)(105,75)
\Photon(84,52.5)(63,60){2}{3}
\Text (74.8914,52.5933)[rt]{$\mathrm{\scriptstyle W^+}$}
\Vertex(84,22.5){2.0}
\ArrowLine(84,22.5)(105,15)
\ArrowLine(105,30)(84,22.5)
\Vertex(21,0){2.0}
\Gluon(0,0)(21,0){3}{3}
\ArrowLine(105,0)(21,0)
\Vertex(42,60){2.0}
\Gluon(0,75)(42,60){3}{6}
\ArrowLine(42,60)(63,60)
\Text (52.5,64)[cb]{$\mathrm{\scriptstyle t}$}
\Vertex(42,22.5){2.0}
\ArrowLine(21,0)(42,22.5)
\Text (25,14.25)[r]{$\mathrm{\scriptstyle b}$}
\ArrowLine(42,22.5)(42,60)
\Text (35,41.25)[r]{$\mathrm{\scriptstyle t}$}
\Photon(84,22.5)(42,22.5){2}{6}
\Text (62.5,18.5)[ct]{$\mathrm{\scriptstyle W^-}$}
\end{picture}
}

\newcommand{\diagramsggveepmumvmxbbxNR}{
\begin{picture}(160,110)(-20,0)
\Text (-5.,0)[r]{$\mathrm{\scriptstyle g}$}
\Text (-5.,75)[r]{$\mathrm{\scriptstyle g}$}
\Text (110,0)[l]{$\mathrm{\scriptstyle \bar{b}}$}
\Text (110,15)[l]{$\mathrm{\scriptstyle \nu_e}$}
\Text (110,30)[l]{$\mathrm{\scriptstyle e^+}$}
\Text (110,45)[l]{$\mathrm{\scriptstyle \mu^-}$}
\Text (110,60)[l]{$\mathrm{\scriptstyle \bar{\nu}_\mu}$}
\Text (110,75)[l]{$\mathrm{\scriptstyle b}$}
\Vertex(79,52.5){2.0}
\ArrowLine(79,52.5)(105,45)
\ArrowLine(105,60)(79,52.5)
\Vertex(79,22.5){2.0}
\ArrowLine(79,22.5)(105,15)
\ArrowLine(105,30)(79,22.5)
\Vertex(53,37.5){2.0}
\Photon(79,22.5)(53,37.5){2}{4}
\Text (69.0011,26.5353)[rt]{$\mathrm{\scriptstyle W^+}$}
\Photon(79,52.5)(53,37.5){2}{4}
\Text (69.0011,48.4647)[rb]{$\mathrm{\scriptstyle W^-}$}
\Vertex(26,0){2.0}
\Gluon(0,0)(26,0){3}{3}
\ArrowLine(105,0)(26,0)
\Vertex(26,75){2.0}
\Gluon(0,75)(26,75){3}{3}
\ArrowLine(26,75)(105,75)
\Vertex(26,37.5){2.0}
\ArrowLine(26,0)(26,37.5)
\Text (22,18.75)[r]{$\mathrm{\scriptstyle b}$}
\ArrowLine(26,37.5)(26,75)
\Text (22,56.25)[r]{$\mathrm{\scriptstyle b}$}
\Photon(53,37.5)(26,37.5){2}{3}
\Text (39.5,33.5)[ct]{$\mathrm{\scriptstyle Z,\gamma}$}
\end{picture}
}

\newcommand{\diagramsggveepmumvmxbbxNRb}{
\begin{picture}(160,110)(-20,0)
\Text (-5.,0)[r]{$\mathrm{\scriptstyle g}$}
\Text (-5.,75)[r]{$\mathrm{\scriptstyle g}$}
\Text (110,0)[l]{$\mathrm{\scriptstyle b}$}
\Text (110,15)[l]{$\mathrm{\scriptstyle \nu_e}$}
\Text (110,30)[l]{$\mathrm{\scriptstyle \mathrm{\scriptstyle e^+}}$}
\Text (110,45)[l]{$\mathrm{\scriptstyle \mu^-}$}
\Text (110,60)[l]{$\mathrm{\scriptstyle \bar{\nu}_\mu}$}
\Text (110,75)[l]{$\mathrm{\scriptstyle \bar{b}}$}
\Vertex(70,52.5){2.0}
\ArrowLine(70,52.5)(105,45)
\ArrowLine(105,60)(70,52.5)
\Vertex(70,22.5){2.0}
\ArrowLine(70,22.5)(105,15)
\ArrowLine(105,30)(70,22.5)
\Vertex(35,0){2.0}
\Gluon(0,0)(35,0){3}{4}
\ArrowLine(35,0)(105,0)
\Vertex(35,75){2.0}
\Gluon(0,75)(35,75){3}{4}
\ArrowLine(105,75)(35,75)
\Vertex(35,22.5){2.0}
\ArrowLine(35,22.5)(35,0)
\Text (31,11.25)[r]{$\mathrm{\scriptstyle b}$}
\Photon(70,22.5)(35,22.5){2}{4}
\Text (52.5,18.5)[ct]{$\mathrm{\scriptstyle W^+}$}
\Vertex(35,52.5){2.0}
\ArrowLine(35,52.5)(35,22.5)
\Text (31,37.5)[r]{$\mathrm{\scriptstyle t}$}
\ArrowLine(35,75)(35,52.5)
\Text (31,63.75)[r]{$\mathrm{\scriptstyle b}$}
\Photon(70,52.5)(35,52.5){2}{4}
\Text (52.5,56.5)[cb]{$\mathrm{\scriptstyle W^-}$}
\end{picture}
}

\newcommand{\diagramsuuxveepmumvmxbbxNR}{
\begin{picture}(160,110)(-20,0)
\Text (-5.,0)[r]{$\mathrm{\scriptstyle u}$}
\Text (-5.,75)[r]{$\mathrm{\scriptstyle \bar{u}}$}
\Text (110,0)[l]{$\mathrm{\scriptstyle \nu_e}$}
\Text (110,15)[l]{$\mathrm{\scriptstyle e^+}$}
\Text (110,30)[l]{$\mathrm{\scriptstyle b}$}
\Text (110,45)[l]{$\mathrm{\scriptstyle \bar{b}}$}
\Text (110,60)[l]{$\mathrm{\scriptstyle \mu^-}$}
\Text (110,75)[l]{$\mathrm{\scriptstyle \bar{\nu}_\mu}$}
\Vertex(70,67.5){2.0}
\ArrowLine(70,67.5)(105,60)
\ArrowLine(105,75)(70,67.5)
\Vertex(70,37.5){2.0}
\ArrowLine(70,37.5)(105,30)
\ArrowLine(105,45)(70,37.5)
\Vertex(70,7.5){2.0}
\ArrowLine(70,7.5)(105,0)
\ArrowLine(105,15)(70,7.5)
\Vertex(35,7.5){2.0}
\ArrowLine(0,0)(35,7.5)
\Photon(70,7.5)(35,7.5){2}{4}
\Text (52.5,3.5)[ct]{$\mathrm{\scriptstyle W^+}$}
\Vertex(35,67.5){2.0}
\ArrowLine(35,67.5)(0,75)
\Photon(70,67.5)(35,67.5){2}{4}
\Text (52.5,71.5)[cb]{$\mathrm{\scriptstyle W^-}$}
\Vertex(35,37.5){2.0}
\ArrowLine(35,7.5)(35,37.5)
\Text (31,22.5)[r]{$\mathrm{\scriptstyle d}$}
\ArrowLine(35,37.5)(35,67.5)
\Text (31,52.5)[r]{$\mathrm{\scriptstyle d}$}
\Gluon(70,37.5)(35,37.5){3}{4}
\Text (52.5,31.5)[ct]{$\mathrm{\scriptstyle g}$}
\end{picture}
}

\newcommand{\diagramsggveepmumvmxbbxNRoff}{
\begin{picture}(160,125)(-20,0)
\Text (-5.,0)[r]{$\mathrm{\scriptstyle g}$}
\Text (-5.,75)[r]{$\mathrm{\scriptstyle g}$}
\Text (110,0)[l]{$\mathrm{\scriptstyle \bar{b}}$}
\Text (110,15)[l]{$\mathrm{\scriptstyle \nu_e}$}
\Text (110,60)[l]{$\mathrm{\scriptstyle e^+}$}
\Text (110,30)[l]{$\mathrm{\scriptstyle \mu^-}$}
\Text (110,45)[l]{$\mathrm{\scriptstyle \bar{\nu}_\mu}$}
\Text (110,75)[l]{$\mathrm{\scriptstyle b}$}
\Vertex(84,37.5){2.0}
\ArrowLine(84,37.5)(105,30)
\ArrowLine(105,45)(84,37.5)
\Vertex(63,45){2.0}
\ArrowLine(105,60)(63,45)
\Photon(84,37.5)(63,45){2}{3}
\Text (80.,39.)[rt]{$\mathrm{\scriptstyle W^-}$}
\Vertex(42,37.5){2.0}
\ArrowLine(42,37.5)(105,15)
\ArrowLine(63,45)(42,37.5)
\Text (51.1547,45.017)[rb]{$\mathrm{\scriptstyle \bar{\nu}_e}$}
\Vertex(21,0){2.0}
\Gluon(0,0)(21,0){3}{3}
\ArrowLine(105,0)(21,0)
\Vertex(21,75){2.0}
\Gluon(0,75)(21,75){3}{3}
\ArrowLine(21,75)(105,75)
\Vertex(21,37.5){2.0}
\ArrowLine(21,0)(21,37.5)
\Text (17,18.75)[r]{$\mathrm{\scriptstyle b}$}
\ArrowLine(21,37.5)(21,75)
\Text (17,56.25)[r]{$\mathrm{\scriptstyle b}$}
\Photon(42,37.5)(21,37.5){2}{3}
\Text (31.5,33.5)[ct]{$\mathrm{\scriptstyle Z}$}
\end{picture}
}
\newcommand{\diagramsuuxveepmumvmxbbxNRoff}{
\begin{picture}(160,110)(-20,0)
\Text (-5.,0)[r]{$\scriptstyle q$}
\Text (-5.,75)[r]{$\scriptstyle \bar{q}$}
\Text (110,0)[l]{$\mathrm{\scriptstyle b}$}
\Text (110,15)[l]{$\mathrm{\scriptstyle \bar{b}}$}
\Text (110,30)[l]{$\mathrm{\scriptstyle \mu^-}$}
\Text (110,75)[l]{$\mathrm{\scriptstyle \bar{\nu}_\mu}$}
\Text (110,45)[l]{$\mathrm{\scriptstyle \nu_e}$}
\Text (110,60)[l]{$\mathrm{\scriptstyle e^+}$}
\Vertex(84,52.5){2.0}
\ArrowLine(84,52.5)(105,45)
\ArrowLine(105,60)(84,52.5)
\Vertex(63,60){2.0}
\ArrowLine(105,75)(63,60)
\Photon(84,52.5)(63,60){2}{3}
\Text (74.1547,55.517)[rt]{$\mathrm{\scriptstyle W^+}$}
\Vertex(42,52.5){2.0}
\ArrowLine(42,52.5)(105,30)
\ArrowLine(63,60)(42,52.5)
\Text (59.1547,58.017)[rb]{$\mathrm{\scriptstyle \mu^+}$}
\Vertex(84,7.5){2.0}
\ArrowLine(84,7.5)(105,0)
\ArrowLine(105,15)(84,7.5)
\Vertex(21,7.5){2.0}
\ArrowLine(0,0)(21,7.5)
\Gluon(84,7.5)(21,7.5){3}{7}
\Text (52.5,1.5)[ct]{$\mathrm{\scriptstyle g}$}
\Vertex(21,52.5){2.0}
\ArrowLine(21,7.5)(21,52.5)
\Text (17,30)[r]{${\scriptstyle q}$}
\ArrowLine(21,52.5)(0,75)
\Photon(42,52.5)(21,52.5){2}{3}
\Text (31.5,56.5)[cb]{$\mathrm{\scriptstyle Z,\gamma}$}
\end{picture}
}

\newcommand{\diagramsuuxveepmumvmxbbxNRoffb}{
\begin{picture}(160,110)(-20,0)
\Text (-5.,0)[r]{${\scriptstyle q}$}
\Text (-5.,75)[r]{${\scriptstyle \bar{q}}$}
\Text (110,0)[l]{$\mathrm{\scriptstyle b}$}
\Text (110,75)[l]{$\mathrm{\scriptstyle \bar{b}}$}
\Text (110,15)[l]{$\mathrm{\scriptstyle \mu^-}$}
\Text (110,60)[l]{$\mathrm{\scriptstyle \bar{\nu}_\mu}$}
\Text (110,30)[l]{$\mathrm{\scriptstyle \nu_e}$}
\Text (110,45)[l]{$\mathrm{\scriptstyle \mathrm{\scriptstyle e^+}}$}

\Vertex(90,37.5){2.0}
\ArrowLine(90,37.5)(105,30)
\ArrowLine(105,45)(90,37.5)

\Vertex(75,45){2.0}
\ArrowLine(105,60)(75,45)
\Photon(90,37.5)(75,45){2}{2}
\Text (69.4111,40.5277)[lt]{$\mathrm{\scriptstyle W^+}$}

\Vertex(60,37.5){2.0}
\ArrowLine(60,37.5)(105,15)
\ArrowLine(75,45)(60,37.5)
\Text (72.7111,42.8277)[rb]{$\mathrm{\scriptstyle \mu^+}$}

\Vertex(45,45){2.0}
\ArrowLine(105,75)(45,45)
\Photon(60,37.5)(45,45){2}{2}
\Text (55.2111,40.3277)[rt]{$\mathrm{\scriptstyle Z,\gamma}$}

\Vertex(30,37.5){2.0}
\ArrowLine(30,37.5)(105,0)
\ArrowLine(45,45)(30,37.5)
\Text (35.7111,44.8277)[rb]{$\mathrm{\scriptstyle \bar{b}}$}

\Vertex(15,37.5){2.0}
\ArrowLine(0,0)(15,37.5)
\ArrowLine(15,37.5)(0,75)
\Gluon(30,37.5)(15,37.5){3}{2}
\Text (22.5,31.5)[ct]{$\mathrm{\scriptstyle g}$}
\end{picture}
}
\newcommand{\FdecayantitopDRtriangleggveepmumvmxbbx}{
\begin{picture}(155,125)(-20,0)
\Text (-5.,0)[r]{$\mathrm{\scriptstyle g}$}
\Text (-5.,75)[r]{$\mathrm{\scriptstyle g}$}
\Text (110,30)[l]{$\mathrm{\scriptstyle \bar{b}}$}
\Text (110,0)[l]{$\mathrm{\scriptstyle \mu^-}$}
\Text (110,15)[l]{$\mathrm{\scriptstyle \bar{\nu}_\mu}$}
\Text (110,45)[l]{$\mathrm{\scriptstyle b}$}
\Text (110,60)[l]{$\mathrm{\scriptstyle \nu_e}$}
\Text (110,75)[l]{$\mathrm{\scriptstyle e^+}$}
\Vertex(79,67.5){2.0}
\ArrowLine(79,67.5)(105,60)
\ArrowLine(105,75)(79,67.5)
\Vertex(53,60){2.0}
\ArrowLine(53,60)(105,45)
\Photon(79,67.5)(53,60){2}{3}
\Text (74.5,68.)[rb]{$\mathrm{\scriptstyle W^+}$}
\Vertex(79,7.5){2.0}
\ArrowLine(79,7.5)(105,0)
\ArrowLine(105,15)(79,7.5)
\Vertex(53,15){2.0}
\ArrowLine(74.9251,21.2117)(53,15)
\Vertex(70.3867,19.6587){2.0}
\ArrowLine(105,30)(74.9251,21.2117)
\Photon(79,7.5)(53,15){2}{3}
\Text (74.5,7.)[rt]{$\mathrm{\scriptstyle W^-}$}
\Vertex(17,15){2.0}
\Gluon(0,0)(17,15){3}{3}
\ArrowLine(53,15)(35,15)
\Text (44,11)[ct]{$\mathrm{\scriptstyle \bar{t}}$}
\Vertex(35,15){2.0}
\ArrowLine(35,15)(17,15)
\Text (26,11)[ct]{$\mathrm{\scriptstyle \bar{t}}$}
\Vertex(17,60){2.0}
\ArrowLine(17,15)(17,60)
\Text (11,37.5)[r]{$\mathrm{\scriptstyle t}$}
\Gluon(0,75)(17,60){3}{3}
\ArrowLine(17,60)(53,60)
\Text (35,65)[cb]{$\mathrm{\scriptstyle t}$}
\GlueArc(53,15)(18,15,180){3}{6}
\Text (56,19)[b]{$\mathrm{\scriptstyle \bar{b}}$}
\end{picture}
}
\newcommand{\FproductionDRboxggveepmumvmxbbx}{
\begin{picture}(155,125)(-20,0)
\Text (-5.,0)[r]{$\mathrm{\scriptstyle g}$}
\Text (-5.,75)[r]{$\mathrm{\scriptstyle g}$}
\Text (110,30)[l]{$\mathrm{\scriptstyle \bar{b}}$}
\Text (110,0)[l]{$\mathrm{\scriptstyle \mu^-}$}
\Text (110,15)[l]{$\mathrm{\scriptstyle \bar{\nu}_\mu}$}
\Text (110,45)[l]{$\mathrm{\scriptstyle b}$}
\Text (110,60)[l]{$\mathrm{\scriptstyle \nu_e}$}
\Text (110,75)[l]{$\mathrm{\scriptstyle e^+}$}
\Vertex(79,67.5){2.0}
\ArrowLine(79,67.5)(105,60)
\ArrowLine(105,75)(79,67.5)
\Vertex(53,60){2.0}
\ArrowLine(53,60)(105,45)
\Photon(79,67.5)(53,60){2}{3}
\Text (74.5,68.)[rb]{$\mathrm{\scriptstyle W^+}$}
\Vertex(79,7.5){2.0}
\ArrowLine(79,7.5)(105,0)
\ArrowLine(105,15)(79,7.5)
\Vertex(53,15){2.0}
\ArrowLine(105,30)(53,15)
\Photon(79,7.5)(53,15){2}{3}
\Text (74.5,7.)[rt]{$\mathrm{\scriptstyle W^-}$}
\Vertex(17,15){2.0}
\Gluon(0,0)(17,15){3}{3}
\ArrowLine(53,15)(35,15)
\Text (44,11)[ct]{$\mathrm{\scriptstyle \bar{t}}$}
\Vertex(35,15){2.0}
\ArrowLine(35,15)(17,15)
\Text (26,11)[ct]{$\mathrm{\scriptstyle \bar{t}}$}
\Vertex(17,60){2.0}
\ArrowLine(17,15)(17,60)
\Text (13,37.5)[r]{$\mathrm{\scriptstyle t}$}
\Gluon(0,75)(17,60){3}{3}
\ArrowLine(17,60)(35,60)
\Text (26,64)[cb]{$\mathrm{\scriptstyle t}$}
\Vertex(35,60){2.0}
\ArrowLine(35,60)(53,60)
\Text (44,64)[cb]{$\mathrm{\scriptstyle t}$}
\Gluon(35,15)(35,60){3}{5}
\end{picture}
}
\newcommand{\FdecaytopDRtriangleggveepmumvmxbbx}{
\begin{picture}(155,125)(-20,0)
\Text (-5.,0)[r]{$\mathrm{\scriptstyle g}$}
\Text (-5.,75)[r]{$\mathrm{\scriptstyle g}$}
\Text (110,30)[l]{$\mathrm{\scriptstyle \bar{b}}$}
\Text (110,0)[l]{$\mathrm{\scriptstyle \mu^-}$}
\Text (110,15)[l]{$\mathrm{\scriptstyle \bar{\nu}_\mu}$}
\Text (110,45)[l]{$\mathrm{\scriptstyle b}$}
\Text (110,60)[l]{$\mathrm{\scriptstyle \nu_e}$}
\Text (110,75)[l]{$\mathrm{\scriptstyle e^+}$}
\Vertex(79,67.5){2.0}
\ArrowLine(79,67.5)(105,60)
\ArrowLine(105,75)(79,67.5)
\Vertex(53,60){2.0}
\ArrowLine(53,60)(74.9251,53.7883)
\Vertex(70.3867,55.3413){2.0}
\ArrowLine(74.9251,53.7883)(105,45)
\Photon(79,67.5)(53,60){2}{3}
\Text (74,68.)[rb]{$\mathrm{\scriptstyle W^+}$}
\Vertex(79,7.5){2.0}
\ArrowLine(79,7.5)(105,0)
\ArrowLine(105,15)(79,7.5)
\Vertex(53,15){2.0}
\ArrowLine(105,30)(53,15)
\Photon(79,7.5)(53,15){2}{3}
\Text (74,7.)[rt]{$\mathrm{\scriptstyle W^-}$}
\Vertex(17,15){2.0}
\Gluon(0,0)(17,15){3}{3}
\ArrowLine(53,15)(17,15)
\Text (35,11)[ct]{$\mathrm{\scriptstyle \bar{t}}$}
\Vertex(17,60){2.0}
\ArrowLine(17,15)(17,60)
\Text (11,37.5)[r]{$\mathrm{\scriptstyle t}$}
\Gluon(0,75)(17,60){3}{3}
\ArrowLine(17,60)(35,60)
\Text (26,64)[cb]{$\mathrm{\scriptstyle t}$}
\Vertex(35,60){2.0}
\ArrowLine(35,60)(53,60)
\Text (44,64)[cb]{$\mathrm{\scriptstyle t}$}
\GlueArc(53,60)(18,180,345){3}{6}
\Text (58,56)[t]{$\mathrm{\scriptstyle b}$}
\end{picture}
}
\newcommand{\NFDRhexagonggveepmumvmxbbx}{
\begin{picture}(155,125)(-20,0)
\Text (-5.,0)[r]{$\mathrm{\scriptstyle g}$}
\Text (-5.,75)[r]{$\mathrm{\scriptstyle g}$}
\Text (110,30)[l]{$\mathrm{\scriptstyle \bar{b}}$}
\Text (110,0)[l]{$\mathrm{\scriptstyle \mu^-}$}
\Text (110,15)[l]{$\mathrm{\scriptstyle \bar{\nu}_\mu}$}
\Text (110,45)[l]{$\mathrm{\scriptstyle b}$}
\Text (110,60)[l]{$\mathrm{\scriptstyle \nu_e}$}
\Text (110,75)[l]{$\mathrm{\scriptstyle e^+}$}
\Vertex(79,67.5){2.0}
\ArrowLine(79,67.5)(105,60)
\ArrowLine(105,75)(79,67.5)
\Vertex(53,60){2.0}
\ArrowLine(79,52.5)(105,45)
\Vertex(79,52.5){2.0}
\ArrowLine(53,60)(79,52.5)
\Text (65,52.5)[ct]{$\mathrm{\scriptstyle b}$}
\Photon(79,67.5)(53,60){2}{3}
\Text (74.5,68.)[rb]{$\mathrm{\scriptstyle W^+}$}
\Vertex(79,7.5){2.0}
\ArrowLine(79,7.5)(105,0)
\ArrowLine(105,15)(79,7.5)
\Vertex(53,15){2.0}
\ArrowLine(105,30)(79,22.5)
\Vertex(79,22.5){2.0}
\ArrowLine(79,22.5)(53,15)
\Text (65,22.5)[cb]{$\mathrm{\scriptstyle \bar{b}}$}
\Gluon(79,22.5)(79,52.5){3}{4}
\Photon(79,7.5)(53,15){2}{3}
\Text (74.5,7.)[rt]{$\mathrm{\scriptstyle W^-}$}
\Vertex(26,15){2.0}
\Gluon(0,0)(26,15){3}{4}
\ArrowLine(53,15)(26,15)
\Text (39.5,11)[ct]{$\mathrm{\scriptstyle \bar{t}}$}
\Vertex(26,60){2.0}
\ArrowLine(26,15)(26,60)
\Text (22,37.5)[r]{$\mathrm{\scriptstyle t}$}
\Gluon(0,75)(26,60){3}{4}
\ArrowLine(26,60)(53,60)
\Text (39.5,64)[cb]{$\mathrm{\scriptstyle t}$}
\end{picture}
}
\newcommand{\NFDRpentagoninfiggveepmumvmxbbx}{
\begin{picture}(155,125)(-20,0)
\Text (-5.,0)[r]{$\mathrm{\scriptstyle g}$}
\Text (-5.,75)[r]{$\mathrm{\scriptstyle g}$}
\Text (110,0)[l]{$\mathrm{\scriptstyle b}$}
\Text (110,15)[l]{$\mathrm{\scriptstyle \nu_e}$}
\Text (110,30)[l]{$\mathrm{\scriptstyle e^+}$}
\Text (110,75)[l]{$\mathrm{\scriptstyle \bar{b}}$}
\Text (110,45)[l]{$\mathrm{\scriptstyle \mu^-}$}
\Text (110,60)[l]{$\mathrm{\scriptstyle \bar{\nu}_\mu}$}
\ArrowLine(88,52.5)(105,45)
\ArrowLine(105,60)(88,52.5)
\Vertex(88,22.5){2.0}
\ArrowLine(88,22.5)(105,15)
\ArrowLine(105,30)(88,22.5)
\Vertex(71,60){2.0}
\ArrowLine(105,75)(88,67.5)
\Vertex(88,67.5){2.0}
\ArrowLine(88,67.5)(71,60)
\Vertex(88,52.5){2.0}
\Photon(88,52.5)(71,60){2}{2}
\Text (85,52)[rt]{$\scriptstyle \mathrm{W^-}$}
\Vertex(71,15){2.0}
\ArrowLine(71,15)(105,0)
\Photon(88,22.5)(71,15){2}{2}
\Text (85,23)[rb]{$\scriptstyle \mathrm{W^+}$}
\Vertex(54,37.5){2.0}
\ArrowLine(54,37.5)(71,15)
\Text (59.3085,23.8387)[rt]{$\scriptstyle \mathrm{t}$}
\ArrowLine(71,60)(54,37.5)
\Text (59.3085,51.1613)[rb]{$\scriptstyle \mathrm{\bar{t}}$}
\Vertex(17,37.5){2.0}
\Gluon(0,0)(17,37.5){3}{5}
\Gluon(0,75)(8.5,56.25){3}{2}
\Vertex(8.5,56.25){2.0}
\Gluon(8.5,56.25)(17,37.5){3}{2}
\Gluon(54,37.5)(17,37.5){3}{4}
\Text (35.5,31)[ct]{$\scriptstyle \mathrm{g}$}
\GlueArc(57.75,-16.75)(89.5161,70.5,123.5){3}{11}
\end{picture}
}
\newcommand{\NFDRpentagonschannelgluonggveepmumvmxbbx}{
\begin{picture}(155,125)(-20,0)
\Text (-5.,0)[r]{$\mathrm{\scriptstyle g}$}
\Text (-5.,75)[r]{$\mathrm{\scriptstyle g}$}
\Text (110,30)[l]{$\mathrm{\scriptstyle b}$}
\Text (110,0)[l]{$\mathrm{\scriptstyle \nu_e}$}
\Text (110,15)[l]{$\mathrm{\scriptstyle e^+}$}
\Text (110,45)[l]{$\mathrm{\scriptstyle \bar{b}}$}
\Text (110,60)[l]{$\mathrm{\scriptstyle \mu^-}$}
\Text (110,75)[l]{$\mathrm{\scriptstyle \bar{\nu}_\mu}$}
\Vertex(88,67.5){2.0}
\ArrowLine(88,67.5)(105,60)
\ArrowLine(105,75)(88,67.5)
\Vertex(88,7.5){2.0}
\ArrowLine(88,7.5)(105,0)
\ArrowLine(105,15)(88,7.5)
\Vertex(71,60){2.0}
\ArrowLine(105,45)(89,52.5)
\Vertex(89,52.5){2.0}
\ArrowLine(89,52.5)(71,60)
\Photon(88,67.5)(71,60){2}{2}
\Text (84.5,68.)[rb]{$\mathrm{\scriptstyle W^-}$}
\Vertex(71,15){2.0}
\ArrowLine(71,15)(88,22.5)
\Vertex(88,22.5){2.0}
\ArrowLine(88,22.5)(105,30)
\Gluon(88,22.5)(88,52.5){3}{4}
\Photon(88,7.5)(71,15){2}{2}
\Text (84.5,7.)[rt]{$\mathrm{\scriptstyle W^+}$}
\Vertex(54,37.5){2.0}
\ArrowLine(54,37.5)(71,15)
\Text (59.3085,23.8387)[rt]{$\mathrm{\scriptstyle t}$}
\ArrowLine(71,60)(54,37.5)
\Text (59.3085,51.1613)[rb]{$\mathrm{\scriptstyle \bar{t}}$}
\Vertex(17,37){2.0}
\Gluon(0,0)(17,37){3}{5}
\Gluon(0,75)(17,37){3}{5}
\Gluon(54,37.5)(17,37){3}{4}
\Text (35.5,31)[ct]{$\scriptstyle \mathrm{g}$}
\Text (77,52.5)[ct]{$\mathrm{\scriptstyle \bar{b}}$}
\Text (77,22.5)[cb]{$\mathrm{\scriptstyle b}$}
\end{picture}
}
\newcommand{\NFSRhexagonqqxveepmumvmxbbx}{
\begin{picture}(155,125)(-20,0)
\Text (-5.,0)[r]{$\scriptstyle q$}
\Text (-5.,75)[r]{$\scriptstyle \bar{q}$}
\Text (110,0)[l]{$\mathrm{\scriptstyle b}$}
\Text (110,15)[l]{$\mathrm{\scriptstyle \nu_e}$}
\Text (110,30)[l]{$\mathrm{\scriptstyle e^+}$}
\Text (110,75)[l]{$\mathrm{\scriptstyle \bar{b}}$}
\Text (110,45)[l]{$\mathrm{\scriptstyle \mu^-}$}
\Text (110,60)[l]{$\mathrm{\scriptstyle \bar{\nu}_\mu}$}
\Vertex(88,52.5){2.0}
\ArrowLine(88,52.5)(105,45)
\ArrowLine(105,60)(88,52.5)
\Vertex(71,60){2.0}
\ArrowLine(105,75)(88.,67.5)
\Vertex(88.,67.5){2.0}
\ArrowLine(88.,67.5)(71,60)
\Photon(88,52.5)(71,60){2}{2}
\Text (84.5,53.)[rt]{$\mathrm{\scriptstyle W^-}$}
\Vertex(88,22.5){2.0}
\ArrowLine(88,22.5)(105,15)
\ArrowLine(105,30)(88,22.5)
\Vertex(54,47.5){2.0}
\Photon(88,22.5)(54,47.5){2}{5}
\Text (71.,36.)[rt]{$\mathrm{\scriptstyle W^+}$}
\ArrowLine(71,60)(54,47.5)
\Text (60.1304,56.9726)[rb]{$\mathrm{\scriptstyle \bar{t}}$}
\Vertex(37,38){2.0}
\ArrowLine(37,38)(105,0)
\ArrowLine(54,47.5)(37,38)
\Text (43.5487,46.2418)[rb]{$\mathrm{\scriptstyle \bar{b}}$}
\Vertex(17,37){2.0}
\ArrowLine(0,0)(17,37)
\ArrowLine(8.5,58)(0,75)
\Vertex(8.5,58){2.0}
\ArrowLine(17,37)(8.5,58)
\Gluon(37,38)(17,37){3}{2}
\Text (27,31)[ct]{$\mathrm{\scriptstyle g}$}
\GlueArc(57.75,-16.75)(89.5161,70.5,123.5){3}{11}
\Text (91,56)[rb]{$\mathrm{\scriptstyle \bar{b}}$}
\end{picture}
}
\newcommand{\NFNRpentagonggveepmumvmxbbx}{
\begin{picture}(155,125)(-20,0)
\Text (-5.,0)[r]{$\mathrm{\scriptstyle g}$}
\Text (-5.,75)[r]{$\mathrm{\scriptstyle g}$}
\Text (110,0)[l]{$\mathrm{\scriptstyle \nu_e}$}
\Text (110,15)[l]{$\mathrm{\scriptstyle e^+}$}
\Text (110,30)[l]{$\mathrm{\scriptstyle b}$}
\Text (110,45)[l]{$\mathrm{\scriptstyle \bar{b}}$}
\Text (110,60)[l]{$\mathrm{\scriptstyle \mu^-}$}
\Text (110,75)[l]{$\mathrm{\scriptstyle \bar{\nu}_\mu}$}
\Vertex(78.75,67.5){2.0}
\ArrowLine(78.75,67.5)(105,60)
\ArrowLine(105,75)(78.75,67.5)
\Vertex(78.75,37.5){2.0}
\ArrowLine(78.75,37.5)(105,30)
\ArrowLine(105,45)(78.75,37.5)
\Vertex(78.75,7.5){2.0}
\ArrowLine(78.75,7.5)(105,0)
\ArrowLine(105,15)(78.75,7.5)
\Vertex(52.5,7.5){2.0}
\Gluon(0,0)(26.25,3.75){3}{4}
\Vertex(26.25,3.75){2.0}
\ArrowLine(26.25,3.75)(52.5,7.5)
\Text (39.375,1.625)[lt]{$\mathrm{\scriptstyle u}$}
\Photon(78.75,7.5)(52.5,7.5){2}{4}
\Text (68.625,3.5)[ct]{$\mathrm{\scriptstyle W^+}$}
\Vertex(52.5,67.5){2.0}
\ArrowLine(52.5,67.5)(26.25,71.25)
\Text (39.375,73.375)[lb]{$\mathrm{\scriptstyle \bar u}$}
\Vertex(26.25,71.25){2.0}
\Gluon(0,75)(26.25,71.25){3}{4}
\Photon(78.75,67.5)(52.5,67.5){2}{4}
\Text (68.625,71.5)[cb]{$\mathrm{\scriptstyle W^-}$}
\Vertex(52.5,37.5){2.0}
\ArrowLine(52.5,7.5)(52.5,37.5)
\Text (48.5,22.5)[r]{$\mathrm{\scriptstyle d}$}
\ArrowLine(52.5,37.5)(52.5,67.5)
\Text (48.5,52.5)[r]{$\mathrm{\scriptstyle d}$}
\Gluon(78.75,37.5)(52.5,37.5){3}{4}
\Text (65.625,31.5)[ct]{$\mathrm{\scriptstyle g}$}
\ArrowLine(26.25,71.25)(26.25,3.75)
\Text (22.5,37.5)[r]{$\mathrm{\scriptstyle u}$}
\end{picture}
}
\newcommand{\NFNRpentagonqqxveepmumvmxbbx}{
\begin{picture}(155,125)(-20,0)
\Text (-5.,0)[r]{$\mathrm{\scriptstyle u}$}
\Text (-5.,75)[r]{$\mathrm{\scriptstyle \bar{u}}$}
\Text (110,0)[l]{$\mathrm{\scriptstyle \nu_e}$}
\Text (110,15)[l]{$\mathrm{\scriptstyle e^+}$}
\Text (110,30)[l]{$\mathrm{\scriptstyle b}$}
\Text (110,45)[l]{$\mathrm{\scriptstyle \bar{b}}$}
\Text (110,60)[l]{$\mathrm{\scriptstyle \mu^-}$}
\Text (110,75)[l]{$\mathrm{\scriptstyle \bar{\nu}_\mu}$}
\Vertex(78.75,67.5){2.0}
\ArrowLine(78.75,67.5)(105,60)
\ArrowLine(105,75)(78.75,67.5)
\Vertex(78.75,37.5){2.0}
\ArrowLine(78.75,37.5)(105,30)
\ArrowLine(105,45)(78.75,37.5)
\Vertex(78.75,7.5){2.0}
\ArrowLine(78.75,7.5)(105,0)
\ArrowLine(105,15)(78.75,7.5)
\Vertex(52.5,7.5){2.0}
\ArrowLine(0,0)(26.25,3.75)
\Vertex(26.25,3.75){2.0}
\ArrowLine(26.25,3.75)(52.5,7.5)
\Text (39.375,73.375)[lb]{$\mathrm{\scriptstyle \bar u}$}
\Photon(78.75,7.5)(52.5,7.5){2}{4}
\Text (68.625,3.5)[ct]{$\mathrm{\scriptstyle W^+}$}
\Vertex(52.5,67.5){2.0}
\ArrowLine(52.5,67.5)(26.25,71.25)
\Vertex(26.25,71.25){2.0}
\ArrowLine(26.25,71.25)(0,75)
\Text (39.375,1.625)[lt]{$\mathrm{\scriptstyle u}$}
\Photon(78.75,67.5)(52.5,67.5){2}{4}
\Text (68.625,71.5)[cb]{$\mathrm{\scriptstyle W^-}$}
\Vertex(52.5,37.5){2.0}
\ArrowLine(52.5,7.5)(52.5,37.5)
\Text (48.5,22.5)[r]{$\mathrm{\scriptstyle d}$}
\ArrowLine(52.5,37.5)(52.5,67.5)
\Text (48.5,52.5)[r]{$\mathrm{\scriptstyle d}$}
\Gluon(78.75,37.5)(52.5,37.5){3}{4}
\Text (65.625,31.5)[ct]{$\mathrm{\scriptstyle g}$}
\Gluon(26.25,3.75)(26.25,71.25){3}{8}
\end{picture}
}

\newcommand{\diagramsuuxveepmumvmxbbxgDRproda}{
\begin{picture}(170,120)(-20,0)
\Text (-5.,0)[r]{$\scriptstyle q$}
\Text (-5.,90)[r]{$\scriptstyle \bar{q}$}
\Text (125,0)[l]{$\mathrm{\scriptstyle g}$}
\Text (125,15)[l]{$\mathrm{\scriptstyle b}$}
\Text (125,30)[l]{$\mathrm{\scriptstyle \nu_e}$}
\Text (125,45)[l]{$\mathrm{\scriptstyle e^+}$}
\Text (125,90)[l]{$\mathrm{\scriptstyle \bar{b}}$}
\Text (125,60)[l]{$\mathrm{\scriptstyle \mu^-}$}
\Text (125,75)[l]{$\mathrm{\scriptstyle \bar{\nu}_\mu}$}

\Vertex(103,67.5){2.0}
\ArrowLine(103,67.5)(120,60)
\ArrowLine(120,75)(103,67.5)

\Vertex(100,37.5){2.0}
\ArrowLine(100,37.5)(120,30)
\ArrowLine(120,45)(100,37.5)

\Vertex(80,75){2.0}
\ArrowLine(120,90)(80,75)
\Photon(100,67.5)(80,75){2}{3}
\Text (97.5955,67.4953)[rt]{$\mathrm{\scriptstyle W^-}$}

\Vertex(80,30){2.0}
\ArrowLine(80,30)(120,15)
\Photon(100,37.5)(80,30){2}{3}
\Text (97.5955,39.4953)[rb]{$\mathrm{\scriptstyle W^+}$}

\Vertex(60,52.5){2.0}
\ArrowLine(60,52.5)(80,30)
\Text (67.0104,38.5925)[rt]{$\mathrm{\scriptstyle t}$}
\ArrowLine(80,75)(60,52.5)
\Text (67.0104,66.4075)[rb]{$\mathrm{\scriptstyle \bar{t}}$}

\Vertex(20,0){2.0}
\ArrowLine(0,0)(20,0)
\Gluon(120,0)(20,0){3}{11}

\Vertex(20,52.5){2.0}
\ArrowLine(20,0)(20,52.5)
\Text (16,26.25)[r]{${\scriptstyle q}$}
\ArrowLine(20,52.5)(0,90)
\Gluon(60,52.5)(20,52.5){3}{5}
\Text (40,58.5)[cb]{$\mathrm{\scriptstyle g}$}
\end{picture}
}
\newcommand{\diagramsuuxveepmumvmxbbxgDRprodb}{
\begin{picture}(170,120)(-20,0)
\Text (-5.,0)[r]{${\scriptstyle q}$}
\Text (-5.,90)[r]{${\scriptstyle \bar{q}}$}
\Text (125,0)[l]{$\mathrm{\scriptstyle g}$}
\Text (125,15)[l]{$\mathrm{\scriptstyle b}$}
\Text (125,30)[l]{$\mathrm{\scriptstyle \nu_e}$}
\Text (125,45)[l]{$\mathrm{\scriptstyle e^+}$}
\Text (125,90)[l]{$\mathrm{\scriptstyle \bar{b}}$}
\Text (125,60)[l]{$\mathrm{\scriptstyle \mu^-}$}
\Text (125,75)[l]{$\mathrm{\scriptstyle \bar{\nu}_\mu}$}
\Vertex(103,67.5){2.0}
\ArrowLine(103,67.5)(120,60)
\ArrowLine(120,75)(103,67.5)
\Vertex(103,37.5){2.0}
\ArrowLine(103,37.5)(120,30)
\ArrowLine(120,45)(103,37.5)
\Vertex(86,75){2.0}
\ArrowLine(120,90)(86,75)
\Photon(103,67.5)(86,75){2}{2}
\Text (99.8854,66.4097)[rt]{$\mathrm{\scriptstyle W^-}$}
\Vertex(86,30){2.0}
\ArrowLine(86,30)(120,15)
\Photon(103,37.5)(86,30){2}{2}
\Text (101.8854,38.4097)[rb]{$\mathrm{\scriptstyle W^+}$}
\Vertex(69,52.5){2.0}
\ArrowLine(69,52.5)(86,30)
\Text (74.3085,38.8387)[rt]{$\mathrm{\scriptstyle t}$}
\ArrowLine(86,75)(69,52.5)
\Text (74.3085,66.1613)[rb]{$\mathrm{\scriptstyle \bar{t}}$}
\Vertex(52,32){2.0}
\Gluon(120,0)(52,32){3}{8}
\Gluon(69,52.5)(52,32){3}{2}
\Text (58.398,50.6532)[rb]{$\mathrm{\scriptstyle g}$}
\Vertex(17,45){2.0}
\ArrowLine(0,0)(17,45)
\ArrowLine(17,45)(0,90)
\Gluon(52,32)(17,45){3}{4}
\Text (34.1584,33)[rt]{$\mathrm{\scriptstyle g}$}
\end{picture}
}
\newcommand{\diagramsggveepmumvmxbbxgDRprodb}{
\begin{picture}(170,120)(-20,0)
\Text (-5.,0)[r]{$\mathrm{\scriptstyle g}$}
\Text (-5.,90)[r]{$\mathrm{\scriptstyle g}$}
\Text (125,0)[l]{$\mathrm{\scriptstyle g}$}
\Text (125,15)[l]{$\mathrm{\scriptstyle b}$}
\Text (125,30)[l]{$\mathrm{\scriptstyle \nu_e}$}
\Text (125,45)[l]{$\mathrm{\scriptstyle e^+}$}
\Text (125,90)[l]{$\mathrm{\scriptstyle \bar{b}}$}
\Text (125,60)[l]{$\mathrm{\scriptstyle \mu^-}$}
\Text (125,75)[l]{$\mathrm{\scriptstyle \bar{\nu}_\mu}$}

\Vertex(103,67.5){2.0}
\ArrowLine(103,67.5)(120,60)
\ArrowLine(120,75)(103,67.5)

\Vertex(103,37.5){2.0}
\ArrowLine(103,37.5)(120,30)
\ArrowLine(120,45)(103,37.5)

\Vertex(86,75){2.0}
\ArrowLine(120,90)(86,75)
\Photon(103,67.5)(86,75){2}{2}
\Text (99.8854,66.4097)[rt]{$\mathrm{\scriptstyle W^-}$}

\Vertex(86,30){2.0}
\ArrowLine(86,30)(120,15)
\Photon(103,37.5)(86,30){2}{2}
\Text (101.8854,38.4097)[rb]{$\mathrm{\scriptstyle W^+}$}

\Vertex(69,52.5){2.0}
\ArrowLine(69,52.5)(86,30)
\Text (74.3085,38.8387)[rt]{$\mathrm{\scriptstyle t}$}
\ArrowLine(86,75)(69,52.5)
\Text (74.3085,66.1613)[rb]{$\mathrm{\scriptstyle \bar{t}}$}

\Vertex(52,32){2.0}
\Gluon(120,0)(52,32){3}{8}
\Gluon(69,52.5)(52,32){3}{2}
\Text (56.398,45.6532)[rb]{$\mathrm{\scriptstyle g}$}

\Vertex(17,45){2.0}
\Gluon(0,0)(17,45){3}{5}
\Gluon(0,90)(17,45){3}{5}
\Gluon(52,32)(17,45){3}{4}
\Text (34.1584,32.5146)[rt]{$\mathrm{\scriptstyle g}$}
\end{picture}
}

\newcommand{\diagramsuuxveepmumvmxbbxgDRinbetween}{
\begin{picture}(170,120)(-20,0)
\Text (-5.,0)[r]{${\scriptstyle q}$}
\Text (-5.,90)[r]{${\scriptstyle \bar{q}}$}
\Text (125,0)[l]{$\mathrm{\scriptstyle \bar{b}}$}
\Text (125,15)[l]{$\mathrm{\scriptstyle \mu^-}$}
\Text (125,30)[l]{$\mathrm{\scriptstyle \bar{\nu}_\mu}$}
\Text (125,45)[l]{$\mathrm{\scriptstyle g}$}
\Text (125,90)[l]{$\mathrm{\scriptstyle b}$}
\Text (125,60)[l]{$\mathrm{\scriptstyle \nu_e}$}
\Text (125,75)[l]{$\mathrm{\scriptstyle e^+}$}

\Vertex(103,67.5){2.0}
\ArrowLine(103,67.5)(120,60)
\ArrowLine(120,75)(103,67.5)

\Vertex(86,75){2.0}
\ArrowLine(86,75)(120,90)
\Photon(103,67.5)(86,75){2}{2}
\Text (97.8854,72.4097)[lb]{$\mathrm{\scriptstyle W^+}$}

\Vertex(103,22.5){2.0}
\ArrowLine(103,22.5)(120,15)
\ArrowLine(120,30)(103,22.5)

\Vertex(69,67.5){2.0}
\Gluon(120,45)(69,67.5){3}{6}
\ArrowLine(69,67.5)(86,75)
\Text (75.8854,74.9097)[rb]{$\mathrm{\scriptstyle t}$}

\Vertex(86,15){2.0}
\ArrowLine(120,0)(86,15)
\Photon(103,22.5)(86,15){2}{2}
\Text (98.8854,24.4097)[rb]{$\mathrm{\scriptstyle W^-}$}

\Vertex(52,50){2.0}
\ArrowLine(86,15)(52,50)
\Text (66.1309,29.7129)[rt]{$\mathrm{\scriptstyle \bar{t}}$}
\ArrowLine(52,50)(69,67.5)
\Text (57.6309,61.5371)[rb]{$\mathrm{\scriptstyle t}$}

\Vertex(17,45){2.0}
\ArrowLine(0,0)(17,45)
\ArrowLine(17,45)(0,90)
\Gluon(52,50)(17,45){3}{4}
\Text (33.9343,53.4598)[rb]{$\mathrm{\scriptstyle g}$}
\end{picture}
}
\newcommand{\diagramsuuxveepmumvmxbbxgDRdecay}{
\begin{picture}(170,120)(-20,0)
\Text (-5.,0)[r]{${\scriptstyle q}$}
\Text (-5.,90)[r]{${\scriptstyle \bar{q}}$}
\Text (125,0)[l]{$\mathrm{\scriptstyle \bar{b}}$}
\Text (125,15)[l]{$\mathrm{\scriptstyle \mu^-}$}
\Text (125,30)[l]{$\mathrm{\scriptstyle \bar{\nu}_\mu}$}
\Text (125,45)[l]{$\mathrm{\scriptstyle \nu_e}$}
\Text (125,60)[l]{$\mathrm{\scriptstyle e^+}$}
\Text (125,90)[l]{$\mathrm{\scriptstyle b}$}
\Text (125,75)[l]{$\mathrm{\scriptstyle g}$}

\Vertex(100,82.5){2.0}
\ArrowLine(100,82.5)(120,90)
\Gluon(120,75)(100,82.5){3}{2}

\Vertex(100,52.5){2.0}
\ArrowLine(100,52.5)(120,45)
\ArrowLine(120,60)(100,52.5)

\Vertex(100,22.5){2.0}
\ArrowLine(100,22.5)(120,15)
\ArrowLine(120,30)(100,22.5)

\Vertex(80,67.5){2.0}
\Photon(100,52.5)(80,67.5){2}{3}
\Text (95.6,54.8)[rt]{$\mathrm{\scriptstyle W^+}$}
\ArrowLine(80,67.5)(100,82.5)
\Text (87.6,78.2)[rb]{$\mathrm{\scriptstyle b}$}

\Vertex(80,15){2.0}
\ArrowLine(120,0)(80,15)
\Photon(100,22.5)(80,15){2}{3}
\Text (98.5955,22.4953)[rb]{$\mathrm{\scriptstyle W^-}$}

\Vertex(60,41.25){2.0}
\ArrowLine(80,15)(60,41.25)
\Text (66.8183,25.7008)[rt]{$\mathrm{\scriptstyle \bar{t}}$}
\ArrowLine(60,41.25)(80,67.5)
\Text (66.8183,56.7992)[rb]{$\mathrm{\scriptstyle t}$}

\Vertex(20,45){2.0}
\ArrowLine(0,0)(20,45)
\ArrowLine(20,45)(0,90)
\Gluon(60,41.25)(20,45){3}{5}
\Text (39.6266,37.1425)[rt]{$\mathrm{\scriptstyle g}$}
\end{picture}
}

\newcommand{\diagramsggveepmumvmxbbxgDRproda}{
\begin{picture}(170,120)(-20,0)
\Text (-5.,0)[r]{$\mathrm{\scriptstyle g}$}
\Text (-5.,90)[r]{$\mathrm{\scriptstyle g}$}
\Text (125,0)[l]{$\mathrm{\scriptstyle b}$}
\Text (125,15)[l]{$\mathrm{\scriptstyle \nu_e}$}
\Text (125,30)[l]{$\mathrm{\scriptstyle e^+}$}
\Text (125,75)[l]{$\mathrm{\scriptstyle \bar{b}}$}
\Text (125,45)[l]{$\mathrm{\scriptstyle \mu^-}$}
\Text (125,60)[l]{$\mathrm{\scriptstyle \bar{\nu}_\mu}$}
\Text (125,90)[l]{$\mathrm{\scriptstyle g}$}

\Vertex(100,52.5){2.0}
\ArrowLine(100,52.5)(120,45)
\ArrowLine(120,60)(100,52.5)

\Vertex(100,22.5){2.0}
\ArrowLine(100,22.5)(120,15)
\ArrowLine(120,30)(100,22.5)

\Vertex(80,60){2.0}
\ArrowLine(120,75)(80,60)
\Photon(100,52.5)(80,60){2}{3}
\Text (97.5955,52.4953)[rt]{$\mathrm{\scriptstyle W^-}$}

\Vertex(80,15){2.0}
\ArrowLine(80,15)(120,0)
\Photon(100,22.5)(80,15){2}{3}
\Text (97.5955,22.4953)[rb]{$\mathrm{\scriptstyle W^+}$}

\Vertex(60,37.5){2.0}
\ArrowLine(60,37.5)(80,15)
\Text (67.0104,23.5925)[rt]{$\mathrm{\scriptstyle t}$}
\ArrowLine(80,60)(60,37.5)
\Text (67.0104,51.4075)[rb]{$\mathrm{\scriptstyle \bar{t}}$}

\Vertex(20,90){2.0}
\Gluon(0,90)(20,90){3}{2}
\Gluon(120,90)(20,90){3}{11}

\Vertex(20,37.5){2.0}
\Gluon(0,0)(20,37.5){3}{5}
\Gluon(20,90)(20,37.5){3}{6}
\Text (14,63.75)[r]{$\mathrm{\scriptstyle g}$}
\Gluon(60,37.5)(20,37.5){3}{5}
\Text (40,31.5)[ct]{$\mathrm{\scriptstyle g}$}
\end{picture}
}

\newcommand{\diagramsuuxveepmumvmxbbxgNRCXCVI}{
\begin{picture}(170,120)(-20,0)
\Text (-5.,0)[r]{${\scriptstyle q}$}
\Text (-5.,90)[r]{${\scriptstyle \bar{q}}$}
\Text (125,0)[l]{$\mathrm{\scriptstyle \nu_e}$}
\Text (125,15)[l]{$\mathrm{\scriptstyle e^+}$}
\Text (125,30)[l]{$\mathrm{\scriptstyle \mu^-}$}
\Text (125,45)[l]{$\mathrm{\scriptstyle \bar{\nu}_\mu}$}
\Text (125,60)[l]{$\mathrm{\scriptstyle g}$}
\Text (125,75)[l]{$\mathrm{\scriptstyle b}$}
\Text (125,90)[l]{$\mathrm{\scriptstyle \bar{b}}$}
\Vertex(90,82.5){2.0}
\ArrowLine(90,82.5)(120,75)
\ArrowLine(120,90)(90,82.5)
\Vertex(90,37.5){2.0}
\ArrowLine(90,37.5)(120,30)
\ArrowLine(120,45)(90,37.5)
\Vertex(90,7.5){2.0}
\ArrowLine(90,7.5)(120,0)
\ArrowLine(120,15)(90,7.5)
\Vertex(60,22.5){2.0}
\Photon(90,7.5)(60,22.5){2}{4}
\Text (79.2111,10.4223)[rt]{$\mathrm{\scriptstyle W^+}$}
\Photon(90,37.5)(60,22.5){2}{4}
\Text (79.2111,32.5777)[rb]{$\mathrm{\scriptstyle W^-}$}
\Vertex(30,82.5){2.0}
\ArrowLine(30,82.5)(0,90)
\Gluon(90,82.5)(30,82.5){3}{7}
\Text (60,89.5)[cb]{$\mathrm{\scriptstyle g}$}
\Vertex(30,60){2.0}
\ArrowLine(30,60)(30,82.5)
\Text (26,71.25)[r]{${\scriptstyle q}$}
\Gluon(120,60)(30,60){3}{10}
\Vertex(30,22.5){2.0}
\ArrowLine(0,0)(30,22.5)
\ArrowLine(30,22.5)(30,60)
\Text (26,41.25)[r]{${\scriptstyle q}$}
\Photon(60,22.5)(30,22.5){2}{4}
\Text (45,18.5)[ct]{$\mathrm{\scriptstyle Z,\gamma}$}
\end{picture}
}
\newcommand{\diagramsuuxveepmumvmxbbxgNRoffCXCVIII}{
\begin{picture}(170,120)(-20,0)
\Text (-5.,0)[r]{${\scriptstyle q}$}
\Text (-5.,90)[r]{${\scriptstyle \bar{q}}$}
\Text (125,0)[l]{$\mathrm{\scriptstyle g}$}
\Text (125,15)[l]{$\mathrm{\scriptstyle b}$}
\Text (125,30)[l]{$\mathrm{\scriptstyle \bar{b}}$}
\Text (125,45)[l]{$\mathrm{\scriptstyle \nu_e}$}
\Text (125,90)[l]{$\mathrm{\scriptstyle e^+}$}
\Text (125,60)[l]{$\mathrm{\scriptstyle \mu^-}$}
\Text (125,75)[l]{$\mathrm{\scriptstyle \bar{\nu}_\mu}$}
\Vertex(96,67.5){2.0}
\ArrowLine(96,67.5)(120,60)
\ArrowLine(120,75)(96,67.5)
\Vertex(72,75){2.0}
\ArrowLine(120,90)(72,75)
\Photon(96,67.5)(72,75){2}{3}
\Text (90.8069,68.5679)[rt]{$\mathrm{\scriptstyle W^-}$}
\Vertex(48,67.5){2.0}
\ArrowLine(48,67.5)(120,45)
\ArrowLine(72,75)(48,67.5)
\Text (63.8069,76.0679)[rb]{$\mathrm{\scriptstyle \bar{\nu}_e}$}
\Vertex(96,22.5){2.0}
\ArrowLine(96,22.5)(120,15)
\ArrowLine(120,30)(96,22.5)
\Vertex(24,0){2.0}
\ArrowLine(0,0)(24,0)
\Gluon(120,0)(24,0){3}{10}
\Vertex(24,22.5){2.0}
\ArrowLine(24,0)(24,22.5)
\Text (20,11.25)[r]{${\scriptstyle q}$}
\Gluon(96,22.5)(24,22.5){3}{8}
\Text (60,16.5)[ct]{$\mathrm{\scriptstyle g}$}
\Vertex(24,67.5){2.0}
\ArrowLine(24,22.5)(24,67.5)
\Text (20,45)[r]{${\scriptstyle q}$}
\ArrowLine(24,67.5)(0,90)
\Photon(48,67.5)(24,67.5){2}{3}
\Text (36,73.5)[cb]{$\mathrm{\scriptstyle Z}$}
\end{picture}
}
\newcommand{\diagramsuuxveepmumvmxbbxgSRprod}{
\begin{picture}(170,120)(-20,0)
\Text (-5.,0)[r]{${\scriptstyle q}$}
\Text (-5.,90)[r]{${\scriptstyle \bar{q}}$}
\Text (125,0)[l]{$\mathrm{\scriptstyle g}$}
\Text (125,15)[l]{$\mathrm{\scriptstyle \bar{b}}$}
\Text (125,30)[l]{$\mathrm{\scriptstyle \mu^-}$}
\Text (125,45)[l]{$\mathrm{\scriptstyle \bar{\nu}_\mu}$}
\Text (125,90)[l]{$\mathrm{\scriptstyle b}$}
\Text (125,60)[l]{$\mathrm{\scriptstyle \nu_e}$}
\Text (125,75)[l]{$\mathrm{\scriptstyle e^+}$}

\Vertex(103,67.5){2.0}
\ArrowLine(103,67.5)(120,60)
\ArrowLine(120,75)(103,67.5)

\Vertex(86,75){2.0}
\ArrowLine(86,75)(120,90)
\Photon(103,67.5)(86,75){2}{2}
\Text (98.8854,68.4097)[rt]{$\mathrm{\scriptstyle W^+}$}

\Vertex(103,37.5){2.0}
\ArrowLine(103,37.5)(120,30)
\ArrowLine(120,45)(103,37.5)

\Vertex(69,62.5){2.0}
\Photon(103,37.5)(69,62.5){2}{5}
\Text (93.6304,46.7774)[lb]{$\mathrm{\scriptstyle W^-}$}
\ArrowLine(69,62.5)(86,75)
\Text (75.1304,71.9726)[rb]{$\mathrm{\scriptstyle t}$}

\Vertex(52,53){2.0}
\ArrowLine(120,15)(52,53)
\ArrowLine(52,53)(69,62.5)
\Text (58.5487,61.2418)[rb]{$\mathrm{\scriptstyle b}$}

\Vertex(35,44.1667){2.0}
\Gluon(120,0)(35,44.1667){3}{10}
\Gluon(52,53)(35,44.1667){3}{2}
\Text (41.6557,54.1328)[rb]{$\mathrm{\scriptstyle g}$}

\Vertex(17,45){2.0}
\ArrowLine(0,0)(17,45)
\ArrowLine(17,45)(0,90)
\Gluon(35,44.1667)(17,45){3}{2}
\Text (25.815,38.5876)[rt]{$\mathrm{\scriptstyle g}$}
\end{picture}
}
\newcommand{\diagramsuuxveepmumvmxbbxgSRdecay}{
\begin{picture}(170,120)(-20,0)
\Text (-5.,0)[r]{${\scriptstyle q}$}
\Text (-5.,90)[r]{${\scriptstyle \bar{q}}$}
\Text (125,0)[l]{$\mathrm{\scriptstyle b}$}
\Text (125,15)[l]{$\mathrm{\scriptstyle \nu_e}$}
\Text (125,30)[l]{$\mathrm{\scriptstyle e^+}$}
\Text (125,45)[l]{$\mathrm{\scriptstyle \mu^-}$}
\Text (125,60)[l]{$\mathrm{\scriptstyle \bar{\nu}_\mu}$}
\Text (125,90)[l]{$\mathrm{\scriptstyle \bar{b}}$}
\Text (125,75)[l]{$\mathrm{\scriptstyle g}$}
\Vertex(100,82.5){2.0}
\ArrowLine(120,90)(100,82.5)
\Gluon(120,75)(100,82.5){3}{3}
\Vertex(100,52.5){2.0}
\ArrowLine(100,52.5)(120,45)
\ArrowLine(120,60)(100,52.5)
\Vertex(80,67.5){2.0}
\Photon(100,52.5)(80,67.5){2}{3}
\Text (92.6,55.8)[rt]{$\mathrm{\scriptstyle W^-}$}
\ArrowLine(100,82.5)(80,67.5)
\Text (87.6,78.2)[rb]{$\mathrm{\scriptstyle \bar{b}}$}
\Vertex(100,22.5){2.0}
\ArrowLine(100,22.5)(120,15)
\ArrowLine(120,30)(100,22.5)
\Vertex(60,52.5){2.0}
\Photon(100,22.5)(60,52.5){2}{6}
\Text (75.6,39.3)[rt]{$\mathrm{\scriptstyle W^+}$}
\ArrowLine(80,67.5)(60,52.5)
\Text (67.6,63.2)[rb]{$\mathrm{\scriptstyle \bar{t}}$}
\Vertex(40,37){2.0}
\ArrowLine(40,37)(120,0)
\ArrowLine(60,52.5)(40,37)
\Text (48.1407,50.7916)[rb]{$\mathrm{\scriptstyle \bar{b}}$}
\Vertex(20,45){2.0}
\ArrowLine(0,0)(20,45)
\ArrowLine(20,45)(0,90)
\Gluon(40,37)(20,45){3}{3}
\Text (29.4066,35.5443)[rt]{$\mathrm{\scriptstyle g}$}
\end{picture}
}
\newcommand{\diagramsggveepmumvmxbbxgSR}{
\begin{picture}(170,120)(-20,0)
\Text (-5.,0)[r]{$\mathrm{\scriptstyle g}$}
\Text (-5.,90)[r]{$\mathrm{\scriptstyle g}$}
\Text (125,0)[l]{$\mathrm{\scriptstyle g}$}
\Text (125,15)[l]{$\mathrm{\scriptstyle \mu^-}$}
\Text (125,30)[l]{$\mathrm{\scriptstyle \bar{\nu}_\mu}$}
\Text (125,75)[l]{$\mathrm{\scriptstyle b}$}
\Text (125,45)[l]{$\mathrm{\scriptstyle \nu_e}$}
\Text (125,60)[l]{$\mathrm{\scriptstyle e^+}$}
\Text (125,90)[l]{$\mathrm{\scriptstyle \bar{b}}$}

\Vertex(100,52.5){2.0}
\ArrowLine(100,52.5)(120,45)
\ArrowLine(120,60)(100,52.5)

\Vertex(80,60){2.0}
\ArrowLine(80,60)(120,75)
\Photon(100,52.5)(80,60){2}{3}
\Text (92.5955,52.4953)[rt]{$\mathrm{\scriptstyle W^+}$}

\Vertex(100,22.5){2.0}
\ArrowLine(100,22.5)(120,15)
\ArrowLine(120,30)(100,22.5)

\Vertex(60,47.5){2.0}
\Photon(100,22.5)(60,47.5){2}{5}
\Text (89.88,32.608)[lb]{$\mathrm{\scriptstyle W^-}$}
\ArrowLine(60,47.5)(80,60)
\Text (67.88,57.142)[rb]{$\mathrm{\scriptstyle t}$}

\Vertex(40,38){2.0}
\Gluon(120,0)(40,38){3}{9}
\ArrowLine(40,38)(60,47.5)
\Text (48.2838,46.3631)[rb]{$\mathrm{\scriptstyle b}$}

\Vertex(20,90){2.0}
\Gluon(0,90)(20,90){3}{3}
\ArrowLine(120,90)(20,90)

\Vertex(20,38){2.0}
\Gluon(0,0)(20,38){3}{5}
\ArrowLine(20,90)(20,38)
\Text (16,64)[r]{$\mathrm{\scriptstyle b}$}
\ArrowLine(20,38)(40,38)
\Text (30,34)[ct]{$\mathrm{\scriptstyle b}$}
\end{picture}
}

\newcommand{\diagramsguveepmumvmxbbxuNR}{
\begin{picture}(170,120)(-20,0)
\Text (-5.,0)[r]{$\mathrm{\scriptstyle g}$}
\Text (-5.,90)[r]{${\scriptstyle q}$}
\Text (125,0)[l]{$\mathrm{\scriptstyle b}$}
\Text (125,15)[l]{$\mathrm{\scriptstyle \bar{b}}$}
\Text (125,30)[l]{${\scriptstyle q}$}
\Text (125,45)[l]{$\mathrm{\scriptstyle \nu_e}$}
\Text (125,60)[l]{$\mathrm{\scriptstyle e^+}$}
\Text (125,75)[l]{$\mathrm{\scriptstyle \mu^-}$}
\Text (125,90)[l]{$\mathrm{\scriptstyle \bar{\nu}_\mu}$}
\Vertex(90,82.5){2.0}
\ArrowLine(90,82.5)(120,75)
\ArrowLine(120,90)(90,82.5)
\Vertex(90,52.5){2.0}
\ArrowLine(90,52.5)(120,45)
\ArrowLine(120,60)(90,52.5)
\Vertex(60,67.5){2.0}
\Photon(90,52.5)(60,67.5){2}{4}
\Text (79.2111,56.4223)[rt]{$\mathrm{\scriptstyle W^+}$}
\Photon(90,82.5)(60,67.5){2}{4}
\Text (79.2111,78.5777)[rb]{$\mathrm{\scriptstyle W^-}$}
\Vertex(90,7.5){2.0}
\ArrowLine(90,7.5)(120,0)
\ArrowLine(120,15)(90,7.5)
\Vertex(30,7.5){2.0}
\Gluon(0,0)(30,7.5){3}{4}
\Gluon(90,7.5)(30,7.5){3}{7}
\Text (60,1.5)[ct]{$\mathrm{\scriptstyle g}$}
\Vertex(30,30){2.0}
\Gluon(30,7.5)(30,30){3}{3}
\Text (23,18.75)[r]{$\mathrm{\scriptstyle g}$}
\ArrowLine(30,30)(120,30)
\Vertex(30,67.5){2.0}
\ArrowLine(30,67.5)(30,30)
\Text (25,48.75)[r]{${\scriptstyle q}$}
\ArrowLine(0,90)(30,67.5)
\Photon(60,67.5)(30,67.5){2}{4}
\Text (45,72.5)[cb]{$\mathrm{\scriptstyle Z,\gamma}$}
\end{picture}
}
\newcommand{\diagramsggveepmumvmxbbxgNR}{
\begin{picture}(170,120)(-20,0)
\Text (-5.,0)[r]{$\mathrm{\scriptstyle g}$}
\Text (-5.,90)[r]{$\mathrm{\scriptstyle g}$}
\Text (125,0)[l]{$\mathrm{\scriptstyle b}$}
\Text (125,75)[l]{$\mathrm{\scriptstyle \bar{b}}$}
\Text (125,15)[l]{$\mathrm{\scriptstyle \nu_e}$}
\Text (125,30)[l]{$\mathrm{\scriptstyle e^+}$}
\Text (125,45)[l]{$\mathrm{\scriptstyle \mu^-}$}
\Text (125,60)[l]{$\mathrm{\scriptstyle \bar{\nu}_\mu}$}
\Text (125,90)[l]{$\mathrm{\scriptstyle g}$}

\Vertex(100,52.5){2.0}
\ArrowLine(100,52.5)(120,45)
\ArrowLine(120,60)(100,52.5)

\Vertex(100,22.5){2.0}
\ArrowLine(100,22.5)(120,15)
\ArrowLine(120,30)(100,22.5)

\Vertex(80,37.5){2.0}
\Photon(100,22.5)(80,37.5){2}{3}
\Text (90.6,30.8)[rt]{$\mathrm{\scriptstyle W^+}$}
\Photon(100,52.5)(80,37.5){2}{3}
\Text (92.6,43.2)[rb]{$\mathrm{\scriptstyle W^-}$}

\Vertex(60,43.125){2.0}
\ArrowLine(120,75)(60,43.125)
\Photon(80,37.5)(60,43.125){2}{3}
\Text (69.3023,38.4343)[rt]{$\mathrm{\scriptstyle Z,\gamma}$}

\Vertex(40,34.5){2.0}
\ArrowLine(40,34.5)(120,0)
\ArrowLine(60,43.125)(40,34.5)
\Text (48.416,42.4855)[rb]{$\mathrm{\scriptstyle \bar{b}}$}

\Vertex(20,90){2.0}
\Gluon(0,90)(20,90){3}{2}
\Gluon(120,90)(20,90){3}{11}

\Vertex(20,34.5){2.0}
\Gluon(0,0)(20,34.5){3}{4}
\Gluon(20,90)(20,34.5){3}{6}
\Text (14,62.25)[r]{$\mathrm{\scriptstyle g}$}
\Gluon(40,34.5)(20,34.5){3}{2}
\Text (30,28.5)[ct]{$\mathrm{\scriptstyle g}$}
\end{picture}
}

\newcommand{\diagramsguveepmumvmxbbxuNRoff}{
\begin{picture}(170,120)(-20,0)
\Text (-5.,0)[r]{$\mathrm{\scriptstyle g}$}
\Text (-5.,90)[r]{${\scriptstyle q}$}
\Text (125,0)[l]{$\mathrm{\scriptstyle b}$}
\Text (125,15)[l]{$\mathrm{\scriptstyle \bar{b}}$}
\Text (125,30)[l]{${\scriptstyle q}$}
\Text (125,45)[l]{$\mathrm{\scriptstyle \bar{\nu}_\mu}$}
\Text (125,90)[l]{$\mathrm{\scriptstyle \mu^-}$}
\Text (125,60)[l]{$\mathrm{\scriptstyle \nu_e}$}
\Text (125,75)[l]{$\mathrm{\scriptstyle e^+}$}
\Vertex(96,67.5){2.0}
\ArrowLine(96,67.5)(120,60)
\ArrowLine(120,75)(96,67.5)
\Vertex(72,75){2.0}
\ArrowLine(72,75)(120,90)
\Photon(96,67.5)(72,75){2}{3}
\Text (84.8069,70.5679)[rt]{$\mathrm{\scriptstyle W^+}$}
\Vertex(48,67.5){2.0}
\ArrowLine(120,45)(48,67.5)
\ArrowLine(48,67.5)(72,75)
\Text (62.8069,75.0679)[rb]{$\mathrm{\scriptstyle \nu_\mu}$}
\Vertex(96,7.5){2.0}
\ArrowLine(96,7.5)(120,0)
\ArrowLine(120,15)(96,7.5)
\Vertex(24,7.5){2.0}
\Gluon(0,0)(24,7.5){3}{3}
\Gluon(96,7.5)(24,7.5){3}{8}
\Text (60,1.5)[ct]{$\mathrm{\scriptstyle g}$}
\Vertex(24,30){2.0}
\Gluon(24,7.5)(24,30){3}{3}
\Text (18,18.75)[r]{$\mathrm{\scriptstyle g}$}
\ArrowLine(24,30)(120,30)
\Vertex(24,67.5){2.0}
\ArrowLine(24,67.5)(24,30)
\Text (20,48.75)[r]{${\scriptstyle q}$}
\ArrowLine(0,90)(24,67.5)
\Photon(48,67.5)(24,67.5){2}{3}
\Text (36,73.5)[cb]{$\mathrm{\scriptstyle Z}$}
\end{picture}
}
\newcommand{\diagramsggveepmumvmxbbxgNRoff}{
\begin{picture}(170,120)(-20,0)
\Text (-5.,0)[r]{$\mathrm{\scriptstyle g}$}
\Text (-5.,90)[r]{$\mathrm{\scriptstyle g}$}
\Text (125,0)[l]{$\mathrm{\scriptstyle b}$}
\Text (125,75)[l]{$\mathrm{\scriptstyle \bar{b}}$}
\Text (125,15)[l]{$\mathrm{\scriptstyle \mu^-}$}
\Text (125,60)[l]{$\mathrm{\scriptstyle \bar{\nu}_\mu}$}
\Text (125,30)[l]{$\mathrm{\scriptstyle \nu_e}$}
\Text (125,45)[l]{$\mathrm{\scriptstyle e^+}$}
\Text (125,90)[l]{$\mathrm{\scriptstyle g}$}

\Vertex(103,37.5){2.0}
\ArrowLine(103,37.5)(120,30)
\ArrowLine(120,45)(103,37.5)

\Vertex(86,45){2.0}
\ArrowLine(120,60)(86,45)
\Photon(103,37.5)(86,45){2}{2}
\Text (99.8854,39.4097)[rt]{$\mathrm{\scriptstyle W^+}$}

\Vertex(69,37.5){2.0}
\ArrowLine(69,37.5)(120,15)
\ArrowLine(86,45)(69,37.5)
\Text (83.8854,42.9097)[rb]{$\mathrm{\scriptstyle \mu^+}$}

\Vertex(52,45){2.0}
\ArrowLine(120,75)(52,45)
\Photon(69,37.5)(52,45){2}{2}
\Text (62.8854,39.4097)[rt]{$\mathrm{\scriptstyle Z,\gamma}$}

\Vertex(35,37.5){2.0}
\ArrowLine(35,37.5)(120,0)
\ArrowLine(52,45)(35,37.5)
\Text (41.8854,44.9097)[rb]{$\mathrm{\scriptstyle \bar{b}}$}

\Vertex(17,90){2.0}
\Gluon(0,90)(17,90){3}{2}
\Gluon(120,90)(17,90){3}{11}

\Vertex(17,37.5){2.0}
\Gluon(0,0)(17,37.5){3}{5}
\Gluon(17,90)(17,37.5){3}{6}
\Text (11,63.75)[r]{$\mathrm{\scriptstyle g}$}
\Gluon(35,37.5)(17,37.5){3}{2}
\Text (26,31.5)[ct]{$\mathrm{\scriptstyle g}$}

\end{picture}
}

\begin{document}
\enlargethispage{2cm}
\thispagestyle{empty}
\def\thefootnote{\fnsymbol{footnote}}
\setcounter{footnote}{1}
\null
\draftdate
\strut\hfill FR-PHENO-2012-007\\
\strut\hfill {ZU-TH 12/12}\\
\vspace{1.5cm}
\begin{center}
{\Large \bf\boldmath
NLO QCD corrections to off-shell top--antitop production with leptonic
decays at hadron colliders
\par} 
\vspace{1.5cm}
{\large
{\sc A.\ Denner$^1$, S.\ Dittmaier$^2$, S.~Kallweit$^3$ 
and S.\ Pozzorini$^3$} } \\[.5cm]
$^1$ {\it Universit\"at W\"urzburg, Institut f\"ur Theoretische Physik und Astrophysik,\\
D-97074 W\"urzburg, Germany}
\\[0.5cm]
$^2$ {\it Albert-Ludwigs-Universit\"at Freiburg, Physikalisches Institut, 
\\
D-79104 Freiburg, Germany}\\[0.5cm]
%
$^3$ {\it Institut f\"ur Theoretische Physik, Universit\"at Z\"urich,
\\
CH-8057 Z\"urich, Switzerland}

\par \vskip 1em
\end{center}\par
\vfill {\bf Abstract:} 
\par 
We present details of a calculation of the cross
section for hadronic top--antitop production in next-to-leading order
(NLO) QCD, including the decays of the top and antitop into bottom
quarks and leptons. 
This calculation is based on matrix
elements for $\leptfs$ production and includes all non-resonant diagrams,
interferences, and off-shell effects of the top quarks. Such contributions
are formally suppressed by the top-quark width
and turn out to be small in the inclusive cross section.
However, they can be 
strongly enhanced  in exclusive observables
that play an important role in Higgs and new-physics searches.
Also non-resonant and off-shell
effects due to the finite W-boson width are investigated in detail, 
but their impact is much smaller than naively expected.
We also introduce a matching approach to improve 
NLO calculations involving intermediate unstable particles.
Using a fixed QCD scale leads to perturbative
instabilities in the high-energy tails of distributions, but an appropriate
dynamical scale stabilises NLO predictions.  
Numerical results for the total cross section,
several distributions, and asymmetries
are presented for Tevatron and the LHC at $7\TeV$, $8\TeV$, and $14\TeV$.

\par
\vskip 2.5cm
\noindent
July 2012
\null
\setcounter{page}{0}
\clearpage
\def\thefootnote{\arabic{footnote}}
\setcounter{footnote}{0}

\section{Introduction}

As the only fundamental known fermion with a mass at the electroweak
scale, the top quark might serve as a key for understanding the 
fermionic mass hierarchy
and is potentially sensitive to physics
beyond the Standard Model.  Since it can be produced copiously at
the LHC in top--antitop pairs, its production cross
section, its decay, and its properties can be studied with high
precision.  Evidently, accurate measurements must be accompanied by 
precise calculations.

The first step towards precise theoretical predictions for
{$\Pt\bar\Pt$} production at hadron colliders was made already about
20~years ago with the calculation of QCD corrections at
next-to-leading-order
(NLO)~\cite{Nason:1989zy,Beenakker:1990maa,Mangano:1991jk,Frixione:1995fj}.
Later also electroweak radiative corrections were
calculated~\cite{Beenakker:1993yr,Moretti:2006nf,Kuhn:2006vh,Bernreuther:2008aw,Hollik:2007sw,Kuhn:2011ri},
and recently important progress has been achieved both in the
resummation of logarithmically enhanced
terms~\cite{Beneke:2009rj,Czakon:2009zw,Ahrens:2010zv,Kidonakis:2010dk}
and towards the inclusion of QCD corrections at
next-to-next-to-leading-order (NNLO)~\cite{Dittmaier:2007wz,Kniehl:2008fd,Anastasiou:2008vd,%
Czakon:2007ej,Czakon:2007wk,Czakon:2008zk,%
Bonciani:2008az,Bonciani:2009nb,Bonciani:2010mn,
GehrmannDeRidder:2009fz,Czakon:2010td}.
First NNLO QCD
results for the quark--antiquark channel have been 
presented in~\citere{Baernreuther:2012ws}.

The above-mentioned predictions are based on the approximation of
stable (on-shell) top quarks, \ie the top-quark decays, which
proceed into pairs of W~bosons and b~quarks in the Standard Model,
are ignored. A thorough LO analysis of hadronic top-pair production and 
decay including off-shell and non-resonant contributions was presented 
in \citere{Kauer:2001sp}. A first important step towards a full NLO 
description of top-pair production and decay was made
in~\citeres{Bernreuther:2004jv,Melnikov:2009dn, Bernreuther:2010ny},
where top-quark decays were treated in a spin-correlated narrow-width
approximation, \ie the top quarks are still on shell, but spin
correlations between production and decay are taken into account.
Meanwhile, NLO top-quark decays in narrow-width approximation were
implemented in MCFM~\cite{Campbell:2012uf}.

Recently, first results at NLO QCD 
for the complete process of
$\PW^+\PW^-\Pb\bar\Pb$ production, with intermediate off-shell top
quarks and including leptonic W-boson decays have been obtained by two
independent groups \cite{Denner:2010jp,Bevilacqua:2010qb}.
Here we give details on one of these calculations
\cite{Denner:2010jp} and extend the presented results in various
directions.

The reaction 
{$\Pp\Pp\to\PW^+\PW^-\Pb\bar\Pb\to\leptfs$} represents
one of the {$2\to4$} particle processes on the Les Houches 
priority list~\cite{Binoth:2010ra}.
While various multi-particle NLO QCD calculations
have been performed in the recent years 
\cite{Bredenstein:2009aj%
,Bevilacqua:2009zn%
,Bevilacqua:2011aa%
,Ellis:2009zw%
,Melia:2011dw%
,Berger:2009zg%
,Berger:2010vm%
,Berger:2010zx%
,Ita:2011wn%
,Bern:2011ep%
,Campanario:2011ud%
,Greiner:2011mp%
,Greiner:2012im%
,Becker:2011vg%
,Bevilacqua:2012em%
}, 
{$\PW^+\PW^-\Pb\bar\Pb$} production involves the treatment of 
resonant particles for the first time in a hadron-collider environment 
on that level of complexity.
The two top resonances can be consistently treated in the complex-mass
scheme that was introduced at the NLO level in the context of the 
calculation of the electroweak corrections to the processes
{$\Pe^+\Pe^-\to\PWp\PWm\to4\,$fermions}~\cite{Denner:2005es,Denner:2005fg},
which was the first full NLO calculation for a {$2\to4$} particle
process.

To compute the virtual corrections, 
we employ explicit diagrammatic representations of the
one-loop amplitudes.  A key feature of our approach is the
factorization of colour structures at the level of individual
diagrams.  This permits to reduce the CPU cost of colour sums
essentially to zero.  Helicity-dependent structures are algebraically
reduced to a common set of so-called standard matrix elements.  In
this way the sums over physical helicities are strongly boosted.
Tensor loop integrals are related to scalar integrals by means of
numerical algorithms that systematically avoid numerical instabilities
from inverse Gram determinants and other spurious singularities
\cite{Denner:2002ii,Denner:2005nn}.  
Scalar loop integrals are required with complex internal masses and are 
evaluated using the results of \citeres{'tHooft:1978xw,Denner:2010tr}.
The real corrections
are handled with the dipole subtraction
method~\cite{Catani:1996vz,Dittmaier:1999mb,Phaf:2001gc,Catani:2002hc},
and the phase-space integration is performed with adaptive
multi-channel techniques
\cite{Berends:1994pv,Denner:1999gp,Dittmaier:2002ap}.  

Our predictions provide a complete description of hadronic $\WWbb $
production, including off-shell {$\Pt\bar\Pt$} intermediate states, as
well as contributions with a single or no top-quark resonance.  This
permits to quantify the accuracy of the narrow-top-width
approximation, which involves only on-shell $\ttb$ contributions and
corresponds to the \mbox{$\Gamma_\Pt\to 0$} limit of our calculation,
at NLO.  For the integrated cross section it turns out that, in
presence of loose cuts, the error of the narrow-top-width
approximation does not exceed 1\%
\cite{Denner:2010jp,Bevilacqua:2010qb,Denner:lh2011}.  This is
perfectly consistent with the theoretically expected
$\ord(\Gamma_\Pt/\Mt)$ suppression of finite-top-width corrections to
inclusive observables.

In more exclusive measurements, such as
precision $\Mt$ determinations, $\ttb$
backgrounds to new physics that are suppressed by vetoing top
resonances, or the $\ttb$ background to \mbox{$\PH\to ll\nu\nu$}
signals in presence of b-jet vetoes,
the investigation of finite-top-width effects is even more important
since their magnitude is not known a priori.
A first
systematic study of finite-top-width effects in exclusive observables, based on a
comparison of our calculation against the narrow-top-width approximation of
\citere{Melnikov:2009dn}, indicates that finite-top-width corrections to
phenomenologically important observables can range from a few per mille
to tens of per cent \cite{Denner:lh2011}.
This motivates us, on the one hand, 
to undertake a more thorough comparison of the narrow- and finite-top-width 
approaches, which will be presented in a forthcoming paper~\cite{ftwpaper}.
On the other hand, it raises the issue of possible non-negligible 
effects resulting from the finite width of intermediate W bosons, which 
is addressed in the present paper.

To study finite-W-width effects we consider two different descriptions 
of the leptonic W-boson decays:
a (spin-correlated) narrow-W-width approximation
and, alternatively, matrix elements for $\leptfs$ production,
including off-shell and non-resonant finite-W-width 
contributions%
\footnote{The virtual corrections are calculated in the double-pole
  approximation for the W~resonances.}.  
While finite-W-width effects have already been included in 
the predictions of~\citere{Bevilacqua:2010qb},
comparing the two above-mentioned approaches
provides the first quantitative assessment of the 
precision of the narrow-W-width approximation in $\leptfs$ production.
In spite of the larger numerical value
of $\Gamma_\PW/\MW$ as compared to  $\Gamma_\Pt/\Mt$,
finite-W-width corrections turn out to be definitely smaller than
finite-top-width effects.  As we discuss, this is due to a
double-suppression mechanism related to subtle cancellations between
finite-W-width contributions to matrix elements and to the
$\Gamma_\Pt$ input parameter.

The paper is organized as follows.  In \refse{se:calculation} we
discuss technical aspects of the calculation, including
finite-top-width and finite-W-width effects (\refse{se:prelim}), as
well as virtual (\refse{se:virtcor}) and real (\refse{se:realcor})
corrections.  In \refse{se:matching} we also introduce a matching
approach that restores exact factorization of $\ttb$ production and
top decays in the \mbox{$\Gamma_\Pt\to 0$} limit.  This permits to
avoid a few-per-cent loss of precision in the non-factorized NLO
description of $\leptfs$ production  while including finite-top-width
effects.
In \refse{se:numres} we present numerical predictions
for Tevatron and the LHC at different collider energies.
The choice of the factorization and renormalization scales
is discussed in some detail in \refse{se:setupB}.
There we introduce a dynamical scale related to the 
transverse energy of the top quarks,
which avoids serious perturbative instabilities
in the high-energy tails of differential distributions.
Moreover, we advocate the use of a collider-dependent QCD scale
adapted to the particular behaviour of the leading partonic channels,
\ie quark--antiquark annihilation at the Tevatron and gluon--gluon
fusion at the LHC. Specifically, we argue that a reduced scale
provides a better description of $\WWbb$ production at the LHC.
Results for the integrated cross sections (\refse{se:xsec}),
asymmetries (\refse{se:asymm}), and several distributions
(\refse{se:distr}), are supplemented by a numerical extrapolation of
the $\leptfs$ cross section to the $\Gamma_\Pt\to 0$ limit
(\refse{se:FtWeffects}).  Finally, in \refse{se:helacnlo} we show
that our results are in good agreement with corresponding
results~\cite{Bevilacqua:2010qb} obtained with {\sc
  HELAC-NLO}~\cite{Bevilacqua:2011xh}.  Our conclusions are presented
in \refse{se:conclusion}, and in \refapp{app:benchmark} we provide
benchmark results for the partonic matrix elements squared in lowest
order and including virtual corrections.

\section{Details of the calculation}
\label{se:calculation}

\subsection{Preliminaries}
\label{se:prelim}
The cross section for the hadronic process $h_1h_2\to \wwfs+X
\to\leptfs+X$ is evaluated according to
\beq
\sigma = \sum_{a,b}\int_0^1\rd x_1 \int_0^1\rd x_2\,
f_{a/h_1}(x_1)\, f_{b/h_2}(x_2)\, \int\rd\hat\sigma_{ab}(x_1 P_1,x_2 P_2),
\eeq
where the parton distribution functions (PDFs) $f_{c/h_i}(x_i)$
describe the probabilities of finding a parton $c$ in hadron $h_i$
with a fraction $x_i$ of the full hadron momentum $P_i$.  In LO the
colliding parton pairs are $ab=\Pg\Pg,q\bar q,\bar qq$, but higher
orders involve contributions from $\Pg q, \Pg\bar q,q\Pg$, and $\bar
q\Pg$ as well. 
We consistently neglect contributions from
initial-state bottom quarks,%
\footnote{We have checked that
bottom-quark-induced contributions to the integrated cross section 
are suppressed at the level of $\lesssim 0.2\%$ at LO.}
\ie $q=\Pu,\Pd,\Pc,\Ps$.  Otherwise we work in the five-flavour scheme
with massless bottom quarks.  In NLO, the partonic cross section
\beq\label{eq:partonic_xs}
\int\rd\hat\sigma_{ab} = 
\int_6\rd\hat\sigma_{0,ab}
+\int_6\rd\hat\sigma_{\virt,ab}
+\int_7\rd\hat\sigma_{\real,ab}
+\int_0^1\rd x\,\int_6\rd\hat\sigma_{\fact,ab}
\eeq
receives contributions from tree diagrams ($\hat\sigma_{0,ab}$),
virtual one-loop corrections ($\hat\sigma_{\virt,ab}$), real parton
emission ($\hat\sigma_{\real,ab}$), and the factorization of collinear
initial-state singularities ($\hat\sigma_{\fact,ab}$) into the PDFs.
The subscript on the integral corresponds to the number of final-state
particles in the reaction $ab\to \wwfs(c)\to\leptfs(c)$, where $c$ is
a possibly emitted parton.
The integral over $x$ indicates that the factorization contribution
involves an additional integration over the fraction $x$ by which
one of the incoming parton momenta is reduced due to 
collinear
initial-state radiation.

In order to ensure the correctness of our results we have evaluated
each ingredient twice and independently, resulting in two independent computer 
codes.

In the following we discuss the treatment of top-quark and W-boson
resonances as well as effects beyond NLO related to the factorization of
$\ttb$ production and top decays in the narrow-top-width limit.  Details of
the calculations of the virtual and real corrections are given in
\refses{se:virtcor} and \ref{se:realcor}.

\subsubsection{Treatment of unstable top quarks}
\label{se:topdec}

Our predictions for the process $h_1h_2\to \wwfs+X \to\leptfs+X$
provide a complete description of hadronic top-quark pair production
and decay, including doubly-resonant contributions where the $\leptfs$
final state results from the decay of a {$\Pt\bar\Pt$} pair, as well
as singly-resonant and non-resonant diagrams, \ie contributions with
only one or no top resonance.  Interferences between doubly-, singly-,
and non-resonant diagrams are consistently taken into account.
\begin{figure}
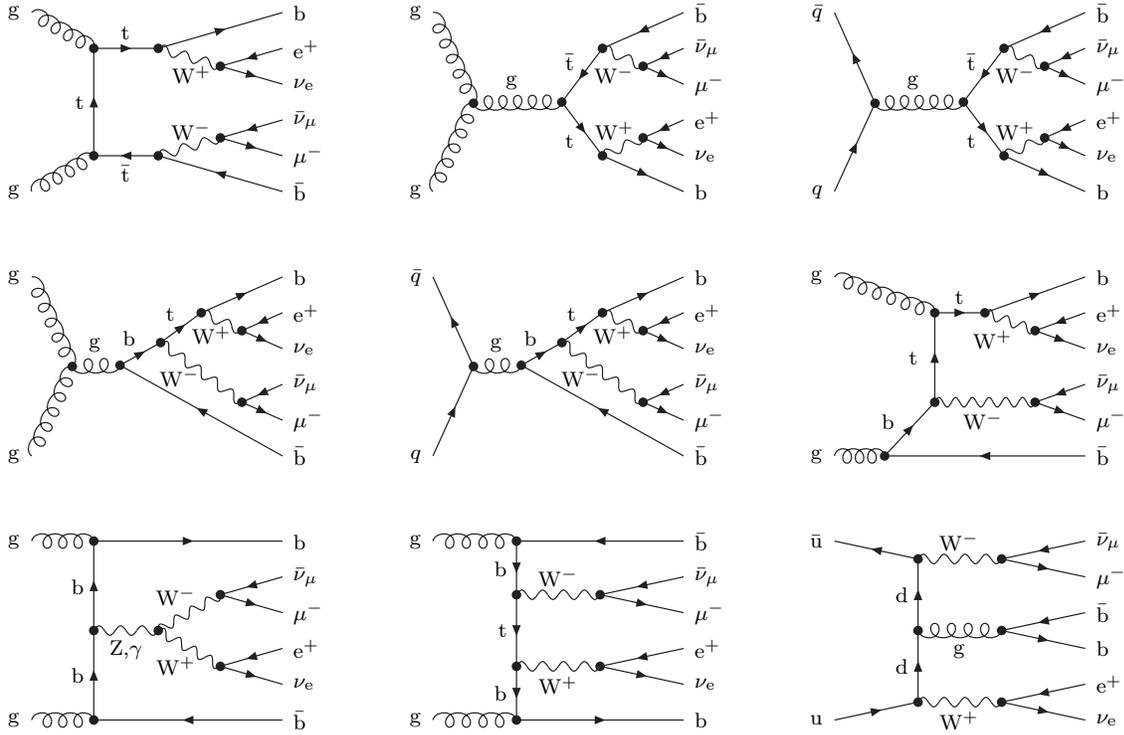

\begin{center}
\unitlength 0.90pt
\SetScale{0.90}
\DRtchanneltopggveepmumvmxbbx
\DRschannelgluonggveepmumvmxbbx
\DRschannelgluonqqxveepmumvmxbbx
\diagramsggveepmumvmxbbxSRa
\diagramsuuxveepmumvmxbbxSR
\diagramsggveepmumvmxbbxSRbmod
\diagramsggveepmumvmxbbxNR
\diagramsggveepmumvmxbbxNRb
\diagramsuuxveepmumvmxbbxNR
\end{center}
\caption{Representative   tree diagrams involving two
  (first line), only one (second line), or no (last line) top-quark
  resonances.}
\label{fig:treegraphs} 
\end{figure}
A few representative LO diagrams are depicted in
\reffi{fig:treegraphs}.  The {$\Pq\bar\Pq$} and gg partonic channels
involve 14 and 31 tree diagrams,
respectively, if only topologies involving two resonant W~bosons are
considered.%
\footnote{Since we treat b quarks as
massless partons there are no Higgs-exchange diagrams at tree level.}  
Additional contributions with less than two W-boson
resonances are discussed in \refse{se:wdec}.

To regularize intermediate top-quark resonances in a gauge-invariant
way we employ the complex-mass scheme~\cite{Denner:2005fg}, where the
top-quark width $\Gamma_\Pt$ is incorporated into the definition of
the (squared) top-quark mass,
\beq
\label{e:tcms}
\mu^2_\Pt=\Mt^2-\ri\Mt\Gamma_\Pt.  
\eeq
In this way,
off-shell-top contributions are consistently 
described by Breit--Wigner distributions,
and all matrix elements are 
evaluated using the complex top mass $\mu_\Pt$.
Technical implications of the 
complex-mass scheme at one loop are discussed in 
\refse{se:mut}.

The inclusive $\leptfs$ cross section is dominated by the
doubly-resonant top-pair contribution and can be described, with
fairly good accuracy, in narrow-top-width approximation.  It is thus
instructive to compare our calculation to this approximation, which
corresponds to the $\Gamma_\Pt\to 0$ limit.  To avoid confusion
between the treatment of top-quark and W-boson decays, in the
following we denote the $\Gamma_\Pt\to 0$ and $\Gamma_\PW\to 0$ limits
as narrow-top-width (NtWA) and narrow-W-width (NwWA) approximations,
respectively.  Contributions that vanish in NtWA and NwWA are called
finite-top-width (FtW) and finite-W-width (FwW) effects, respectively.
Our treatment of FwW effects is discussed in \refse{se:wdec}.

For what concerns top resonances, we point out that FtW contributions
are included everywhere in our calculation, \ie we never make use of
the NtWA.  Nevertheless, in the following we briefly introduce this
approximation in order to discuss the origin of FtW effects and other
features of our predictions. In the NtWA, each top-quark resonance
leads to
\begin{equation}
\label{eq:delta}
\lim_{\Gamma_\Pt\to 0}\frac{1}{(p_\Pt^2-\Mt^2)^2+\Mt^2\Gamma_\Pt^2} = \frac{\pi}{\Mt\Gamma_\Pt}\;\delta\big(p_\Pt^2-\Mt^2\big),
\end{equation}
where the $\delta$-function that forces the top quark on its mass
shell is accompanied by a $1/\Gamma_\Pt$ factor.  In NtWA the
$\leptfs$ cross section includes only contributions involving two top
resonances, which are proportional to $1/\Gamma^2_\Pt$. Singly- and
non-resonant diagrams, as well as their interference with
doubly-resonant diagrams, are neglected due to their suppression in
the $\Gamma_\Pt\to 0$ limit.  As a result of these approximations, the
differential $h_1h_2\to \ttb\to\lvlvbb$ cross section is factorized
into the $h_1h_2\to \ttb$ production cross section times the $\Pt\to
\PW\Pb\to l\nu_l\Pb$ partial decay widths,
\beqar
\label{eq:NtWA}
\rd\sigma_\NtWA&=& {\Gamma^{-2}_{\Pt}} \left(\rd\sigma_\ttb\,\pw{i}{}\,\pwb{j}{}\right),
\eeqar
where the subscripts $i,j$ refer to the (anti)top-decay final states
$\nu_\Pe\Pep\Pb$ and $\mu^-\bar\nu_\mu\bar\Pb$, and the total
top-quark width is obtained by summing over all relevant decay
channels,
\beqar
\label{eq:GTtot}
\Gamma_\Pt= \sum_k \int \pw{k}{}.
\eeqar
Top-quark spin correlations in \refeq{eq:NtWA} are implicitly understood.

In NtWA, LO and NLO partonic cross sections can be schematically written as
\beqar
\label{eq:NtWALO}
\rd\sigma_\NtWA^\LO &=& (\GtLO)^{-2} \Bigl[\rd\sigma_{\ttb}^{\LO}\,\pw{i}{\LO}\,\pwb{j}{\LO}\Bigr],\nl
\label{eq:NtWANLO}
\rd\sigma_\NtWA^\NLO &=& {(\GtNLO)^{-2}} \Bigl[
\left(\rd\sigma_{\ttb}^{0}+\rd\sigma_{\ttb}^{1}\right)
\,\pw{i}{0}\,\pwb{j}{0}
+
\rd\sigma_{\ttb}^{0}\,\left(
\pw{i}{1}\,\pwb{j}{0}+
\pw{i}{0}\,\pwb{j}{1}
\right)
\Bigr],\nl
\eeqar
where the superscripts $0$ and $1$ indicate LO and correction contributions
to NLO quantities, \ie $\pw{k}{\NLO}=\pw{k}{0}+\pw{k}{1}$ and
$\rd\sigma_{\ttb}^{\NLO}=\rd\sigma_{\ttb}^{0}+\rd\sigma_{\ttb}^{1}$.
Note that $\rd\sigma_{\ttb}^{0}\neq \rd\sigma_{\ttb}^{\LO}$, since the
ingredients of $\rd\sigma_\NtWA^\LO$ and $\rd\sigma_\NtWA^\NLO$ have
to be evaluated with input parameters at the corresponding
perturbative order, \ie $\rd\sigma_{\ttb}^{\LO}$ and
$\rd\sigma_{\ttb}^{0}$ are evaluated with PDFs and $\alphas$ in
LO and NLO, respectively.  Since the top decay does not involve
$\alphas$ at LO we have $\pw{k}{0}=\pw{k}{\LO}$.  Note also that LO
and NLO predictions must be computed using total decay widths $\GtLO$
and $\GtNLO$, respectively.  Naturally, this also holds for our
calculation, not only in NtWA.

As a result of the truncation of the perturbative expansion, the NtWA of the NLO cross section \refeq{eq:NtWANLO} 
involves only three contributions, 
where the corrections are  applied either to $\rd\sigma_\ttb$ or to one of the decays. 
A pure fixed-order 
NLO calculation does not include contributions 
like $\rd\sigma_{\ttb}^{1}\pw{i}{1}\pwb{j}{0}$,
where the NLO correction is simultaneously applied to the production and decay parts of the process.
Such contributions are formally of NNLO, but 
given their non-negligible numerical impact it is
desirable to take them into account.
As discussed in \refse{se:matching}, this can be achieved by means of a simple prescription
which  is applicable also in presence of FtW and FwW corrections and
is derived by matching Eq.~\refeq{eq:NtWALO} to Eq.~\refeq{eq:NtWA} at the level of
the fully inclusive cross section. 

As compared to the NtWA, our calculation includes two types of
additional effects owing to the FtW: corrections
resulting from the top-quark off-shellness in doubly-resonant channels
and contributions from singly- as well as
non-resonant diagrams. At NLO, FtW effects manifest themselves also
in the form of non-factorizable corrections to doubly-resonant diagrams, 
which originate from
one-loop topologies or interferences between real-emission
diagrams, where top--antitop production and top decays
are linked via exchange of QCD partons.
Technically, the calculation of non-factorizable, singly-, and
non-resonant one-loop contributions involves pentagon and hexagon
diagrams, which represent a much higher level of complexity as
compared to the four- and lower-point diagrams that appear in NtWA.

The non-factorizable contributions of virtual and real origin are
enhanced by large logarithms of $\Gamma_\Pt/\Mt$ originating from soft
gluons.  However, it is well known that---in {\it sufficiently
  inclusive} {$\Pt\bar\Pt$} observables---such logarithms cancel, and
the remaining FtW effects yield rather small contributions of order
$\Gamma_\Pt/\Mt\simeq
0.9\%$~\cite{Fadin:1993dz,Fadin:1993kt,Melnikov:1993np}.
Nevertheless, a reliable quantitative assessment of FtW effects is
important in order to achieve per-cent-level precision in the
(inclusive and differential) description of {$\Pt\bar\Pt$} production.
Comparing our predictions to the NtWA, 
one can obtain a precise
determination of FtW contributions to any infrared-safe observable.
For the case of the integrated cross section, we performed this
comparison by means of a numerical extrapolation of our predictions to
the $\Gamma_\Pt\to 0$ limit \cite{Denner:2010jp}.  The results,
discussed in \refse{se:FtWeffects}, demonstrate that the NtWA for the
integrated cross section agrees with the full calculation at the
sub-per-cent level.

As already mentioned in the introduction,
in more exclusive phenomenological studies, such as precision
$\Mt$~measurements or Higgs- and new-physics searches,
the size of FtW corrections cannot be anticipated a priori and turns out to 
range from  a few per mille to tens of per cent  \cite{Denner:lh2011,ftwpaper}.

\subsubsection{Matching to NLO inclusive $\Pt\bar\Pt$ cross section}
\label{se:matching}
Let us now discuss effects related to the truncation of the
perturbative expansion at NLO in the presence of unstable intermediate
particles. To start with, we consider the fully inclusive cross
section in NtWA,
\beqar
\label{eq:inclcs}
\int\rd\sigma_\NtWA
&=&
\sigma_\ttb
\,\BR{i}{}\,\BRb{j}{},
\eeqar
which is obtained by integrating \refeq{eq:NtWA} over the full phase space and 
is given by the on-shell inclusive $\ttb$ cross section,
\beq
\sigma_{\Pt\bar\Pt}=
\int\rd\sigma_{\Pt\bar\Pt},
\eeq
times the branching fractions
\beq
\label{eq:BR}
\BR{k}{}
=
\frac{\tw{k}{}}{\Gamma_\Pt}
=
\frac{\int\pw{k}{}}{\Gamma_\Pt},
\eeq
with $k=i,j$. 
Apart from Coulomb effects near threshold, the
above relation between the $\Pp\Pp\to\ttb\to ij$ and
the on-shell $\ttb$ cross sections is valid to all orders of perturbation theory~\cite{Fadin:1993kt}.
However, due to  missing higher-order terms, the NLO approximation \refeq{eq:NtWALO} does not 
fulfil \refeq{eq:inclcs} exactly.
The mismatch can be expressed in terms of products of NLO contributions as follows,
\beqar
\label{eq:trunca}
\Delta\sigma^\NLO_\trunc
&=&
\sigma_{\ttb}^{\NLO}\,\BR{i}{\NLO}\,\BRb{j}{\NLO} -\int \rd\sigma_\NtWA^\NLO
\nl
&=& 
(\GtNLO)^{-2}
\left(\sigma_{\ttb}^{0}+\sigma_{\ttb}^{1}\right)
\left(\tw{i}{0}+\tw{i}{1}\right) 
\left(\twb{j}{0}+\twb{j}{1}\right) 
-\int \rd\sigma_\NtWA^\NLO \nl
&=& 
(\GtNLO)^{-2}
\Bigl[
\left(\sigma_{\ttb}^{0}+\sigma_{\ttb}^{1}\right)
\,\tw{i}{1}\,\twb{j}{1}
+
\sigma_{\ttb}^{1}\,\left(
\tw{i}{1}\,\twb{j}{0}+
\tw{i}{0}\,\twb{j}{1}
\right)
\Bigr].
\end{eqnarray}
Rewriting \refeq{eq:trunca} as a relative correction 
to \refeq{eq:inclcs} yields
\beqar
\label{eq:truncb}
\delta^\NLO_\trunc
&=&
\frac
{\Delta\sigma^\NLO_\trunc}
{\sigma_{\ttb}^{\NLO}\,\BR{i}{\NLO}\,\BRb{j}{\NLO}}
=
\left[x_i(1-x_j)+(1-x_i)x_j\right]
\de_\ttb
+x_ix_j,
\end{eqnarray}
where the factors
\beq
\label{eq:matchingcoeff}
\de_\ttb=
\frac{\sigma_{\ttb}^{1}}{\sigma_{\ttb}^{\NLO}}=
1-\frac{\sigma_{\ttb}^{0}}{\sigma_{\ttb}^{\NLO}}
\eeq
and
\beq
\label{eq:matchingcoeffb}
x_k=\frac{\tw{k}{1}}{\tw{k}{\NLO}}=
1-\frac{\tw{k}{0}}{\tw{k}{\NLO}}
\eeq
represent NLO corrections to $\ttb$ production and decay, respectively.
For the case of a di-lepton final state, where
$x_i=x_j=x$, Eq.~\refeq{eq:truncb} simplifies to
\beqar
\label{eq:truncc}
\delta^\NLO_\trunc
&=&
2x(1-x)\de_\ttb
+x^2.
\end{eqnarray}
Since $x\simeq 10\%$ and $\de_\ttb\simeq 10{-}30\%$,
the correction $\delta^\NLO_\trunc$ can reach 
a few per cent and should thus be taken into account.

\newcommand{\pWw}[2]{\Gamma_{\PW\to #1}^{#2}}
\newcommand{\GaW}[1]{\Gamma_{\PW}^{#1}}
\newcommand{\Gat}[1]{\Gamma_{\Pt}^{#1}}
In the case of di-leptonic decays of the $\ttb$ system, this can be
achieved by using the approach of
\citeres{Melnikov:2009dn,Campbell:2012uf}, where the factor
$(\GtNLO)^{-2}=(\Gamma_\Pt^0 +\Gamma_\Pt^1)^{-2}$ in \refeq{eq:NtWALO}
is expanded and truncated at NLO.  The corresponding differential NLO
cross section reads
\beqar
\label{eq:MS}
\rd\sigma_\MS^\NLO =
{(\Gat{0})^{-2}} &&\Bigl[
\left(
\rd\sigma_{\ttb}^{0}
+\rd\sigma_{\ttb}^{1}\right)
\,\pw{i}{0}\,\pwb{j}{0}
\nl&&{}
+
\rd\sigma_{\ttb}^{0}\,\left(
\pw{i}{1}\,\pwb{j}{0}+
\pw{i}{0}\,\pwb{j}{1}
-2\,\frac{\Gat{1}}{\Gat{0}}
\,\pw{i}{0}\,\pwb{j}{0}
\right)
\Bigr].\qquad
\eeqar
Integrating over the full phase space, and expressing $\pw{k}{0}$ and
$\pw{k}{1}$ in terms of the total top width and W-decay branching
fractions via
\beqar
\label{eq:ptwLO}
\tw{k}{0}&=& \Gamma_\Pt^0\,\frac{\pWw{k}{0}}{\GaW{0}},
\qquad
\frac{\tw{k}{1}}{\tw{k}{0}}=
\frac{\Gat{1}}{\Gat{0}}
+
\frac{\pWw{k}{1}}{\pWw{k}{0}}
-
\frac{\GaW{1}}{\GaW{0}},
\eeqar
one obtains the following expression, where the total and partial top-decay widths cancel out,
\beqar
\label{eq:MSinta}
\int\rd\sigma_\MS^\NLO &=& 
\left[
\sigma_{\ttb}^{0}+\sigma_{\ttb}^{1}
+
\sigma_{\ttb}^{0}\,\sum_{k=i,j}
\left(
\frac{\tw{k}{1}}{\tw{k}{0}}
-\frac{\Gat{1}}{\Gat{0}}
\right)
\right]
\prod_{k=i,j}
\frac{\tw{k}{0}}{\Gat{0}}
\nl&=& 
\left[
\sigma_{\ttb}^{0}+\sigma_{\ttb}^{1}
+
\sigma_{\ttb}^{0}\,\sum_{k=i,j}
\left(
\frac{\pWw{k}{1}}{\pWw{k}{0}}
-\frac{\GaW{1}}{\GaW{0}}
\right)
\right]
\prod_{k=i,j}
\frac{\pWw{k}{0}}{\GaW{0}}.
\eeqar
Note that, consistently with the treatment of the top width in
\refeq{eq:MS}, we expanded the NLO W-width term in \refeq{eq:ptwLO} as
$1/\GaW{\NLO}=1/\GaW{0}-\GaW{1}/(\GaW{0})^2$.  Moreover, to be fully
general, in \refeq{eq:ptwLO} we also included corrections $\pWw{k}{1}$
to the W decays.  Comparing \refeq{eq:MSinta} to the factorized
expression \refeq{eq:inclcs} with NLO branching fractions
\beqar
\label{eq:brs}
\BR{k}{\NLO}&=&
\frac{\tw{k}{\NLO}}{\Gat{\NLO}}=
\frac{\Gat{\NLO}\BRw{k}{\NLO}}{\Gat{\NLO}}=
\frac{\pWw{k}{0}+\pWw{k}{1}}{\GaW{0}+\GaW{1}}
\eeqar
yields
\beqar
\label{eq:MSdiff}
\lefteqn{
\sigma_{\ttb}^{\NLO}\,\BR{i}{\NLO}\,\BRb{j}{\NLO}-\int\rd\sigma_\MS^\NLO
=\sigma_{\ttb}^{1}
\left[
\prod_{k=i,j}
\frac{\pWw{k}{0}+\pWw{k}{1}}{\GaW{0}+\GaW{1}}-
\prod_{k=i,j}
\frac{\pWw{k}{0}}{\GaW{0}}
\right]}
\nl&&{}+
\sigma_{\ttb}^{0}
\left\{
\prod_{k=i,j}
\frac{\pWw{k}{0}+\pWw{k}{1}}{\GaW{0}+\GaW{1}}-
\left[1+
\sum_{k=i,j}
\left(
\frac{\pWw{k}{1}}{\pWw{k}{0}}
-\frac{\GaW{1}}{\GaW{0}}
\right)
\right]
\prod_{k=i,j}
\frac{\pWw{k}{0}}{\GaW{0}}
\right\}.
\eeqar
This indicates that, in general, the inclusive cross section resulting
from \refeq{eq:MS} is not identical to \refeq{eq:inclcs}.  The
mismatch \refeq{eq:MSdiff} is formally of NNLO and is due to the
corrections $\pWw{k}{1}$ to the exclusive W decays and to the
fixed-order expansion of $1/(\GaW{0}+\GaW{1})$.  Thus, if the W decays
do not receive NLO corrections the fixed-order approach \refeq{eq:MS}
can be reconciled with \refeq{eq:inclcs} by avoiding the fixed-order
expansion of the W-width term, \ie using
$1/\GaW{\NLO}=1/(\GaW{0}+\GaW{1})$ everywhere.  This corresponds to
replacing $1/\GaW{0}\to 1/(\GaW{0}+\GaW{1})$ and omitting the
$\GaW{1}/\GaW{0}$ term in \refeq{eq:MSdiff}.

In practice, this is applicable to di-lepton final states, but not to 
hadronically decaying top quarks or in presence of 
electroweak corrections. Therefore we prefer to ensure the validity of \refeq{eq:inclcs} at 
NLO  in a different way.
Instead of removing higher-order contributions by expanding and
truncating the term $(\GtNLO)^{-2}$, which originates from the Dyson
resummation of the top-quark self-energy in \refeq{eq:delta}, we keep
the exact NLO top-quark width everywhere and supplement our
calculation by the missing higher-order contributions
\refeq{eq:trunca}.  This is done by correcting the normalization of
the NLO $\Pp\Pp\to\ttb\to ij$ cross section in NtWA,
\beq
\label{eq:NLOm}
\rd\sigma_{\NtWA}^\NLOm=
(1+\delta^\NLOm)\rd\sigma_{\NtWA}^\NLO,
\eeq
with a matching factor
\beq
\label{eq:matching}
\delta^\NLOm=\frac{\delta_\trunc^\NLO}{1-\delta_\trunc^\NLO}
\eeq
that restores the  consistency with the on-shell $\ttb$ total cross section, \ie
\beqar
\int\rd\sigma_{\NtWA}^{\NLOm}
&=&
\sigma_\ttb^\NLO
\,\BR{i}{\NLO}\,\BRb{j}{\NLO}.
\eeqar
For the total cross section, the approximations \refeq{eq:MS} and
\refeq{eq:NLOm} are fully equivalent (for di-lepton final states and if
$\GaW{\NLO}$ is used everywhere).
However, they yield different predictions for exclusive observables,
in which case we expect \refeq{eq:NLOm} to be more accurate, thanks to the 
presence of higher-order contributions.

The matching correction \refeq{eq:matching} guarantees a consistent
inclusive cross section also in presence of FtW (and FwW) effects. In
this case, for the corrected cross section
\beqar
\label{eq:fullmatching}
\rd\sigma_{\leptfs}^\NLOm=(1+\delta^\NLOm)\,\rd\sigma_{\leptfs}^\NLO,
\eeqar
we have
\beqar
\label{eq:inclcsfw}
\int\rd\sigma_{\leptfs}^\NLOm
&=&
\sigma_\ttb^\NLO
\,\left(\BR{l\nu_l\Pb}{\NLO}\right)^2
+ \De\si_{\FtW}^\NLO,
\eeqar
which remains consistent with \refeq{eq:inclcs} for $\Gamma_\Pt\to 0$.
Moreover, the matching procedure does not disturb the 
$\ord(\Gamma_\Pt/\Mt)$ finite-width contributions $\De\si_{\FtW}^\NLO$
to $\sigma_{\leptfs}$, 
since the interplay of these two corrections
represents a suppressed NNLO effect of 
order $\delta^\NLOm\Gamma_\Pt/\Mt$.

Finally, let us note that in order to determine the correction factor
\refeq{eq:matchingcoeff} that enters $\delta^\NLOm$, instead of performing
an explicit on-shell $\ttb$ calculation we use
\beqar
\label{eq:matchingcoeffapp}
1-\de_\ttb&=&
\frac{\sigma_{\ttb}^{0}}{\sigma_{\ttb}^{\NLO}}
\simeq
\frac{\int\rd\sigma_{\ttb}^{0}\,\theta_\cuts}{\int\rd\sigma_{\ttb}^{\NLO}\,\theta_\cuts}
\simeq
\frac{\int\rd\sigma_{\leptfs}^{0}\,\theta_\cuts}{\int\rd\sigma_{\leptfs}^{\NLO}\,\theta_\cuts}
\left(\frac{\GtNLO}{\GtLO}\right)^2,
\eeqar
\ie we replace the fully inclusive $\sigma_\ttb$ cross section by the
$\leptfs$ cross section in presence of the (fairly inclusive) cuts
specified in \refse{se:numres}.  By definition, the various $\sigma^0$
and $\sigma^\NLO$ cross sections in \refeq{eq:matchingcoeffapp} must
be evaluated using NLO PDFs, NLO $\alphas$, and $\GtNLO$, and the
factor $\left({\GtNLO}/{\GtLO}\right)^2$ on the r.h.s.~of
\refeq{eq:matchingcoeffapp} compensates for the fact that, in order to
match the l.h.s, $\rd\sigma_{\leptfs}^{0}$ should be computed with a
LO top-quark width.  We expect the above approximation to be quite
precise since the mismatch between the $\ttb$ and $\leptfs$ NLO cross
sections in \refeq{eq:matchingcoeffapp} induces a strongly suppressed
error of order $(\delta^\NLOm)^2$ in $\sigma_{\leptfs}^\NLOm$.
Moreover, cut and finite-width effects in \refeq{eq:matchingcoeffapp}
cancel to a large extent in the ratios and are further suppressed by
the factor $2x(1-x)$ in \refeq{eq:truncc}.

In the derivation of the above $\NLOm$ correction factor we implicitly assumed that 
the top-quark width input parameter and the matrix elements are evaluated at the same 
QCD scale. However, while $\Mt$ represents the natural scale choice to compute 
$\GtNLO$, in general the matrix elements for $\leptfs$ production might be evaluated using 
a different scale $\mu$, like the dynamical scale proposed in \refse{se:setupB}, which 
adapts to the hard scale of the $\ttb$ production part of the process.
Since the scale $\mu$ also enters the top-decay part of the matrix elements, 
a scale choice $\mu\neq\Mt$
implies that partial and total top-decay widths are evaluated at different scales.
This mismatch can be compensated by redefining the
NLO differental cross section as
\newcommand{\mucorr}{\mathrm{PWC}}
\beq\label{scalecorr}
\left.
\rd\sigma_{\leptfs}^\NLO\,
\right|_{\mucorr} = \left(\frac{\GtNLO(\Mt)}{\GtNLO(\mu)}\right)^2 \rd\sigma_{\leptfs}^\NLO,
\eeq
where the overall factor $({\GtNLO(\Mt)}/{\GtNLO(\mu)})^2$
effectively restores $\mu\to\Mt$ in the partial decay widths.
The (dynamical or fixed) scale $\mu$ is the one used in the matrix elements,
including variations in scale-dependence studies.
The relative shift of the NLO cross section induced by the correction \refeq{scalecorr} 
amounts to about $-0.02\times(\mu/\Mt-1)$ for \mbox{$\mu/\Mt \sim 1$}.
In principle this correction can be applied to all NLO predictions.
However, we decided to present fixed-order NLO results in a more conventional 
way (scale $\mu$ in matrix elements and $\GtNLO$ at the fixed scale $\Mt$)
and to include the above correction  only in $\NLOm$ predictions.
To this end, we evaluate \refeq{eq:fullmatching} and \refeq{eq:matchingcoeffapp}
using ``partial-width corrected''  NLO predictions \refeq{scalecorr}.
In presence of a dynamical scale, $\mu = \mudyn$, 
instead of a phase-space-dependent correction we apply a constant factor
${\GtNLO(\Mt)}/{\GtNLO(\bar\mudyn)}$ evaluated at the logarithmic average
of $\mudyn$ (see \refta{tab:scales}).

\subsubsection{Treatment of unstable W bosons}
\label{se:wdec}
To describe the leptonic W-boson decays, \mbox{$\PW^+\to\nu_\Pe\Pe^+$}
and \mbox{$\PW^-\to\mu^-\bar\nu_\mu$}, we employ two different
approaches: a spin-correlated NwWA and, alternatively, the full set of
diagrams contributing to $\leptfs$ production.  Analogously to the
case of top-quark decays, the NwWA includes only contributions with
two on-shell W~bosons, while the FwW calculation involves additional
effects from off-shell W~bosons and from singly-resonant
\mbox{$\nu_e\Pep\mu^-\bar\nu_\mu$} production.\footnote{All channels
  contributing to $\leptfs$ production involve at least one resonant
  W~boson.}  In NwWA, the effect of the W-boson width is retained only
in the W~propagators---approximated by
$\pi/(\MW\Gamma_\PW)\delta(p_\PW^2-\MW^2)$--- while elsewhere we set
$\Gamma_\PW=\Gamma_\PZ=0$ and use real-valued parameters $\MW$ and
$\MZ$.  
In the alternative calculation with the full set of diagrams,
to include FwW effects we employ the
complex-mass scheme, \ie we use complex W- and Z-boson masses and a
corresponding complex-valued weak mixing angle,
\beqar
\label{e:WZcms}
\mu^2_\PW=\MW^2-\ri\MW\Gamma_\PW,\quad
\mu^2_\PZ=\MZ^2-\ri\MZ\Gamma_\PZ,\quad
\cw=\sqrt{1-\sw^2}=\frac{\mu_\PW}{\mu_\PZ}.
\eeqar

\begin{figure}
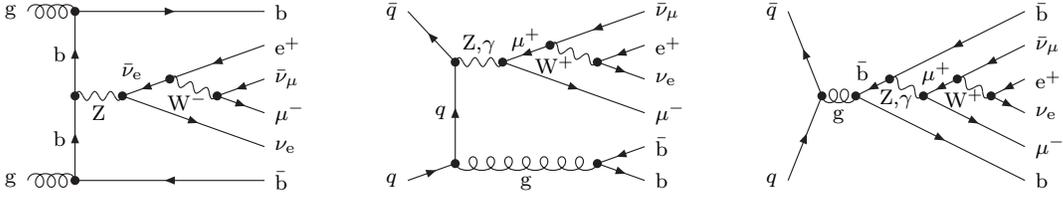

\begin{center}
\diagramsggveepmumvmxbbxNRoff
\diagramsuuxveepmumvmxbbxNRoff
\diagramsuuxveepmumvmxbbxNRoffb
\end{center}
\caption{Examples of LO diagrams that involve a single W-boson resonance and 
contribute to the FwW corrections.}
\label{fig:SRtreegraphs} 
\end{figure}
Examples of tree diagrams in NwWA and additional singly-resonant 
diagrams contributing to the FwW calculation are shown in
\reffi{fig:treegraphs} and \reffi{fig:SRtreegraphs}, respectively.  In
all singly-resonant diagrams, the four-lepton final state originates
from $\PZ,\gamma\to {\nu_e\Pep\mu^-\bar\nu_\mu}$ subtopologies
involving one intermediate $s$-channel $\PWp$ or $\PWm$ propagator.
We note that such topologies do not involve any top-quark resonance
and are thus expected to be strongly suppressed.  Their inclusion
increases the number of LO diagrams in the $q\bar q$ (gg) channel from
14 (31) to 38 (79).

At NLO, a resonant W-boson pair decaying into ${\nu_e\Pep\mu^-\bar\nu_\mu}$
does not receive  (factorizable or non-factorizable) QCD corrections.
This renders the inclusion of FwW effects fairly straightforward, as compared to the
case of FtW effects.  
Since FwW contributions are strongly suppressed already at LO
(see \refse{se:numres}), 
we adopt a double-pole approximation 
to include FwW effects in the 
{\it infrared-finite part} of the virtual QCD corrections, which is defined by
the sum of the virtual corrections and the contribution of the $I$-operator
of the real corrections in the dipole subtraction 
approach~\cite{Catani:1996vz,Catani:2002hc}. 
This approach,
which is discussed in \refse{se:LPA},
involves only doubly-W-resonant diagrams.
Similarly as in NwWA, the
matrix elements in double-pole approximation are 
evaluated with on-shell W~bosons and  using 
$\Gamma_\PW=\Gamma_\PZ=0$ and real W- and Z-boson masses
everywhere, apart from the W~propagator.
The latter reads $1/(p_\PW^2-\MW^2+\ri\Gamma_\PW\MW)$
and takes into account off-shell-W effects.
Apart from the (finite part of the) virtual corrections, 
in all other NLO contributions FwW effects are included exactly,
using the  complex-mass scheme \refeq{e:WZcms}.

For what concerns the expected magnitude of FwW contributions to inclusive $\Pt\bar\Pt$ observables, one might 
naively expect an impact of order $\Gamma_\PW/\MW\simeq2.5\%$, \ie three times larger
than the typical size of FtW effects.
However, more careful considerations indicate that the consistent inclusion of
FwW corrections in matrix elements and input parameters 
leads to doubly-suppressed contributions of
$\mathcal{O}(\frac{\Gamma_\PW\Gamma_\Pt}{\MW\Mt})$ in the inclusive cross section.
This is due to the fact that, in the $\Gamma_\Pt\to 0$ limit, 
the only contribution to
the inclusive $\leptfs$ cross section \refeq{eq:inclcsfw} 
that involves W-boson decays, and is thus sensitive to $\Gamma_\PW$, 
is the branching ratio
\beq
\label{eq:BRlept}
\BR{l\nu_l\Pb}{\NLO}
=
\frac{\int\pw{l\nu_l\Pb}{\NLO}
}{\GtNLO}.
\eeq
Moreover, owing to the identical $\Gamma_\PW$ dependence of the
numerator and denominator of \refeq{eq:BRlept}---which correspond to
the integrated top-decay matrix elements and the $\Gamma_\Pt$ input
parameter in our calculation---FwW corrections cancel in
the branching ratio.  Thus the consistent inclusion of FwW
corrections---in the $ab\to \leptfs(c)$ matrix elements and in the
calculation of the $\Gamma_\Pt$ input parameter---does not affect the
leading contribution to \refeq{eq:inclcsfw} and yields only tiny
corrections of $\mathcal{O}(\frac{\Gamma_\PW\Gamma_\Pt}{\MW\Mt})$ to
the fully inclusive cross section.  These considerations do not depend
on the inclusion of the matching correction \refeq{eq:fullmatching}
and remain valid also in presence of contributions of type
\refeq{eq:trunca}, which violate \refeq{eq:inclcs}.  We also point out
that omitting FwW effects in the top-quark width used as input
parameter would induce a fake FwW shift of $\ord(\Gamma_\PW/\MW)$ in
the integrated cross section.

The validity of the double-suppression mechanism
is restricted to the case 
where the top-decay phase spaces are fully integrated over,
and the presence of cuts is expected to lead to additional
FwW effects resulting from the incomplete cancellation of $\ord(\Gamma_\PW/\MW)$ corrections 
to the numerator and denominator of \refeq{eq:BRlept}.
However, as long as the cuts are rather inclusive,
also such FwW effects  are expected to remain well below
$\ord(\Gamma_\PW/\MW)$.
This justifies the approach of~\citere{Denner:2010jp},
where off-shell and non-resonant contributions of the top quark
to $\wwfs$ production
were combined with \mbox{W-boson} decays in NwWA.

For very exclusive observables, in contrast, 
it is important to investigate 
whether FwW corrections can become non-negligible.
To this end, in this paper we study FwW
effects by comparing several distributions 
obtained in NwWA and with the FwW variant of our calculation.
While FwW effects have already been included in 
the predictions of~\citere{Bevilacqua:2010qb}, 
our study provides the first quantitative assessment of the 
precision of the NwWA in $\leptfs$ production.

\subsection{Virtual corrections}
\label{se:virtcor}

\subsubsection{Diagram-by-diagram approach}

The sum of the LO and virtual contributions to the partonic cross section
is derived from the LO and one-loop matrix elements $\M_0$ and $\M_1$
according to
\beq
\int_6\rd\hat\sigma_{0,ab} + \int_6\rd\hat\sigma_{\virt,ab} = 
\frac{1}{2\hat s} \int\rd\Phi_6\, \biggl\{
\sum_{\mathrm{col}}\sum_{\mathrm{pol}} |\M_{0,ab}|^2 
+\sum_{\mathrm{col}}\sum_{\mathrm{pol}} 2\Re\Bigl(\M_{1,ab}\;\M_{0,ab}^*\Bigr)
\biggr\},
\eeq
where the squared partonic centre-of-mass (CM) energy $\hat s$ appears
in the overall flux factor $1/(2\hat s)$ and $\rd\Phi_6$ is the
phase-space element of the six-particle final state.
The sums run over all colour and helicity states and 
implicitly
include the averaging
over initial-state colours and helicities.
We calculate the amplitudes $\M_0$ and $\M_1$ for
{$\Pq\bar\Pq/\Pg\Pg\to\PWp\PWm\Pb\bar\Pb\to \leptfs$} in terms of explicit
Feynman diagrams and algebraically reduce the one-loop diagrams to
spin structures (``standard matrix elements'') and standard one-loop
tensor integrals, which are subsequently evaluated numerically, as described 
in more detail below.

The {$\Pq\bar\Pq$} and {$\Pg\Pg$} channels comprise 
294 and 795 one-loop diagrams, respectively 
(see
examples in \reffi{fig:loops}).%
\footnote{While our calculation includes 
the contributions of all three quark generations, 
we count diagrams with closed fermion loops only for a single generation.}
The most complicated ones are the 96
pentagons and 21 hexagons that contribute to the gg channel  and involve tensor integrals up to
rank five.
\begin{figure} 
\begin{center}
\unitlength 0.90pt
\SetScale{0.90}
\FdecayantitopDRtriangleggveepmumvmxbbx
\FproductionDRboxggveepmumvmxbbx
\FdecaytopDRtriangleggveepmumvmxbbx
\NFDRhexagonggveepmumvmxbbx
\NFDRpentagoninfiggveepmumvmxbbx
\NFDRpentagonschannelgluonggveepmumvmxbbx
\NFSRhexagonqqxveepmumvmxbbx
\NFNRpentagonggveepmumvmxbbx
\NFNRpentagonqqxveepmumvmxbbx
\end{center}
\caption{Examples of one-loop diagrams contributing to
  $\Pq\bar\Pq/\Pg\Pg\to\PWp\PWm\Pb\bar\Pb\to\leptfs$:
  doubly-top-resonant diagrams with corrections to $\ttb$ production
  or decay (first line), non-factorizable pentagons and hexagons with
  two top-quark resonances (second line), pentagons and hexagons with
  less than two top resonances (third line).}
\label{fig:loops}
\end{figure}
Feynman diagrams are generated with two independent versions of {\sc
  FeynArts}~\cite{Kublbeck:1990xc,Hahn:2000kx}, and one-loop
amplitudes are reduced as already described for
$\Pp\Pp\to\Pt\bar\Pt\Pb\bar\Pb$ in
\citeres{Bredenstein:2008zb,Bredenstein:2010rs} using two in-house
{\sc Mathematica} programs, one of which relies on {\sc
  FormCalc}~\cite{Hahn:1998yk} for preliminary manipulations.

The employed approach strongly mitigates the complexity inherent in
Feynman diagrams by exploiting factorization of colour matrices,
reduction of helicity structures to compact spinor chains, and
recycling a multitude of common subexpressions.  The reduced
expressions are automatically converted into {\sc Fortran77} programs
that evaluate colour/helicity summed quantities with very high CPU
efficiency.

The virtual corrections are obtained from the interference of the
one-loop and LO matrix elements summed over external-state colours and
helicities on a diagram-by-diagram basis.

\subsubsection{Colour factorization}
 
One of the key features of the diagram-by-diagram approach is that the
cost related to the large number of diagrams is compensated by the
possibility to perform colour sums very efficiently.  This is a
consequence of colour factorization: individual (sub)diagrams, $\Gamma$, consist
of a single colour-stripped amplitude ${\cal A}^{(\Gamma)}$ multiplied
by a single colour factor ${\cal C}^{(\Gamma)}$,
\beqar\label{eq:colfact}
\M^{(\Gamma)}
&=&
\A^{(\Gamma)}
{\cal C}^{(\Gamma)}.
\eeqar
More precisely, each diagram gives rise to $3^{n_4}$ colour-factorized
contributions of type \refeq{eq:colfact}, where $n_4$ is the number of
quartic gluon vertices in the diagram.  These terms are handled as
separate subdiagrams.  However, most diagrams do not involve quartic
couplings, and their colour structures factorize completely.
The colour factor ${\cal C}^{(\Gamma)}$ can be linearly decomposed in terms
of a fixed basis of colour structures ${\cal C}_k$,
\beq
{\cal C}^{(\Gamma)} = \sum_k c_k^{(\Gamma)} {\cal C}_k,
\eeq
where the coefficients $c_k^{(\Gamma)}$ are just a set of numbers for each diagram $\Ga$.
For the {$\Pq\bar\Pq$} and {$\Pg\Pg$} channels there are only 2 and 3 different colour
structures, respectively, which may be taken to be
\beqar
{\cal C}^{\Pq\bar\Pq}_1 &=& T^c_{c_{\bar\Pq}c_\Pq} T^c_{c_\Pb c_{\bar\Pb}},\qquad
{\cal C}^{\Pq\bar\Pq}_2 = \de_{c_{\bar\Pq}c_\Pq} \de_{c_\Pb c_{\bar\Pb}},
\nn\nl
{\cal C}^{\Pg\Pg}_1 &=& \delta_{ab} \de_{c_\Pb c_{\bar\Pb}}, \qquad
{\cal C}^{\Pg\Pg}_2 = \ri f_{abc} T^c_{c_\Pb c_{\bar\Pb}}, \qquad
{\cal C}^{\Pg\Pg}_3 = d_{abc} T^c_{c_\Pb c_{\bar\Pb}}. 
\eeqar
Here $a,b$ and $c_i$ denote the colour indices of the two incoming gluons and the fermion $i$, 
respectively, $T^c$ are the group generators in the fundamental (quark) representation,
and $f_{abc}$, $d_{abc}$ are the usual SU(3) constants.

The summation over external colours is performed once and for all at
the level of the colour basis and the LO matrix element.  To this end,
we compute the colour-interference matrix
\beqar
I_{kl}=
\sum_{\mathrm{col}} 
{\cal C}_k
{\cal C}^*_{l},
\eeqar
and, reducing the full tree matrix element in colour space,
\beqar
\M_0
&=&
\sum_{l}
\M_{l}^{(\LO)}
{\cal C}_{l},
\eeqar
we build the interference of $\M_0$ with the elements of the
colour basis as
\beqar\label{coltreeint}
\tilde\M_{k}^{(\LO)}
&=&
\sum_{\mathrm{col}} 
{\cal C}_k \M_0^* =
\sum_{l}{I}_{kl}\left(\M_{l}^{(\LO)}\right)^*.
\eeqar
Then, upon reduction of the factorized colour structure of the 
loop diagrams, 
\beqar\label{colofactb}
\M^{(\Gamma)}
&=&
\A^{(\Gamma)}
{\cal C}^{(\Gamma)}
=
\A^{(\Gamma)}
\left(\sum_k c_k^{(\Gamma)}{\cal C}_{k}\right),
\eeqar
we obtain the colour-summed interference 
between $\M_0^*$ and the complete one-loop amplitude
$\M_1 = \sum_\Ga \M^{(\Gamma)}$
as
\beq\label{colofactc}
\sum_{\mathrm{col}} \M_1 \M_0^* =
\sum_{\mathrm{col}} \sum_\Ga \M^{(\Gamma)}\M_0^* =
\sum_\Ga \A^{(\Gamma)}
\left(\sum_{k} c_k^{(\Gamma)}
\tilde\M_{k}^{(\LO)}\right).
\eeq
The colour-summed result is given by a combination of previously computed
colour--Born interference terms \refeq{coltreeint}.
For each phase-space point,
this requires
{\em a single evaluation} of the non-trivial colour-stripped
amplitude ${\cal A}^{(\Gamma)}$ of each (sub)diagram.

\subsubsection{Algebraic reduction of helicity structures and helicity sums}
The helicity structures encountered in the explicit evaluation of all
Feynman diagrams are algebraically reduced to a common basis of
standard matrix elements (SMEs).  The general form of SMEs for the
$a(k_1)b(k_2)\to\PW^+(k_3)\PW^-(k_4)\Pb(k_5)\bar\Pb(k_6)$ channel for
the initial states $ab=\Pq\bar\Pq/\Pg\Pg$ is
\beqar\label{SMEs}
{\hat\M}^{\Pq\bar\Pq}_{m,\sigma\tau} &=&
Q_{m;\mu_3\mu_4}^{\nu_1\dots \nu_l}
\Bigl[\bar v_{\bar\Pq}(k_1) \gamma_{\nu_1}\dots \gamma_{\nu_k}\omega_\sigma u_{\Pq}(k_2)\Bigr]
\varepsilon^{\mu_3 *}_{\PW^+}(k_3)
\varepsilon^{\mu_4 *}_{\PW^-}(k_4)
\nn\nl
&& {}\times
\Bigl[\bar v_{\Pb}(k_5) \gamma_{\nu_{k+1}}\dots
\gamma_{\nu_l}\omega_\tau u_{\bar\Pb}(k_6)\Bigr]
,
\nn\nl
{\hat\M}^{\Pg\Pg}_{m,\tau} &=&
Q_{m;\mu_1\dots\mu_4}^{\nu_1\dots \nu_l}
\varepsilon^{\mu_1}_{\Pg}(k_1)
\varepsilon^{\mu_2}_{\Pg}(k_2)
\varepsilon^{\mu_3 *}_{\PW^+}(k_3)
\varepsilon^{\mu_4 *}_{\PW^-}(k_4)
\left[\bar v_{\Pb}(k_5) \gamma_{\nu_{1}}\dots
\gamma_{\nu_l}\omega_\tau u_{\bar\Pb}(k_6)\right]
,\qquad
\eeqar
where $Q_{m;\mu_3\mu_4}^{\nu_1\dots \nu_l}$ and
$Q_{m;\mu_1\dots\mu_4}^{\nu_1\dots \nu_l}$ consist of combinations of
metric tensors and external momenta, and $\sigma,\tau=\pm$ refer to
the chirality projectors $\omega_\pm=(1\pm\gamma_5)/2$.  In the
double-pole approximation (see \refse{se:LPA}), W-boson decays are
described via effective polarisation vectors
\beqar
\varepsilon^{\mu *}_{\PW^+}(k_3)
&=&
\frac{e\,
\bar u(k_{\nu_\Pe}) \gamma^{\mu}\omega_- v(k_\Pep)
}{\sqrt{2}\sw\Bigl((k_{\nu_\Pe}+k_\Pep)^2-\MW^2+\ri\MW\Gamma_\PW\Bigr)},
\nl
\varepsilon^{\mu *}_{\PW^-}(k_4)
&=&
\frac{e\,
\bar u(k_{\mu^-}) \gamma^{\mu}\omega_- v(k_{\bar\nu_\mu})
}{\sqrt{2}\sw\Bigl((k_{\mu^-}+k_{\bar\nu_\mu})^2-\MW^2+\ri\MW\Gamma_\PW\Bigr)},
\eeqar
which include the left-handed lepton currents
and the W-boson propagators. 
In our calculation we encounter about 800 and 2000 SMEs for the
$\Pq\bar\Pq$ and $\Pg\Pg$ channels, respectively.
These compact spinor chains
permit to decouple helicity information from the remnant parts of the
diagrams, so that helicity sums can be performed in a
diagram-independent and efficient way.  In practice, the
colour-stripped part of each loop diagram [see \refeq{colofactb}] is
expressed as a linear combination of SMEs and tensor integrals,
\beqar\label{SMEdecomp}
\A^{(\Gamma)}
&=&
\sum_m {\cal F}^{(\Gamma)}_m {\hat \M}_m,
\nn\\
{\cal F}^{(\Gamma)}_m &=& \sum_{P} 
\sum_{j_1,\ldots,j_P=0}^{N-1} 
{\cal K}^{(\Gamma)}_{m;j_1\dots j_P}
{T^{N}_{j_1\dots j_P}} 
\;+\; \mbox{rational parts},
\eeqar
where the index $m$ here also includes the chirality indices $\sigma,\tau$.
The coefficients ${\cal K}^{(\Gamma)}_{m;j_1\dots j_P}$ are rational
functions of the kinematic invariants.  These functions involve only
denominators from intermediate-particle propagators and are free
of spurious poles that might generate numerical instabilities.
The functions ${T^{N}_{j_1\dots j_P}}$ are the coefficients of the
one-loop tensor integrals, whose evaluation is briefly described below.

Helicity sums are performed at the level of the interference of the
diagram-independent SMEs with the colour-projected Born amplitude
\refeq{coltreeint},
\beqar
M_{km}
&=&
\sum_{\mathrm{hel}}
{\hat\M}_m \tilde\M_k^{(\LO)}
=
\sum_{l}
I_{kl}
\sum_{\mathrm{hel}}
 {\hat\M}_m \left(\M^{(\LO)}_{l}\right)^*.
\eeqar
This matrix is computed only once per phase-space point employing the
Weyl--van der Waerden spinor formalism of \citere{Dittmaier:1998nn}.
Using $M_{km}$ one can directly obtain the colour- and helicity-summed
contributions of each loop diagram in terms of its colour- and
helicity-independent form factors ${\cal F}^{(\Gamma)}_m$ and the
coefficients ${c}^{(\Gamma)}_k$ of its factorized colour structure
\refeq{colofactb},
\beq\label{colhelsum}
\sum_{\mathrm{col}}
\sum_{\mathrm{hel}}
\M_1 \M_0^* 
=
\sum_{\mathrm{col}}
\sum_{\mathrm{hel}}
\sum_\Gamma
\M^{(\Gamma)} \M_0^* 
=
\sum_\Ga \sum_m {\cal F}_m^{(\Gamma)}
\left(\sum_{k}
c_k^{(\Gamma)} 
M_{km}\right).
\eeq

\subsubsection{Covariant decomposition and numerical reduction of
tensor integrals}  
Tensor one-loop integrals with $N$ propagators and $P$ Lorentz indices
are expressed in terms of totally symmetric covariant structures
$\{g\dots g p\dots p\}^{\mu_1\dots\mu_P}_{j_1\dots j_P}$ involving
$g^{\mu\nu}$ and the external momenta $p_1,\dots,p_{N-1}$,
\beq
\frac{(2\pi\mu)^{4-D}}{\ri\pi^{2}}\int \rd^{D}q\,
\frac{q^{\mu_1}\dots q^{\mu_P}}
{\prod_{i=0}^{N-1}\left[(q+p_{i})^2-m_i^2+\ri 0\right]}
=
\sum_{j_1,\dots,j_P=0}^{N-1} 
{T^{N}_{j_1\dots j_P}}\; 
{\{g\dots g p\dots p\}^{\mu_1\dots\mu_P}_{j_1\dots j_P}},
\eeq
with $D$ denoting the number of space--time dimensions.  For details
of the notation we refer to \citere{Denner:2005nn}.  To describe
$N$-point integrals with $N\ge 5$, tensor structures with only four
external momenta would be sufficient.  However, in order to avoid
potential instabilities due to inverse Gram determinants we use a
redundant set of structures, including the metric tensor and $N-1$
momenta.

The virtual corrections to $q\bar q\to\PWp\PWm\Pb\bar\Pb$ and
$\Pg\Pg\to\PWp\PWm\Pb\bar\Pb$ involve tensor integrals up to ranks
$P=4$ and $P=5$, respectively.  
As sketched above, the one-loop amplitudes are expressed
as linear combinations of tensor-integral coefficients
$T^{N}_{j_1,\dots,j_P}$.  The latter are evaluated by {\em
  numerical\/} libraries that recursively reduce them to master
integrals using the methods of \citeres{Denner:2002ii,Denner:2005nn}.
Avoiding an explicit reduction of analytic expressions to master
integrals, this numerical approach prevents prohibitively large
expressions and permits to adapt the reduction strategy to the
specific numerical problems that appear in different phase-space
regions.

Tensor $N$-point integrals with $N\ge 5$ are expressed in terms of
lower-rank and lower-point integrals exploiting the
four-dimensionality of
space--time~\cite{Denner:2002ii,Denner:2005nn}.%
\footnote{Similar reductions are described in
  \citeres{Binoth:2005ff,Fleischer:2010sq}.}  The tensor rank and the
number of propagators are simultaneously
reduced without introducing inverse Gram determinants.  Consequently,
the maximal power of inverse Gram determinants resulting from the
entire reduction is given by the maximal rank of four-point integrals,
which never exceeds four in renormalizable gauges.  Scalar hexagons
and pentagons are reduced to boxes using Melrose's method~\cite{Melrose:1965kb}.
Tensor 4-point and 3-point integrals are reduced to scalar integrals
with the Passarino--Veltman algorithm \cite{Passarino:1978jh} as long
as no small Gram determinant appears in the reduction.  
If small Gram determinants occur, alternative
schemes are applied \cite{Denner:2005nn}.%
\footnote{Similar procedures based on numerical evaluations of
  specific one-loop integrals~\cite{Binoth:2005ff,Ferroglia:2002mz} or
  expansions in small determinants~\cite{Giele:2004ub} have also been
  proposed by other authors.}  More precisely, we make use of expansions of
the tensor coefficients about the limit of vanishing Gram determinants
and possibly other kinematical determinants.  
 One- and two-point tensor integrals 
are obtained with numerically stable analytic expressions.

Ultraviolet (UV) divergences are regularized dimensionally throughout,
but infrared (IR) divergences are treated in different variants, which
comprise pure dimensional regularization with strictly massless light
quarks and a hybrid scheme with small quark masses 
and massless gluons.  
The corresponding
scalar integrals are evaluated using the methods and results of
\citeres{'tHooft:1978xw,Beenakker:1988jr,Denner:2010tr}, and different
regularization schemes are translated into each other as described in
\citere{Dittmaier:2003bc}.

The calculation of tensor integrals is implemented in two independent
{\sc Fortran} libraries. This permits to perform detailed cross
checks, which confirm the excellent numerical stability of the
reduction procedure.  An automatic cache system is implemented that
strongly boosts the reduction by recycling a multitude of tensor
integrals among Feynman diagrams with common sub-topologies.

\subsubsection{Rational parts}

In $D=4-2\epsilon$ dimensions, UV-singular tensor integrals
give rise to $1/\epsilon_{\mathrm{UV}}$ poles and UV-finite
remainders $\hat{T}^{N}_{j_1\dots j_P}$,
\beqar\label{rationala}
T^{N}_{j_1\dots j_P}
&=&
\hat{T}^{N}_{j_1\dots j_P}
+
\frac{R^{N}_{j_1\dots j_P}}{\epsilon_{\mathrm{UV}}}.
\eeqar
Consequently, their $D$-dependent prefactors $f(D)$
need to be expanded in $D-4$, 
\beqar\label{rationalb}
f(D) T^{N}_{j_1\dots j_P}
&=&
f(4) T^{N}_{j_1\dots j_P}
-2 f'(4) R^{N}_{j_1\dots j_P},
\eeqar
resulting in so-called rational terms that are
proportional to the pole residues $R^{N}_{j_1\dots j_P}$.  Rational
contributions originate from $D$-dependent terms in tensor-reduction
identities and in the loop-momentum-independent part of the diagram
numerators.  

We employ the treatment of rational terms of ultraviolet or infrared
origin as described in Appendix~A of \citere{Bredenstein:2008zb}.  We
use the fact (proven in Appendix~A of \citere{Bredenstein:2008zb})
that in the 't Hooft--Feynman gauge and similar gauge
fixings rational terms of IR origin cancel in truncated one-loop
amplitudes and only contribute via wave-function renormalization
factors. Rational terms of UV origin are obtained in a straightforward
way by performing the relevant expansions about $D=4$ automatically
by means of a catalogue of residues of UV poles 
(see e.g.~\citere{Denner:2005nn}).

\subsubsection{Treatment of unstable top quarks}
\label{se:mut}
The presence of intermediate unstable top quarks in
{$h_1h_2\to\PW^+\PW^-\Pb\bar\Pb\to\leptfs$} represents a non-trivial
new aspect as compared to previous NLO QCD studies of multi-particle
processes.  To regularize intermediate top-quark resonances in a
gauge-invariant way, we employ the complex-mass
scheme~\cite{Denner:2005fg}.  In this approach the top-quark width
$\Gamma_\Pt$ is incorporated into the definition of the renormalized
(squared) top-quark mass \refeq{e:tcms}.  This complex parameter
$\mu_\Pt^2$ is identified with the position of the pole of the
top-quark propagator, and the top-mass counterterm $\delta\mu_\Pt$ is
related to the top-quark self-energy at {$p_\Pt^2=\mu^2_\Pt$} via
\cite{Denner:2005fg}
\beqar
\label{eq:mtct} 
\frac{\delta\mu_\Pt}{\mu_\Pt}&=&\frac{1}{2}\left[
\Sigma^{\Pt,\rR}(\mu^2_\Pt) +\Sigma^{\Pt,\rL}(\mu^2_\Pt)
+2\Sigma^{\Pt,\rS}(\mu^2_\Pt) \right],
\eeqar
where
$\Sigma^{\Pt,\rR}$, $\Sigma^{\Pt,\rL}$, and
$\Sigma^{\Pt,\rS}$ are the left-handed, right-handed, and scalar parts
of the top self-energy, respectively.
This yields the one-loop counterterm
\beqar
\label{eq:mtctexp} 
\frac{\delta\mu_\Pt}{\mu_\Pt}&=&
-\frac{\alphas}{\pi}\left[
\frac{(4\pi)^\epsilon\Gamma(1+\epsilon)}{\epsilon}+\ln\left(\frac{\mu^2}{\mu_\Pt^2}\right)
+\frac{4}{3}
\right]
\eeqar
in $D=4-2\epsilon$ dimensions, where $\mu$ is the reference scale of dimensional regularization.
The evaluation of one-loop scalar box integrals in the presence of
complex masses represents another non-trivial aspect of the
complex-mass scheme.  In our calculation we employ the results of
\citere{Denner:2010tr}, where explicit analytic continuations 
of box integrals have
been presented for all kinematic configurations that are relevant
for physical processes.

\subsubsection{On-shell projection for off-shell W bosons}
\label{se:LPA}
As discussed in \refse{se:wdec}, we treat the leptonic W-boson decays 
in (spin-correlated) NwWA and, alternatively, including also
FwW effects. In the latter case, we employ exact LO and real-emission amplitudes---including
all non-resonant and off-shell effects resulting from the FwW---together with
the one-loop matrix elements and the $I$-operator from dipole subtraction
in ``double-pole approximation'' (DPA).
The DPA represents the leading contribution in an expansion of the corresponding
matrix elements around the resonance poles in the W-boson propagators.

The pole expansion is based on the idea~\cite{Stuart:1991xk,Aeppli:1993rs}
to separate the resonant part 
of an amplitude $\M$
which peaks at $p^2\sim M^2$ 
in the following way,
\beq\label{eq:poleexpansion}
\M(p) = \frac{R(p^2)}{p^2-M^2} + N(p^2) =
\frac{R(M^2)}{p^2-M^2} + \frac{R(p^2)-R(M^2)}{p^2-M^2} + N(p^2),
\eeq
where $R(M^2)$ is the gauge-independent residue of the resonance 
and the non-resonant contributions from $R(p^2)-R(M^2)$ and $N(p^2)$ 
are neglected in DPA.
A consistent
implementation of this idea, however, involves some complications, in
particular, if two resonances instead of one are involved.
Nevertheless this concept was very successfully applied, for instance,
to W-pair production in $\Pep\Pem$ annihilation at LEP2, as e.g.\ 
reviewed in \citere{Grunewald:2000ju}. Here we follow the ``hybrid
concept'' of \citere{Denner:2000bj}, where only the IR-finite 
virtual
corrections are treated in DPA, while keeping the lowest-order and
the remaining
real-emission matrix elements fully off shell.  One of the mentioned
subtleties concerns the appearance of so-called non-factorizable
corrections, which involve
non-analytic terms like $\ln(p^2-M^2)$ in
the resonance region. Such terms are ignored in the naive equation
\refeq{eq:poleexpansion}; 
they are caused by soft photon or gluon exchange
between production and decay processes that are linked by the
resonance. In our case, such non-factorizable corrections do not
appear, since we consider QCD corrections in combination with purely
leptonic W-boson decays, \ie there is no gluon exchange between
W~production and decay.  Thus, only so-called factorizable corrections
are relevant for us, which are just the corrections to the residue
$R(M^2)$.

In this context, a second subtlety arises in the evaluation of $R(M^2)/(p^2-M^2)$
concerning the kinematics. The residue $R(p^2)$ typically depends on the momenta
of all involved particles, and the consistent substitution $p^2\to M^2$ requires
a deformation of momenta on the full off-shell phase space of the reaction.
This ``on-shell projection'' is needed in order to define a
gauge-invariant set of corrections and maps each phase-space point
into an associated phase-space point with on-shell W bosons.
This procedure actually involves some freedom, but this ambiguity
changes the results only at the level of non-resonant terms.

Technically, we proceed as follows to obtain the 
virtual
corrections in DPA for 
the W-boson resonances in
$q\bar q/\Pg\Pg\to\leptfs$.
We select those diagrams that involve two resonant W
bosons and perform a projection of the final-state momenta 
that puts W~bosons onto their mass shells.
Since we deal with corrections of ${\cal O}(\alphas)$ in DPA only,
the intrinsic ambiguity leads to differences
of the order of $\alphas\GW/(\pi\MW)$ for different versions of the
projection, 
provided that the
projection does not induce large changes in the matrix elements.
To this end, 
we must keep the invariant masses of the top and antitop
quarks fixed. In fact, we do not modify the four-momenta of the top and
antitop and define the on-shell projected momenta $\pon_i$ in terms of the
original momenta $p_i$ as follows:
\beqar
\pon_{\Pb}&=&p_{\Pb}\frac{p^2_{\Pt}-\MW^2}{2p_{\Pt}p_{\Pb}}, \qquad
\pon_{\PWp}=p_{\Pt}-\pon_{\Pb},\nl
\pon_{\Pep}&=&p_{\Pep}\frac{\MW^2}{2\pon_{\PWp}p_{\Pep}}, \qquad
\pon_{\Pne}=\pon_{\PWp}-\pon_{\Pep},\nl
\pon_{\Pbbar}&=&p_{\Pbbar}\frac{p^2_{\Ptbar}-\MW^2}{2p_{\Ptbar}p_{\Pbbar}}, \qquad
\pon_{\PWm}=p_{\Ptbar}-\pon_{\Pbbar},\nl
\pon_{\Pmum}&=&p_{\Pmum}\frac{\MW^2}{2\pon_{\PWm}p_{\Pmum}}, \qquad
\pon_{\Pnmubar}=\pon_{\PWm}-\pon_{\Pmum}.
\eeqar
Each top decay is treated separately, the momenta of the bottom quarks
and leptons are rescaled such that $\pon_{\PW^\pm}^2=\MW^2$ and
$\pon_{\Pne}^2=0=\pon_{\Pnmubar}^2$, and the momenta of the neutrinos
and W~bosons are determined from four-momentum conservation.  
The above-mentioned ambiguity in the on-shell projection, e.g., manifests itself in our choice to rescale the charged-lepton momenta and to fix the neutrino
momenta via momentum conservation; the analogous on-shell projection with the roles of charged leptons and neutrinos interchanged would be equally good.
Apart from the W~propagators,
the matrix elements are computed using 
on-shell projected momenta,
physical (\ie real) W- and Z-boson masses,
and $\Gamma_\PW=\Gamma_\PZ=0$.
The phase space 
and the W~propagators are evaluated using the 
original off-shell momenta and including the W-boson width.

This procedure accounts for the most important off-shell effects
in spite of the use of the DPA for the one-loop matrix element.
This fact is strongly supported by the results on W-pair production
in $\Pep\Pem$ annihilation, where comparisons between DPA and full
off-shell calculations exist for LO~\cite{Beenakker:1998gr}
and NLO electroweak corrections~\cite{Denner:2005es,Denner:2005fg}.

\subsection{Real corrections}
\label{se:realcor}

\subsubsection{Matrix-element evaluation}
\begin{figure}
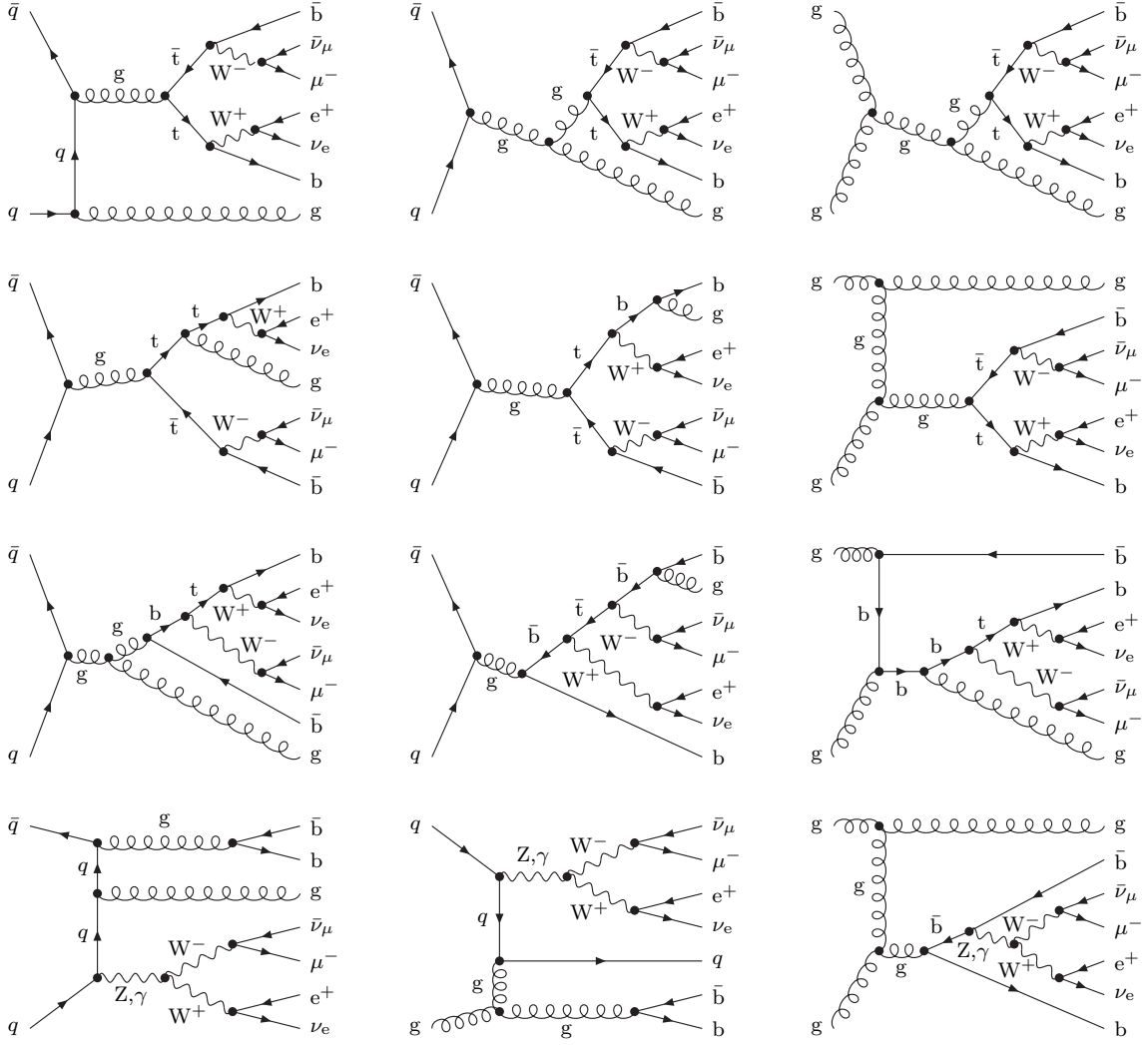
 
\begin{center}
\unitlength 0.85pt
\SetScale{0.85}
\diagramsuuxveepmumvmxbbxgDRproda
\diagramsuuxveepmumvmxbbxgDRprodb
\diagramsggveepmumvmxbbxgDRprodb
\diagramsuuxveepmumvmxbbxgDRinbetween
\diagramsuuxveepmumvmxbbxgDRdecay
\diagramsggveepmumvmxbbxgDRproda
\diagramsuuxveepmumvmxbbxgSRprod
\diagramsuuxveepmumvmxbbxgSRdecay
\diagramsggveepmumvmxbbxgSR
\diagramsuuxveepmumvmxbbxgNRCXCVI
\diagramsguveepmumvmxbbxuNR
\diagramsggveepmumvmxbbxgNR
\end{center}
\caption{Examples of real-emission diagrams with
two (first two lines),
one (third line) or no (last line)
top-quark resonances. All depicted diagrams involve two W-boson resonances.}
\label{fig:nlotrees}
\end{figure}

\begin{figure}
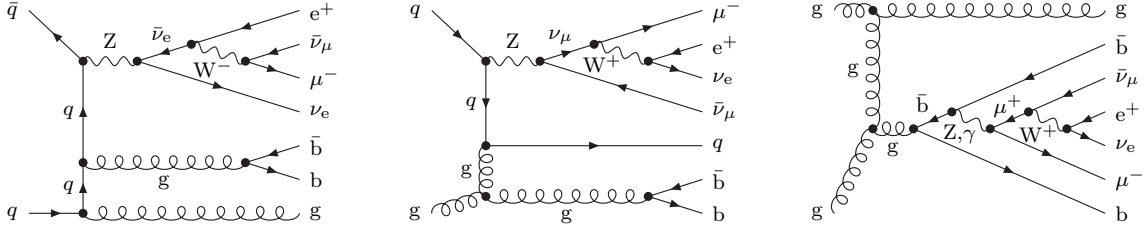
 
\begin{center}
\unitlength 0.85pt
\SetScale{0.85}
\diagramsuuxveepmumvmxbbxgNRoffCXCVIII
\diagramsguveepmumvmxbbxuNRoff
\diagramsggveepmumvmxbbxgNRoff
\end{center}
\caption{Examples of real-emission diagrams that
involve a single W-boson resonance and 
contribute to the FwW corrections.}
\label{fig:nlotreesNR}
\end{figure}

The real corrections receive contributions from the {$2\to 7$}
partonic processes {$\Pg\Pg\to\leptfs\Pg$},
{$\Pq\bar\Pq\to\leptfs\Pg$}, $\Pg q\to\leptfs q$, and 
$\Pg\bar q\to\leptfs\bar q$.  In the NwWA the $\Pg\Pg$ channel
involves 208 tree diagrams, while the $q\bar q$, $\Pg q$, and $\Pg\bar q$
channels, which are related by crossing symmetry, are described by 90
tree diagrams each. 
Examples of doubly-W-resonant real-emission diagrams 
are depicted in \reffi{fig:nlotrees}.
When including FwW effects, additional singly-W-resonant diagrams must
be taken into account (for examples see \reffi{fig:nlotreesNR})
resulting in a total number of 508 and 234 diagrams  for the $\Pg\Pg$
channel and the $q\bar q$ channel, respectively. 
The {$2\to 7$} matrix elements are evaluated with {\sc
  Madgraph}~\cite{Alwall:2007st} and, alternatively, using the
Weyl--van-der-Waerden formalism of \citere{Dittmaier:1998nn}.  
In addition,
in presence of FwW effects we used {\sc OpenLoops} \cite{Cascioli:2011va}
to evaluate the real-emission matrix elements  in the complex-mass scheme.   All real
bremsstrahlung amplitudes have been checked by comparing different
calculations for several phase-space points.

\subsubsection{Treatment of IR singularities}
Owing to IR singularities all the NLO pieces in \refeq{eq:partonic_xs}
are singular, but their sum is finite. When combining virtual and
real corrections, 
soft singularities as well as singularities connected to collinear configurations
in the final state cancel for ``infrared-safe'' observables after
applying a jet algorithm. The remaining singularities connected to the collinear
initial-state splittings 
factorize and are removed via $\overline{\mathrm{MS}}$ PDF redefinitions, 
\ie they are compensated by 
the term $\rd\hat\sigma_{\fact,ab}$ in \refeq{eq:partonic_xs}.

To isolate the IR divergences and cancel them analytically, we employ
the subtraction formalism. To this end, we rewrite
\refeq{eq:partonic_xs} as
\beqar\label{eq:partonic_xs_sub}
\int\rd\hat\sigma_{ab} &=& 
\int_6\rd\hat\sigma_{0,ab}
+\int_6\left[\rd \hat\sigma_{\virt,ab}-\rd \hat\sigma_{\virt,ab}^{\sub}\right]
+\int_7\left[\rd \hat\sigma_{\real,ab}-\rd \hat\sigma_{\real,ab}^{\sub}\right]
\nl&&{}
+\int_0^1\rd x\,\int_6\left[\rd\hat\sigma_{\fact,ab}-\rd\hat\sigma_{\fact,ab}^{\sub}\right]
\nl&=&
\int_6\rd\hat\sigma_{0,ab}
+\int_6\rd \hat\sigma_{\virt,ab}^{\fin}
+\int_7\rd \hat\sigma_{\real,ab}^{\fin}
+\int_0^1\rd x\,\int_6\rd\hat\sigma_{\fact,ab}^{\fin}\,.
\eeqar
Here the added subtraction terms $\rd \hat\sigma_{\virt,ab}^{\sub}$,
$\rd \hat\sigma_{\real,ab}^{\sub}$, and
$\rd\hat\sigma_{\fact,ab}^{\sub}$ are constructed in such a way that
their sum vanishes exactly and that they individually cancel all IR
singularities of the corresponding original terms locally in phase
space. 
As discussed in \refse{se:realcor},
when including FwW effects, the DPA is applied to the
subtracted virtual corrections, \ie the second term on the r.h.s.\ of
\refeq{eq:partonic_xs_sub},
otherwise there would be a mismatch in the IR structure between virtual
corrections and corresponding subtraction terms
(see \citere{Denner:2000bj} for the same reasoning
concerning photon radiation).

For the subtraction terms, we choose in-house implementations of the
dipole subtraction
formalism~\cite{Catani:1996vz,Dittmaier:1999mb,Phaf:2001gc,Catani:2002hc}
for NLO QCD calculations.  Specifically this is done in dimensional
regularization with strictly massless light quarks (including b
quarks) and, alternatively, in a hybrid scheme with small quark masses
and massless gluons with the respective dipole subtraction terms from
\citere{Catani:2002hc}; 
the typical agreement between the results from the two alternative
treatments is at the level of 12 digits.

Soft and collinear singularities in the ``endpoint part'' (the $I$
operator of \citeres{Catani:1996vz,Catani:2002hc}) of the
subtraction function, $\rd \hat\sigma_{\virt,ab}^{\sub}$,  are regularized
using the same regularization prescription (dimensional or with small
quark masses) as the corresponding virtual corrections.  No
regularization is needed in the subtraction terms for the real
corrections, $\rd \hat\sigma_{\real,ab}^{\sub}$. For both the $q\bar
q$ and $\Pg\Pg$ channels 12 different dipole subtraction terms need to
be included while each $\Pg\qparbar$ channel requires only 6, since
we demand b~quarks with finite transverse momentum in the final state.

{\it Version~1} of the real corrections employs 
a fully automatized implementation of dipole subtraction terms, 
which has been developed and tested during the calculation of VV+jet 
production~\cite{Binoth:2010ra,Dittmaier:2009un,Campanario:2010hp}.
The underlying matrix elements are based on helicity amplitudes 
employing the Weyl--van-der-Waerden formalism of \citere{Dittmaier:1998nn}, 
which facilitates the treatment of helicity correlations. The colour 
structure of the dipole terms is generated automatically from the inserted 
colour operators and the underlying Born amplitudes.

In {\it version~2} all dipole subtraction terms have been implemented
by hand  into the Monte Carlo generator.  Colour and helicity
correlations that enter the subtraction procedure are generated by
means of {\sc AutoDipole}~\cite{Hasegawa:2009tx} 
in NwWA
and {\sc OpenLoops} \cite{Cascioli:2011va} for the FwW case.

The cancellation between real matrix elements and dipole subtraction
terms has been verified numerically in all soft and collinear regions.
The individual dipole subtraction terms, the subtracted real matrix
elements, and the integrated subtraction terms ($P$ and $K$ terms of
\citeres{Catani:1996vz,Catani:2002hc}) have been compared point-wise
between the two independent calculations. The agreement was generally
at the level of $13{-}14$ digits.

\subsubsection{Phase-space integration and numerical performance}
To achieve sufficient numerical stability, we perform the
17-dimensional phase-space integration using multi-channel Monte Carlo
techniques~\cite{Berends:1994pv} with adaptive  weight
optimization~\cite{Kleiss:1994qy}.  
For each of the bremsstrahlung
Feynman diagrams a corresponding channel is taken into account in
the Monte Carlo integration.
In addition, 
the integration of the
dipole-subtracted {$2\to 7$} contributions is optimized by means of
extra channels corresponding to the dipole kinematics.
For each of the 12 different dipoles and
each tree-level diagram a new channel is introduced resulting in
$12\times 31=372$ and $12\times 14=168$ 
extra channels in the NwWA in the
$\Pg\Pg$ and $\Pq\Pqbar$ channels, respectively. When including FwW
effects the number of additional channels grows to $12\times 79=948$
and $12\times 38=456$, respectively.  In the $\Pg\qparbar$ channel 
dipoles based on both $\Pg\Pg$ and $\Pq\Pqbar$ Born subprocess are needed. 
So $3\times 31 + 3\times 14=135$ and $3\times 79 + 3\times 38=351$ 
channels are added here in the NwWA and with FwW effects, respectively.
These additional channels lead to some improvement in the convergence
of the Monte Carlo integration.

\begin{sloppypar}
In {\it version~1}, the phase-space integration, implemented in {\sc C++}, 
is based on {\sc RacoonWW} \cite{Denner:1999gp}, but the phase-space mappings 
are built up in a more generic way very similar to the approach of 
{\sc Lusifer}~\cite{Dittmaier:2002ap}.  
\end{sloppypar}

The Monte Carlo generator of {\it version~2} is a further development
of the one used in {\sc COFFER$\ga\ga$} \cite{Bredenstein:2005zk} and
for the calculation of the NLO corrections to $\Pp\Pp\to\PH\jet\jet+X$
\cite{Ciccolini:2007jr} and $\Pp\Pp\to\Pt\Ptbar\Pb\Pbbar+X$
\cite{Bredenstein:2010rs}.  

For the complete NLO cross section we found
agreement between the two versions of our code at the few-per-mille  level.
We also performed a detailed comparison of all differential distributions 
presented in this paper and found good agreement.

For a typical run we generate $10^8$ phase-space points,
and the fraction of points within cuts is roughly 50\%.
This yields an accuracy for the NLO cross section of
about 0.2--0.3\%.  For the LHC, the contributions of the
$\Pg\raisebox{.6em}{\tiny $(-)$}\hspace{-.83em}\Pu$,
$\Pg\raisebox{.6em}{\tiny $(-)$}\hspace{-.83em}\Pd$, $\Pu\Pubar$,
and $\Pd\Pdbar$ channels were calculated for every 2nd, 5th, 4th, and
5th event, respectively. The virtual corrections for the $\Pg\Pg$,
$\Pu\Pubar$, and $\Pd\Pdbar$ channels were evaluated for every 100th,
200th, and 500th event, respectively. The resulting runtime on a
3\,GHz Intel Xeon processor using the {\tt ifort} compiler is about
$170\,\mathrm{h}$.  The bulk of the runtime is taken by the real
corrections to the $\Pg\Pg$ channel.  For the virtual corrections the
CPU time is dominated by the $\Pg\Pg$ channel and amounts to $180\,$ms
per event.  For the Tevatron, the weights of the different
contributions are different, the runtime is somewhat smaller for a
comparable accuracy.  For the numerical results shown in the following
almost $10^9$ events were generated.

\section{Numerical results}
\label{se:numres}
In the following we present predictions for various observables
that are relevant for top-pair production,
either as signal or as background to Higgs production or new physics.
Neither the following setup nor the considered observables
are deliberately chosen to enhance the calculated off-shell and
finite-top-width effects. Our aim rather is to quantify these effects
for the standard observables.

\subsection{Input parameters}
\label{se:setup}
We study the process $h_1h_2\to\leptfs+X$ at
Tevatron ($\Pp\bar\Pp$ collisions) with $\sqrt{s}=1.96\TeV$ and at the
LHC ($\Pp\Pp$ collisions) for the collider energies $7\TeV$, $8\TeV$, and
$14\TeV$.  The input parameters and the default setup are basically
the same as in \citere{Denner:2010jp}, where first results of our
calculation have been presented. For completeness we specify the input
and discuss modifications.

In NLO\,(LO) QCD we employ the MSTW2008NLO\,(LO) parton
distributions~\cite{Martin:2009iq} and use the running of the strong
coupling constant $\alphas$ with two-loop\,(one-loop) accuracy as
provided by the LHApdf library.  In the renormalization of the strong
coupling constant the top-quark loop in the gluon self-energy is
subtracted at zero momentum,   and the number of active flavours is
$N_{\mathrm{F}}=5$. In this scheme, the running of
$\alpha_{\mathrm{s}}$ is generated solely by the contributions of the
light-quark and gluon loops.  Contributions induced by the strongly
suppressed bottom-quark parton density are neglected.  For the
gauge-boson and top-quark masses we use
\beq
\Mt=172.0\GeV, \quad \MW=80.399\GeV, \quad \MZ=91.1876\GeV.
\eeq  
The masses of all other quarks, including b quarks, are neglected.  In
view of the negligibly small Higgs-mass dependence we adopt the
{$\MH\to\infty$} limit, \ie we neglect closed fermion loops
involving top quarks coupled to Higgs bosons.
The electroweak coupling is 
derived from the Fermi constant in the $\GF$-scheme,
\beq\label{eq:GFscheme}
\GF=1.16637\times10^{-5}\GeV^{-2},\quad
\alpha=\frac{\sqrt{2}}{\pi}\GF\/\MW^2\left(1-\frac{\MW^2}{\MZ^2}\right).
\eeq
While in narrow-W-width approximation (NwWA) we employ the usual real-valued
weak mixing angle, $\sw^2=1-\MW^2/\MZ^2$, when including finite-W-width (FwW) effects 
we employ the complex-valued mixing angle \refeq{e:WZcms}.
To derive the electromagnetic coupling \refeq{eq:GFscheme},
we always use the real W- and Z-boson masses.

The leptonic decays of W~bosons are treated in two different ways:
in the (spin-correlated) NwWA and including FwW effects. In the latter case, as
discussed in \refse{se:wdec}, FwW contributions have to be included
both in the matrix elements and in the top-quark width used as input
parameter.  The top-quark width for unstable W~bosons in NLO QCD was
given in \citere{Jezabek:1988iv}.  Neglecting the bottom-quark mass,
as we do throughout this paper, the LO top-quark width reads
\newcommand{\gaw}{\gamma_\PW}
\beq
\label{eq:topwidthLO}
\GtLO=\frac{\GF\Mt^5}{16\sqrt{2}\pi^2\MW^2}\int_0^1 \frac{\rd y\,
  \gaw}{(1- y/\bar y)^2+\gaw^2}\,F_0(y)
\eeq
with $\gaw=\GW/\MW$, $\bar y=(\MW/\Mt)^2$, and
\beq
F_0(y)=2(1-y)^2(1+2y).
\eeq
Including NLO QCD corrections, the top-quark width is given by
\beq
\GtNLO=\frac{\GF\Mt^5}{16\sqrt{2}\pi^2\MW^2}\int_0^1 \frac{\rd
  y\,\gaw}{(1-y/\bar y)^2+\gaw^2}\biggl[F_0(y)-\frac{2\alphas}{3\pi}F_1(y)\biggr]
\eeq
with
\beqar
\label{eq:topwidthF1}
F_1(y)&=&2(1-y)^2(1+2y)\left[
           \pi^2+2\Li(y)-2\Li(1-y)\right]
\nl&&{}
           +4y(1-y-2y^2)\ln(y)
           +2(1-y)^2(5+4y)\ln(1-y)
\nl&&{}
           -(1-y)(5+9y-6y^2).
\eeqar
In NwWA, \ie for $\gaw\to0$, the top width follows from \refeq{eq:topwidthLO}--\refeq{eq:topwidthF1}
by the replacement
\beq
 \frac{\gaw}{(1- y/\bar y)^2+\gaw^2}\to\pi\bar y\,\delta(y-\bar y).
\eeq

With the above formulas and our input parameter set we obtain
\beq\label{eq:GtNwWA}
\GtLO=1.4655\GeV, \qquad \GtNLO=1.3376\GeV
\eeq
in the NwWA and
\beq\label{eq:GtfwW}
\GtLO=1.4426\GeV, \qquad \GtNLO=1.3167\GeV
\eeq
including FwW corrections.  
Since the leptonic W-boson
decays do not receive NLO QCD corrections and 
$\leptfs$ production does not involve $\PZ\to f\bar f$ subprocesses,
for the gauge-boson widths we use the NLO QCD values%
\footnote{Using the measured Z-boson width instead has no significant
  effect on our results.}
\beq
\label{eq:GammaWZ}
\Gamma_\PW=2.09974\GeV,\qquad\Gamma_\PZ=2.50966\GeV
\eeq
everywhere, \ie for LO as well as for NLO matrix elements.

\subsection{Jet definition, cuts, and scale choice}
\label{se:setupB}
We now turn to the event selection. Final-state quarks and gluons with
pseudo-rapidity {$|\eta|<5$} are converted into infrared-safe jets
using the anti-$\kT$ algorithm~\cite{Cacciari:2008gp}.  The
jet-resolution parameter $R$ is set to $R=0.4$ and $R=0.5$ for the
Tevatron and the LHC, respectively.  After recombination, we impose
cuts on the transverse momenta and pseudo-rapidities of the leptons
and b~jets, and on the missing transverse momentum. For Tevatron
we choose
\beqar\label{eq:Tevatroncuts}
p_{\mathrm{T,\Pb}}&>&20\GeV,\qquad
{|\eta_{\Pb}|<2.5},\qquad
{p_{\mathrm{T,miss}}>25\GeV},\\\nn
p_{\mathrm{T},\Pl}&>&20\GeV, \qquad
{|\eta_{\Pl}|<2.5},
\eeqar
and for the LHC
\beqar\label{eq:LHCcuts}
p_{\mathrm{T,\Pb}}&>&30\GeV,\qquad
{|\eta_{\Pb}|<2.5},\qquad
{p_{\mathrm{T,miss}}>20\GeV},\\\nn
p_{\mathrm{T},\Pl}&>&20\GeV, \qquad
{|\eta_{\Pl}|<2.5},
\eeqar
where $p_{\mathrm{T,miss}}$ is obtained from the vector sum of all visible
transverse momenta after jet recombination.

For the factorization ($\mu_\rF$) and renormalization ($\mu_\rR$)
scales we have considered different choices. A common approach, which
was adopted in \citeres{Denner:2010jp,Bevilacqua:2010qb}, is to set
$\mu_\rR=\mu_\rF=\Mt$.  However, as shown in \refse{se:distr}, this
choice leads to perturbative instabilities in the high-energy tails of
differential distributions. The use of a dynamical scale, which adapts
to the hard scattering energy, guarantees a much better convergence of
the perturbative expansion.  We thus consider an alternative scale
choice, based on the kinematic variable
\beq
\label{eq:dynscale}
\dynscale = \sqrt{\sqrt{\Mt^2+p_{\rT,\Pt}^2}\sqrt{\Mt^2+p_{\rT,\Ptbar}^2}},
\eeq
which corresponds to the geometric average of the top- and
antitop-quark transverse energies.  Similar scales had already been
used in early papers
\cite{Nason:1989zy,Beenakker:1990maa,Frixione:1995fj}.  The top
transverse energy $\dynscale$ coincides with $\Mt$ for vanishing
transverse momenta and adapts to the higher scattering energy at large
transverse momenta.  While we use the scales $\mufix=\Mt$ or
$\mudyn=\dynscale$ to describe $\leptfs$ production at Tevatron, for
the LHC we use half of these scales, \ie $\mufix=\Mt/2$ or
$\mudyn=\dynscale/2$.  The different scale choice is motivated by the
fact that $\Pt\Ptbar$ production at the Tevatron is dominated by
$s$-channel quark--antiquark annihilation, while the dominant
$\Pt\bar\Pt$ production mechanism at the LHC is $t$-channel gluon
fusion, which prefers smaller scales.  Moreover, in
\citere{Bonciani:1998vc} it has been demonstrated that the
contributions beyond NLO in an NLL soft-gluon resummation are smaller
for $\mu=\Mt/2$ than for $\mu=\Mt$.
\begin{table}
$$\arraycolsep10pt
\begin{array}{c|c|c|c|c}
\mathrm{collider} &  \mufix & \mudyn & \sqrt{s}\,[\TeV] &\bar\mudyn\,[\GeV] \\
\hline
\mathrm{Tevatron} &   \Mt   &  \dynscale   &  1.96 & 203.1  \\   
\hline
\mathrm{LHC}      &   \Mt/2 &  \dynscale/2 &  7  & 105.8  \\    
                  &         &              &  8  & 106.5  \\    
                  &         &              &   14 & 109.2  \\   
\end{array}
$$
\caption{Fixed ($\mufix$) and dynamical ($\mudyn$) scales used for 
Tevatron and LHC predictions. The last column shows the logarithmic
average of the dynamical scale,  as defined in \refeq{eq:logaver}.}
\label{tab:scales}
\end{table}
Our different scale choices are summarized in \refta{tab:scales}.
There we also show the logarithmic average $\bar\mudyn$ of the
dynamical scale, defined via
\beq\label{eq:logaver}
\ln\bar\mudyn =\frac{\int\ln{(\mudyn)}\rd\sigma}{\int\rd\sigma}.
\eeq
The numerical values of $\bar\mudyn$ indicate that, for what concerns
the integrated cross section, using the dynamical scale corresponds to
an effective increase of the fixed scale by roughly $18{-}27\%$,
depending on the collider energy. 

The scale uncertainty of our LO and NLO predictions is determined by
uniform variations of the renormalization and factorization scales,
\beq
\muF=\muR=\mu,
\eeq
around the central values
\beq
\mu_0=\mufix\qquad \mbox{or}\qquad \mu_0=\mudyn.
\eeq
When varying the renormalization scale in PDFs and matrix elements,
we keep it fixed in the top-quark width, which is always evaluated at the scale $\Mt$.
The mismatch between the scales used in partial and total top-decay widths 
is compensated by the partial-width correction \refeq{scalecorr},
which we include in $\NLOm$ predictions as discussed in \refse{se:matching}.
In \refse{se:xsec}, to investigate the scale dependence of the LO and NLO
integrated cross section we vary $\mu$ up and down by a factor eight.
For all other results
we provide LO and NLO predictions with
uncertainties corresponding to factor-two scale variations. 
More
precisely, the observables are evaluated at three different scales,
$\mu/\mu_0=0.5,1,2$; the central value is obtained for
$\mu=\mu_0$, and the error band is determined by 
the envelope of the three scales.

\subsection{Total cross section and scale dependence}
\label{se:xsec}
We first present 
Tevatron and LHC total cross sections
with fixed 
and dynamical scales (see \refta{tab:scales}), both
in NwWA and including FwW effects.
The results, listed in \refta{tab:xsec}, correspond to the 
standard cuts defined in \refeqs{eq:Tevatroncuts} and \refeqf{eq:LHCcuts}.
\begin{table}
$$
\arraycolsep 5pt
\begin{array}{cccccccc}
\sqrt{s}
& \mu_0  & \Ga_\PW & \sigma_{\LO} & K & \si_{\NLO} 
&\frac{\sigma_\NLOm}{\si_\NLO} & \sigma_\NLOm \\
{}[\mathrm{TeV}] &&&[\mathrm{fb}]&& [\mathrm{fb}]&& [\mathrm{fb}]  \\
\hline
1.96 & \Mt & \mathrm{\FwW} &     44.197(2)^{    +44.4 \%}_{    -28.2 \%} & 0.95 &      41.77(2)^{     -8.8 \%}_{     -5.2 \%} & 0.98 &      40.93(2)^{     +2.3 \%}_{    -10.2 \%}\\
1.96 & \Mt & \mathrm{\NwWA} &     44.304(2)^{    +44.4 \%}_{    -28.2 \%} & 0.94 &      41.78(2)^{     -9.1 \%}_{     -5.1 \%} & 0.98 &      40.96(2)^{     +2.1 \%}_{    -10.0 \%}\\
1.96 & \dynscale & \mathrm{\FwW} &     40.480(2)^{    +42.8 \%}_{    -27.5 \%} & 1.03 &      41.77(2)^{     -3.3 \%}_{     -7.1 \%} & 0.96 &      40.23(2)^{     +5.7 \%}_{    -11.3 \%}\\
1.96 & \dynscale & \mathrm{\NwWA} &     40.580(2)^{    +42.8 \%}_{    -27.5 \%} & 1.03 &      41.79(3)^{     -3.6 \%}_{     -7.0 \%} & 0.96 &      40.27(3)^{     +5.4 \%}_{    -11.2 \%}\\
\hline
7 & \Mt/2 & \mathrm{\FwW} &     922.22(3)^{    +44.9 \%}_{    -28.5 \%} & 0.93 &      862.1(8)^{    -13.1 \%}_{     -2.9 \%} & 1.01 &      870.4(8)^{     -0.2 \%}_{     -8.7 \%}\\
7 & \Mt/2 & \mathrm{\NwWA} &     925.77(3)^{    +44.8 \%}_{    -28.5 \%} & 0.93 &      864.1(8)^{    -13.4 \%}_{     -2.8 \%} & 1.01 &      872.7(8)^{     -0.4 \%}_{     -8.6 \%}\\
7 & \dynscale/2 & \mathrm{\FwW} &     824.00(3)^{    +42.8 \%}_{    -27.6 \%} & 1.05 &      866.9(6)^{     -5.5 \%}_{     -5.7 \%} & 0.99 &      854.0(7)^{     +4.2 \%}_{    -10.4 \%}\\
7 & \dynscale/2 & \mathrm{\NwWA} &     827.22(3)^{    +42.8 \%}_{    -27.6 \%} & 1.05 &      867.0(8)^{     -5.9 \%}_{     -5.5 \%} & 0.99 &      854.7(8)^{     +3.9 \%}_{    -10.3 \%}\\
\hline
8 & \Mt/2 & \mathrm{\FwW} &    1278.20(4)^{    +43.1 \%}_{    -27.8 \%} & 0.95 &      1219(1)^{    -10.9 \%}_{     -3.2 \%} & 1.01 &      1226(1)^{     +0.9 \%}_{     -8.8 \%}\\
8 & \Mt/2 & \mathrm{\NwWA} &    1283.06(5)^{    +43.1 \%}_{    -27.8 \%} & 0.95 &      1221(1)^{    -11.2 \%}_{     -3.1 \%} & 1.01 &      1229(1)^{     +0.7 \%}_{     -8.7 \%}\\
8 & \dynscale/2 & \mathrm{\FwW} &    1141.69(4)^{    +41.1 \%}_{    -26.9 \%} & 1.07 &      1221(1)^{     -4.1 \%}_{     -5.8 \%} & 0.98 &      1200(1)^{     +5.0 \%}_{    -10.3 \%}\\
8 & \dynscale/2 & \mathrm{\NwWA} &    1146.17(4)^{    +41.1 \%}_{    -26.9 \%} & 1.07 &      1225(1)^{     -4.3 \%}_{     -5.7 \%} & 0.98 &      1203(1)^{     +4.8 \%}_{    -10.3 \%}\\
\hline
14 & \Mt/2 & \mathrm{\FwW} &     4416.8(2)^{    +36.3 \%}_{    -24.8 \%} & 1.01 &      4468(4)^{     -3.1 \%}_{     -4.4 \%} & 0.99 &      4439(4)^{     +5.2 \%}_{     -9.0 \%}\\
14 & \Mt/2 & \mathrm{\NwWA} &     4433.3(2)^{    +36.3 \%}_{    -24.8 \%} & 1.01 &      4473(4)^{     -3.3 \%}_{     -4.3 \%} & 0.99 &      4445(5)^{     +5.0 \%}_{     -8.9 \%}\\
14 & \dynscale/2 & \mathrm{\FwW} &     3953.1(2)^{    +34.6 \%}_{    -24.0 \%} & 1.12 &      4420(4)^{     +0.8 \%}_{     -6.1 \%} & 0.97 &      4301(4)^{     +7.6 \%}_{     -10. \%}\\
14 & \dynscale/2 & \mathrm{\NwWA} &     3968.1(2)^{    +34.6 \%}_{    -24.0 \%} & 1.12 &      4430(4)^{     +0.7 \%}_{     -6.0 \%} & 0.97 &      4311(4)^{     +7.5 \%}_{     -9.9 \%}\\
\hline

\end{array}
$$
\caption{Total cross section for
  $\Pp\Pp/\Pp\bar\Pp\to\leptfs+X$
  within cuts for Tevatron and LHC at different CM energies in the
  NwWA and including FwW effects
  both with fixed and dynamical scales.
  In the cross-section numbers the upper variation corresponds to the scale
  $\mu=0.5\mu_0$, the lower to $\mu=2\mu_0$.}
\label{tab:xsec}
\end{table}%
Besides the LO and NLO $\leptfs$ cross sections,
we also present improved results
that include the $\NLOm$ matching corrections of \refse{se:matching}
and can be regarded as our best predictions.
These $\sigma_\leptfs^\NLOm$ results
incorporate higher-order effects determined by the
relation \refeq{eq:inclcsfw}
between the narrow-top-limit of the inclusive $\leptfs$ cross section and the on-shell $\ttb$ cross section,
as well as the ``partial-with correction'' \refeq{scalecorr}. 

\begin{figure}
\centerline{
\includegraphics[width=.45\textwidth]{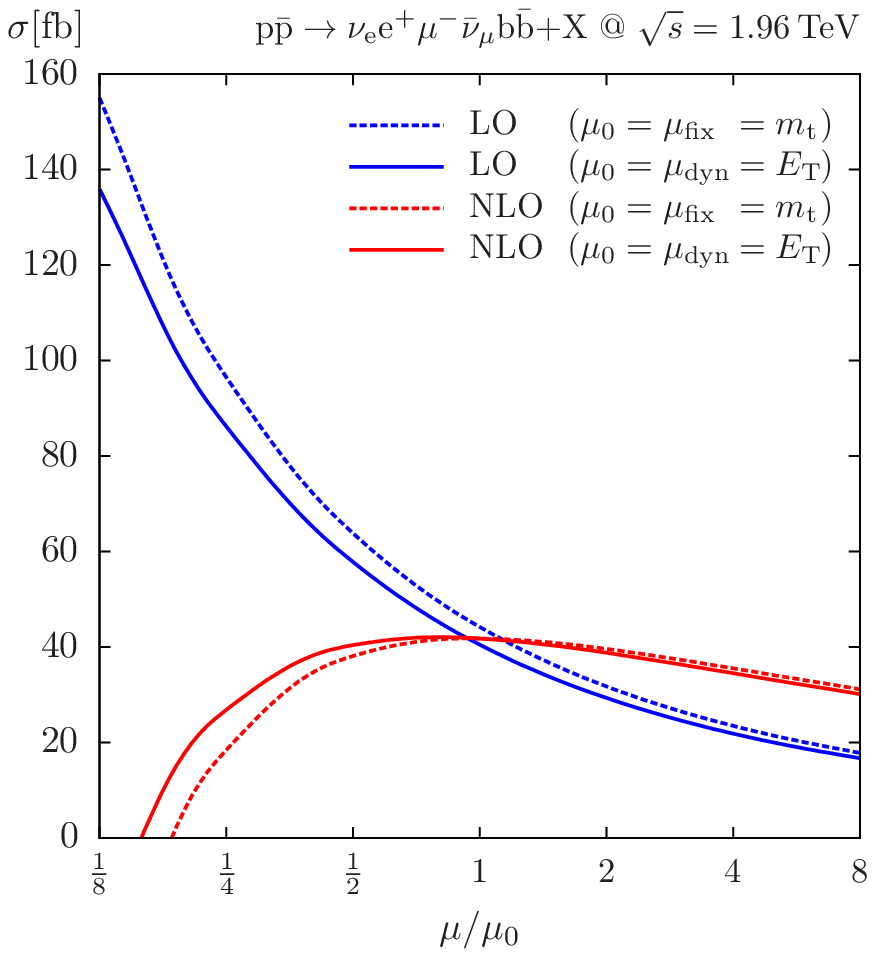}
\qquad
\includegraphics[width=.45\textwidth]{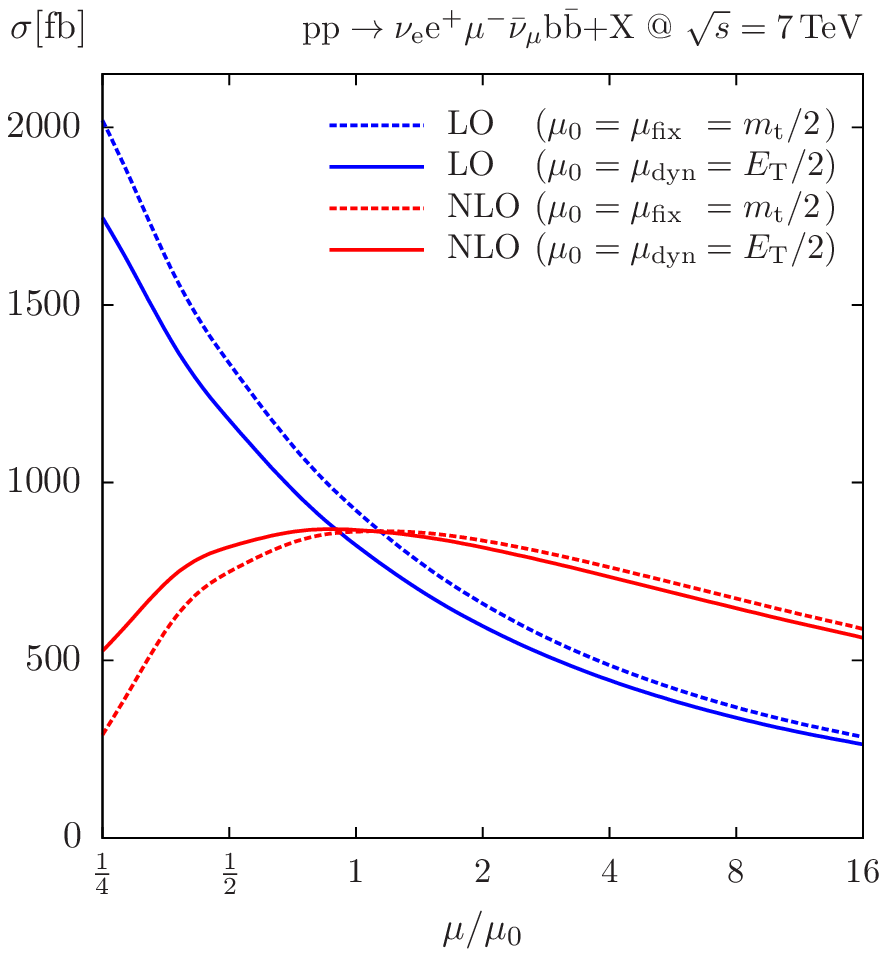}}
\caption{Scale dependence of the LO and NLO cross sections at Tevatron (left)
  and the LHC with $\sqrt{s}=7\TeV$ (right). The renormalization and factorization scales are 
varied around the fixed ($\mu_0=\mufix$) or dynamical ($\mu_0=\mudyn$) central values  defined in \refta{tab:scales}.}
\label{fi:scaledep1}
\vspace*{1cm}
\centerline{
\includegraphics[width=.45\textwidth]{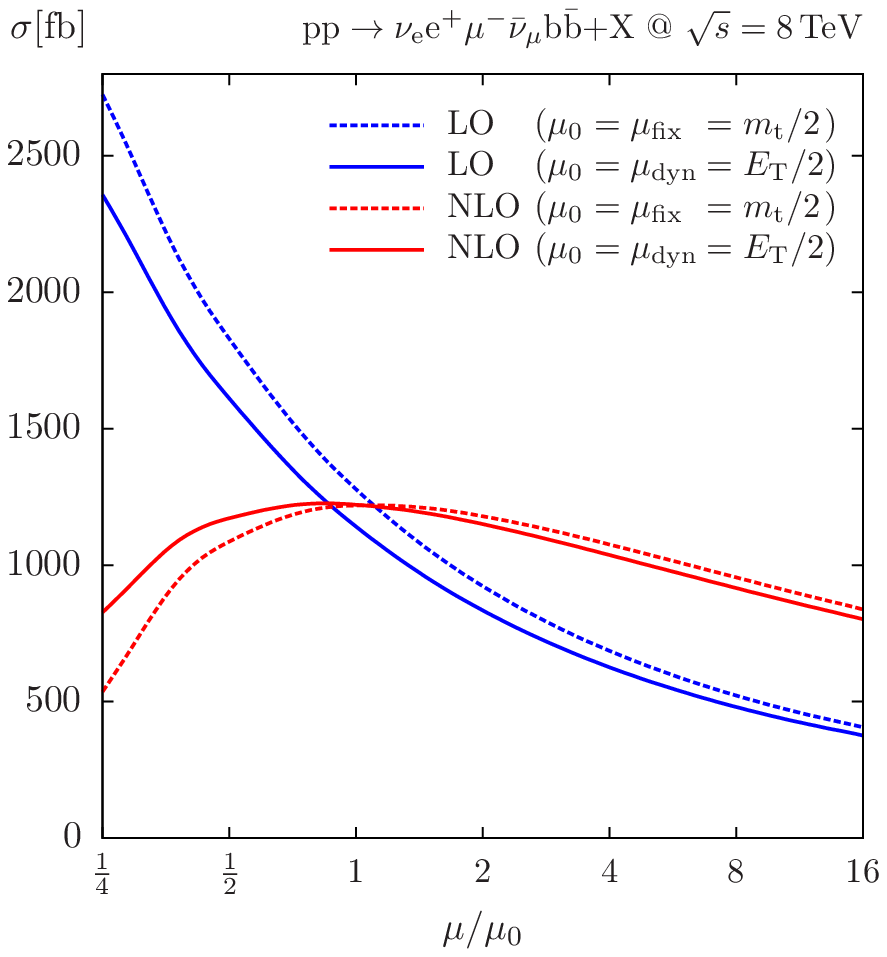}
\qquad
\includegraphics[width=.45\textwidth]{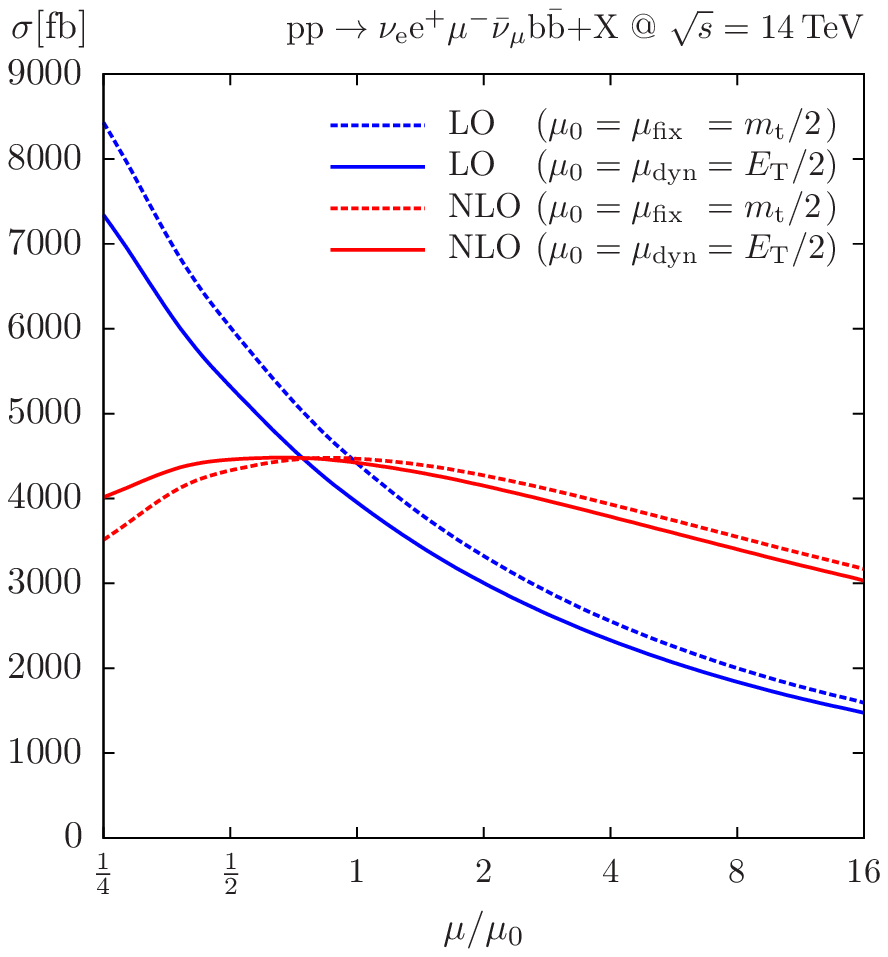}}
\caption{Scale dependence of the LO and NLO cross sections at the LHC with $\sqrt{s}=8\TeV$ (left) and
$14\TeV$ (right). The renormalization and factorization scales are 
varied around the fixed or dynamical central values $\mu_0$ defined in \refta{tab:scales}.}
\label{fi:scaledep2}
\end{figure}

At the Tevatron, the $K$ factor is $0.95$ for the fixed scale and
$1.03$ for the dynamical scale. Similarly, for the LHC we find $K$
factors slightly below one for the fixed scale and $5{-}12\%$ above
one for the dynamical scale. If we had not reduced the central
scale by a factor $2$, the $K$ factors at the LHC would be around
$1.27{-}1.38$. The use of the dynamical scale increases the $K$
factor by about $10\%$ and changes the NLO predictions by less than $1\%$.
The inclusion of the NLO corrections reduces the
scale dependence from $36{-}45\%$ to $4{-}13\%$ at the LHC and from 
$44\%$ to $9\%$ at the Tevatron.
In most cases, the NLO cross section goes down irrespectively of the 
direction of the scale variation. This
indicates the presence of a maximum of 
$\sigma_\NLO(\mu)$ in the vicinity of the central scale $\mu_0$ 
(see \reffis{fi:scaledep1}--\ref{fi:scaledep2}).
The residual scale
dependence is comparable for fixed and dynamical scales. 
Including the partial-width correction \refeq{scalecorr}
in the fixed-order NLO predictions would result in 
a shift of roughly $+2\%$ in the central values of the LHC cross sections 
(due to the reduced central scale),
while the upper ($\mu=0.5\mu_0$) and lower ($\mu=2\mu_0$)
NLO scale variations at Tevatron and the LHC 
would move by $+2\%$ and $-2\%$, respectively.

The FwW effects amount to $-0.3\%$ at the Tevatron and $-0.4\%$ at the
LHC rather independent of the collider energy and the scale choice.
These per-mille-level FwW effects confirm the strong suppression
anticipated in \refse{se:wdec}.  If we did not use the top width
calculated with FwW effects, the inconsistency of the branching
fraction \refeq{eq:BR} would lead to a fake FwW effect of roughly $-3\%$.
The $\NLOm$ correction, which matches our NLO results to the
on-shell $\ttb$ cross section and includes the partial-width correction
\refeq{scalecorr}, ranges from $-4\%$ to $+1\%$, depending on the scale 
choice and the collider energy. The scale dependence of $\sigma_{\NLOm}$ 
is at the level of $10\%$.

\reffis{fi:scaledep1} and \ref{fi:scaledep2} display the
 dependence of the LO and NLO cross sections 
under uniform variations of the fixed and dynamical
QCD scales at Tevatron and the LHC with $\sqrt{s}=7\TeV$, $8\TeV$, and $14\TeV$.
The fact that the central scale $\mu_0$ is shifted to the left
in the LHC plots reflects the collider-dependent scale choice
(see \refta{tab:scales}).
For Tevatron, the maximum of the NLO cross section is obtained for
$\mu\simeq \mu_0=\Mt$ or $\dynscale$, and the $K$ factor is close to one in this
region.  For the LHC, the maximum of the NLO cross section is shifted
to smaller values near 
$\mu\simeq\mu_0=\Mt/2$  or $\dynscale/2$, and also the $K$ factor
reaches a minimum there. These features are quite independent of the
LHC energy and can be attributed to the different dominant
production channels at Tevatron and the LHC. 
The improved stability of the cross section against NLO corrections and scale variations
supports the use of the ``reduced scales'' $\mufix=\Mt/2$ and $\mudyn=\dynscale/2$
at the LHC.

\subsection{Asymmetries}
\label{se:asymm}
Asymmetries in top--antitop production have found particular interest
recently, triggered by experimental results from Tevatron
\cite{Aaltonen:2011kc,Abazov:2011rq,Aaltonen:2012}.  The situation is
different for $\Pp\bar\Pp$ collisions at Tevatron and $\Pp\Pp$
collisions at the LHC.  In this section we discuss FwW effects and
compare asymmetries obtained with fixed and dynamical scale choices.
Results based on dynamical scales and including FwW effects should be
regarded as our ``best'' NLO QCD predictions.  
For a recent review, including a discussion of
electroweak correction effects 
\cite{Hollik:2011ps,Kuhn:2011ri},
we refer to \citere{Bernreuther:2012sx}.

\subsubsection{Asymmetries for Tevatron}
At LO, on-shell top--antitop production is totally symmetric under
the exchange $\Pt\leftrightarrow\bar\Pt$
both for $\Pq\Pqbar$ and $\Pg\Pg$ production
mechanisms.  
For the $\Pg\Pg$ channel this is a consequence of C~invariance of QCD,
for $\Pq\Pqbar$ the symmetry is accidental.
Consequently, the LO angular distributions of the top and
antitop quarks are forward--backward symmetric for
$\Pp\bar\Pp$ collisions. 
This accidental symmetry is violated by
NLO corrections, in particular by interference terms in
$\Pq\Pqbar\to\Pt\Ptbar\Pg$ and $\Pg\Pq\to\Pt\Ptbar\Pq$ radiative
processes \cite{Halzen:1987xd} and by interference terms in the
$\Pq\Pqbar$ channel between virtual corrections and the leading order
\cite{Kuhn:1998kw}. Since top--antitop production at Tevatron is
dominated by the $\Pq\Pqbar$ channel, an observable forward--backward
asymmetry emerges. In the laboratory frame this can be defined as
\beq\label{eq:afbt}
A^{\FB}_{\Pt,\lab}=\frac{\si(y_{\Pt}>0)-\si(y_{\Pt}<0)}
 {\si(y_{\Pt}>0)+\si(y_{\Pt}<0)},
\eeq
where $y_{\Pt}$ and $y_{\Ptbar}$ are the rapidities of the top and
antitop quark, respectively. 
In our predictions we set $\si=\si_{\NLO}$ both in numerator
and denominator. Owing to CP symmetry, the asymmetry of the antitop is
oppositely equal to the asymmetry of the top quark.  
Our numerical results for the asymmetry \refeq{eq:afbt} are given in the
second column of the upper part of \refta{tab:afb_t_woff} including
FwW effects.  Compared to the result for the fixed scale $\mu_0=\Mt$,
the asymmetry is lowered by $10{-}15$\% for the dynamical scale
$\mu_0=\dynscale$.  
The scale uncertainty is large, since the numerator is effectively known
only in leading order and has a scale dependence like $\alphas^3$, while the
NLO cross section in the denominator is rather stable against scale
variations.
Within the integration accuracy, the
results for on-shell and off-shell W~bosons coincide.

In the literature, the asymmetry $A^{\FB}_{\Pt,\lab}$ has often been
calculated by using the LO cross section in the denominator
\cite{Kuhn:2011ri,Bernreuther:2010ny,Hollik:2011ps,Kuhn:1998kw}. This
reduces the scale dependence by at least a factor of $4$, 
while using NLO predictions in the numerator and the denominator of \refeq{eq:afbt}, as in the
present paper, provides a more conservative estimate of theoretical uncertainties.
In any case, this choice does not really affect the
size of the finite-width effects that we are after. We do not include
QED or electroweak corrections.

We note that for off-shell top quarks an asymmetry already exists in
LO, e.g.\ from non-resonant diagrams where \PW~bosons are emitted from
initial-state quarks. This asymmetry is, however, only at the level of
$10^{-4}$ and thus negligible.

An analogous asymmetry can be defined for the charged leptons.  
Results for the positron including FwW effects are given in the third
column of the upper part of \refta{tab:afb_t_woff}, both for fixed and
dynamical scales.  Our results at the fixed scale $\mu_0=\Mt$ for both
$A^{\FB}_{\Pt,\lab}$ and $A^{\FB}_{\Pep,\lab}$ agree well with the
corresponding numbers given in
\citeres{Bernreuther:2010ny,Bevilacqua:2010qb,Antunano:2007da}.

Alternatively, asymmetries can be defined in the CM frame of the
$\Pt\Ptbar$ system as
\beq
A^{\FB}_{\Pt,\CM}=\frac{\si(y_{\Pt}>y_{\Ptbar})-\si(y_{\Pt}<y_{\Ptbar})}
{\si(y_{\Pt}>y_{\Ptbar})+\si(y_{\Pt}<y_{\Ptbar})}.
\eeq
At the CDF experiment, these asymmetries have been measured separately 
in two regions of the $\Pt\Ptbar$ invariant mass $M_{\Pt\Ptbar}$, viz.\ for
$M_{\Pt\Ptbar} >450\GeV$ and $M_{\Pt\Ptbar} <450\GeV$, 
with the result \cite{Aaltonen:2012}
\beqar
A^{\FB,\mathrm{CDF}}_{\Pt,\CM}&=&0.162(47)
,\nl
A^{\FB,\mathrm{CDF}}_{\Pt,\CM}(M_{\Pt\Ptbar} <450\GeV)&=&0.078(54)
,\nl
A^{\FB,\mathrm{CDF}}_{\Pt,\CM}(M_{\Pt\Ptbar} >450\GeV)&=&0.296(67)   
.
\eeqar
The D0 experiment has measured \cite{Abazov:2011rq}
\beqar
A^{\FB,\mathrm{D0}}_{\Pt,\CM}&=&0.196(65).
\eeqar
From our calculation, using $\mu_0=\dynscale$ and including FwW effects, we find 
the results given in the lower part of \refta{tab:afb_t_woff}.
While these results agree well with theoretical calculations of other
groups \cite{Bernreuther:2010ny,Antunano:2007da}, the measured
asymmetry is significantly higher, in particular for $M_{\Pt\Ptbar}
>450\GeV$.

\begin{table}
$$
\arraycolsep 5pt
\begin{array}{ccc}
\mu_0 & A^{\FB}_{\Pt,\lab} & A^{\FB}_{\Pep,\lab} \\
\hline
\Mt &     0.0499(5)^{   +0.0342}_{   -0.0143} &     0.0361(5)^{   +0.0256}_{   -0.0107} \\
\dynscale &     0.0454(5)^{   +0.0259}_{   -0.0119} &     0.0321(5)^{   +0.0190}_{  -0.0087} \\
\hline

\end{array}
$$
\vspace*{1em}
$$
\arraycolsep 5pt
\begin{array}{cccc}
\mu_0 & 
A^{\FB}_{\Pt,\CM} &
A^{\FB}_{\Pt,\CM}(M_{\Pt\Ptbar} <450\GeV) &
A^{\FB}_{\Pt,\CM}(M_{\Pt\Ptbar} >450\GeV) \\
\hline
\Mt &     0.0749(5)^{   +0.0514}_{   -0.0214} &     0.0491(6)^{   +0.0267}_{   -0.0126} &      0.1281(11)^{    +0.1286}_{   -0.0421} \\
\dynscale &     0.0683(5)^{   +0.0391}_{   -0.0181} &     0.0486(6)^{   +0.0250}_{   -0.0122} &     0.1078(9)^{   +0.0737}_{   -0.0307} \\
\hline

\end{array}
$$
\caption{Forward--backward asymmetries for the top quark and the charged
  lepton at  the Tevatron including FwW effects, using a
  dynamical $\mu_0=\dynscale$ or a fixed $\mu_0=\Mt$ scale.}
\label{tab:afb_t_woff}
\end{table}%

\subsubsection{Asymmetries for the LHC}
At the LHC, the forward-backward asymmetry of the 
quark- and antiquark-induced partonic subprocesses does not show
up in observables,
since all distributions are forward--backward symmetric for the 
proton--proton  initial state. However, 
as a result of the dominance of valence quarks 
over sea (anti)quarks at large momentum fractions,
the partonic asymmetry 
manifests itself as a hadronic central--edge asymmetry.
In practice,  in the laboratory frame, antitop quarks tend to be produced more centrally than top quarks.
The ATLAS collaboration \cite{ATLAS-Atop:2011} measures an asymmetry based on the
LAB-frame rapidity of the top quarks,
\beq
A^{\CE,y}_{\Pt}=\frac{\si(|y_{\Pt}|>|y_{\Ptbar}|)-\si(|y_{\Pt}|<|y_{\Ptbar}|)}
{\si(|y_{\Pt}|>|y_{\Ptbar}|)+\si(|y_{\Pt}|<|y_{\Ptbar}|)},
\eeq
while CMS \cite{CMS-Atop:2010} prefers to investigate an asymmetry
based on the pseudo-rapidity,
\beq
A^{\CE,\eta}_{\Pt}=\frac{\si(|\eta_{\Pt}|>|\eta_{\Ptbar}|)-\si(|\eta_{\Pt}|<|\eta_{\Ptbar}|)}
{\si(|\eta_{\Pt}|>|\eta_{\Ptbar}|)+\si(|\eta_{\Pt}|<|\eta_{\Ptbar}|)}.
\eeq
Again these asymmetries can be defined for charged leptons as well,
\beq
A^{\CE,y}_{\Pl}=\frac{\si(|y_{\Pep}|>|y_{\Pmum}|)-\si(|y_{\Pep}|<|y_{\Pmum}|)}
{\si(|y_{\Pep}|>|y_{\Pmum}|)+\si(|y_{\Pep}|<|y_{\Pmum}|)}.
\eeq
Since the charged leptons can be considered massless, $A^{\CE,y}_{\Pl}
= A^{\CE,\eta}_{\Pl}$. 
Results for the top-quark and charged-lepton central--edge asymmetries
for the dynamical $\mu_0=\dynscale/2$ and fixed $\mu_0=\Mt/2$ scale
and with FwW effects are given in \refta{tab:ace_t_woff}.  Also in
this case FwW effects turn out to be strongly suppressed (at the
sub-per-mille level).
\begin{table}
$$
\arraycolsep 5pt
\begin{array}{ccccc}
E_{\CM} [\mathrm{TeV}] & \mu_0 & A^{\CE,y}_{\Pt} & A^{\CE,\eta}_{\Pt} &
A^{\CE,y}_{\Pl} = A^{\CE,\eta}_{\Pl} \\
\hline
7 & \Mt/2 &     0.0078(9)^{  +0.0050}_{  -0.0020} &     0.0100(9)^{  +0.0068}_{  -0.0026} &     0.0050(9)^{  +0.0034}_{  -0.0013} \\
7 & \dynscale/2 &     0.0075(7)^{  +0.0038}_{  -0.0017} &     0.0093(7)^{  +0.0049}_{  -0.0022} &     0.0045(7)^{  +0.0022}_{ -0.0010} \\
\hline
8 & \Mt/2 &     0.0063(8)^{  +0.0039}_{  -0.0016} &     0.0087(8)^{  +0.0058}_{  -0.0024} &     0.0027(8)^{  +0.0012}_{ -0.0005} \\
8 & \dynscale/2 &     0.0050(7)^{  +0.0021}_{  -0.0010} &     0.0070(8)^{  +0.0033}_{  -0.0016} &     0.0037(7)^{  +0.0014}_{ -0.0007} \\
\hline
14 & \Mt/2 &     0.0024(9)^{  +0.0016}_{ -0.0007} &     0.0034(9)^{  +0.0021}_{  -0.0010} &     0.0021(9)^{  +0.0013}_{ -0.0006} \\
14 & \dynscale/2 &     0.0032(8)^{  +0.0015}_{ -0.0008} &     0.0047(8)^{  +0.0022}_{  -0.0012} &     0.0021(8)^{ +0.0009}_{ -0.0005} \\
\hline

\end{array}
$$
\caption{Central--edge asymmetries for the top quark and the charged
  lepton  for the LHC at 
  different CM energies including FwW effects, using a
  dynamical $\mu_0=\dynscale/2$ or a fixed $\mu_0=\Mt/2$ scale.}
\label{tab:ace_t_woff}
\end{table}%

\subsection{Differential distributions}
\label{se:distr}
In the following we present various distributions 
obtained by applying the cuts specified in \refse{se:setupB}.
For each of the observables illustrated in 
\reffis{fi:ptep_lhc7_fs}--\ref{fi:ht_tev_vs}
we present three plots. The left plots display
absolute LO (blue, dashed) and NLO (red, solid) predictions together with corresponding
uncertainty bands resulting from scale variations within \mbox{$0.5<\mu/\mu_0<2$}.
In the upper--right plots we show the same LO and NLO predictions normalized to LO results
at the central scale, \ie 
$K_\LO(\mu)=\rd\sigma_\LO(\mu)/\rd\sigma_\LO(\mu_0)$
and
$K_\NLO(\mu)=\rd\sigma_\NLO(\mu)/\rd\sigma_\LO(\mu_0)$. Here
the blue band illustrates the relative scale uncertainty of the LO cross section,
and the central curve of the red band corresponds to the usual NLO correction factor,
$K=K_\NLO(\mu_0)$.
These first two plots always include FwW effects, whose impact is illustrated in the lower--right plots.
There we display the FwW correction factor
\beq
\Delta_{\scriptsize\mbox{FwW}}=\frac{\rd\sigma^{\scriptsize\mbox{FwW}}}
{\rd\sigma^{\scriptsize\mbox{NwWA}}}-1,
\eeq
obtained by comparing the NwWA and the FwW variants of our calculation.

All results in \reffis{fi:ptep_lhc8_vs}--\ref{fi:ht_tev_vs}
are obtained with the dynamical scales  $\mu_0=\dynscale/2$ at the LHC 
and $\mu_0=\dynscale$ at Tevatron.
To motivate this choice, we first show, in \reffis{fi:ptep_lhc7_fs} and \ref{fi:ptt_lhc7_fs},
that using a fixed scale $\mu=\Mt/2$ at the LHC  
leads to serious perturbative instabilities
in the high-energy tails of distributions.

\subsubsection{Differential distributions for the LHC at $8\TeV$}
\begin{figure}
\includegraphics[width=\textwidth]{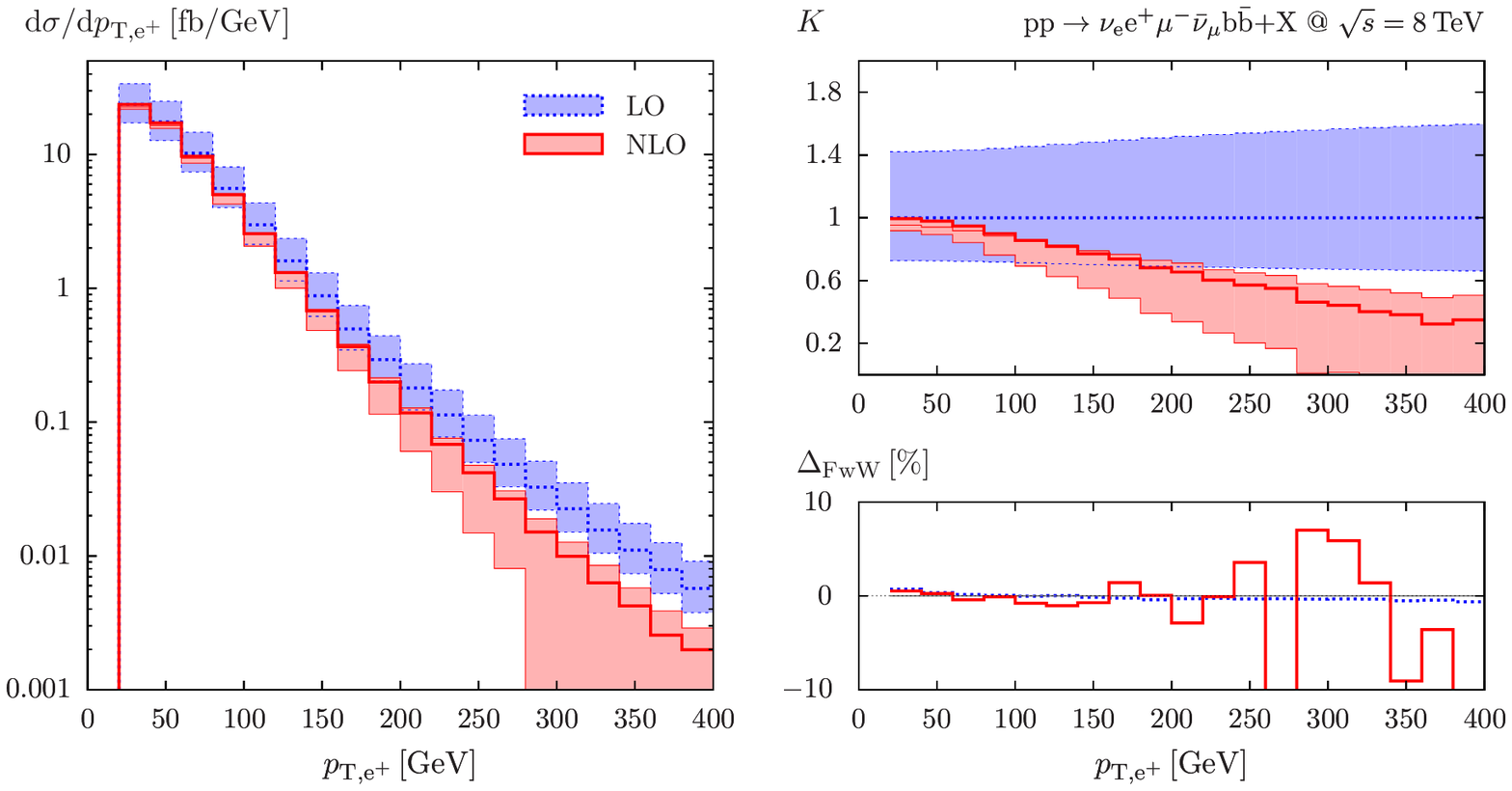}
\caption{Transverse-momentum distribution of the positron with standard cuts for the LHC at $\sqrt{s}=8\TeV$ for fixed scale
  $\mu_0=\Mt/2$.}
\label{fi:ptep_lhc7_fs}
\vspace*{4ex}
\includegraphics[width=\textwidth]{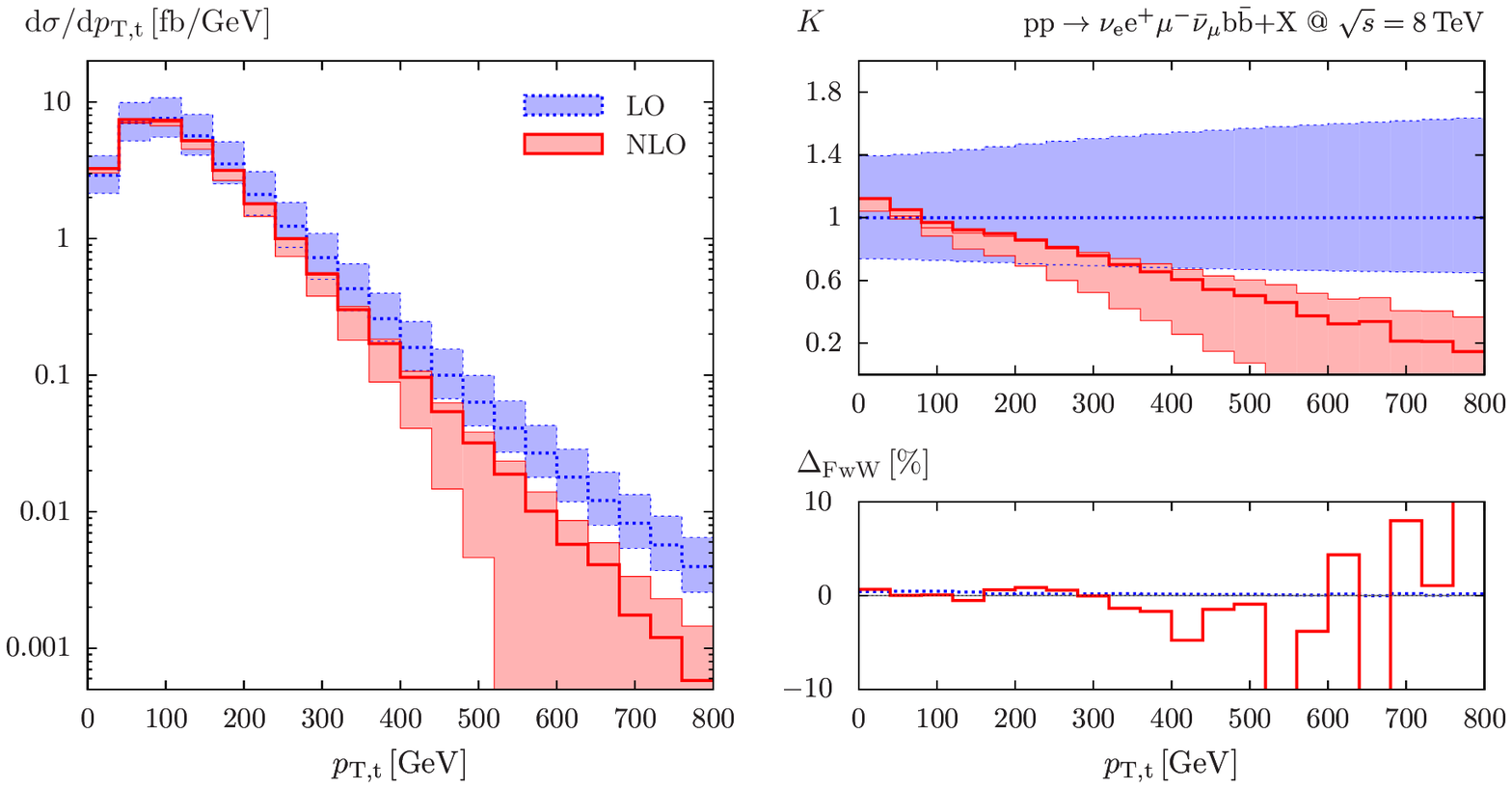}
\caption{Transverse-momentum distribution of the top quark with standard cuts for the LHC at $\sqrt{s}=8\TeV$ for fixed scale
  $\mu_0=\Mt/2$.}
\label{fi:ptt_lhc7_fs}
\end{figure}

\newcommand{\distrfsdir}{LHC8.fs2.distributionCV.next.cp05KfactorCV.cp03Delta}
\newcommand{\distr}{distributionCV.next.cp05KfactorCV.cp03Delta}
\newcommand{\distrdir}{LHC8.vs2.distributionCV.next.cp05KfactorCV.cp03Delta}
We first provide results for the LHC at $8\TeV$.
As examples for typical $\pt$ distributions we present the
distributions in the transverse momenta of the positron and of the
top quark for the fixed scale $\mu=\Mt/2$ in \reffis{fi:ptep_lhc7_fs} and
\ref{fi:ptt_lhc7_fs}. 
The tails of these distributions are relevant for new-physics
searches based on boosted top quarks, 
while the lepton-$p_\rT$ distribution plays an important role for the 
$\ttb$ cross-section acceptance.
The typical lepton $\pt$ is below $100\GeV$ and the average $\pt$ of
the top quark is around $100\GeV$. The FwW effects are within
integration errors.
In the plotted range, the cross section falls by almost four orders of
magnitude, while the $K$ factor drops by 
75\%.
In the high-$\pt$ tails, the  corrections become so large 
that NLO predictions move outside the LO band and 
their scale variation exceeds $-100\%$.
This pathologic behaviour at large $\pt$ can (at least in part) be attributed
to large logarithms associated with the running of $\alphas$,
which can be effectively resummed by adapting the QCD renormalization scale to 
the characteristic scattering energy.
This motivates us to adopt the dynamical scale $\mu_0=\dynscale/2$
as our standard LHC scale choice, which is applied to 
all LHC observables presented in the following.
 
\begin{figure}
\includegraphics[width=\textwidth]{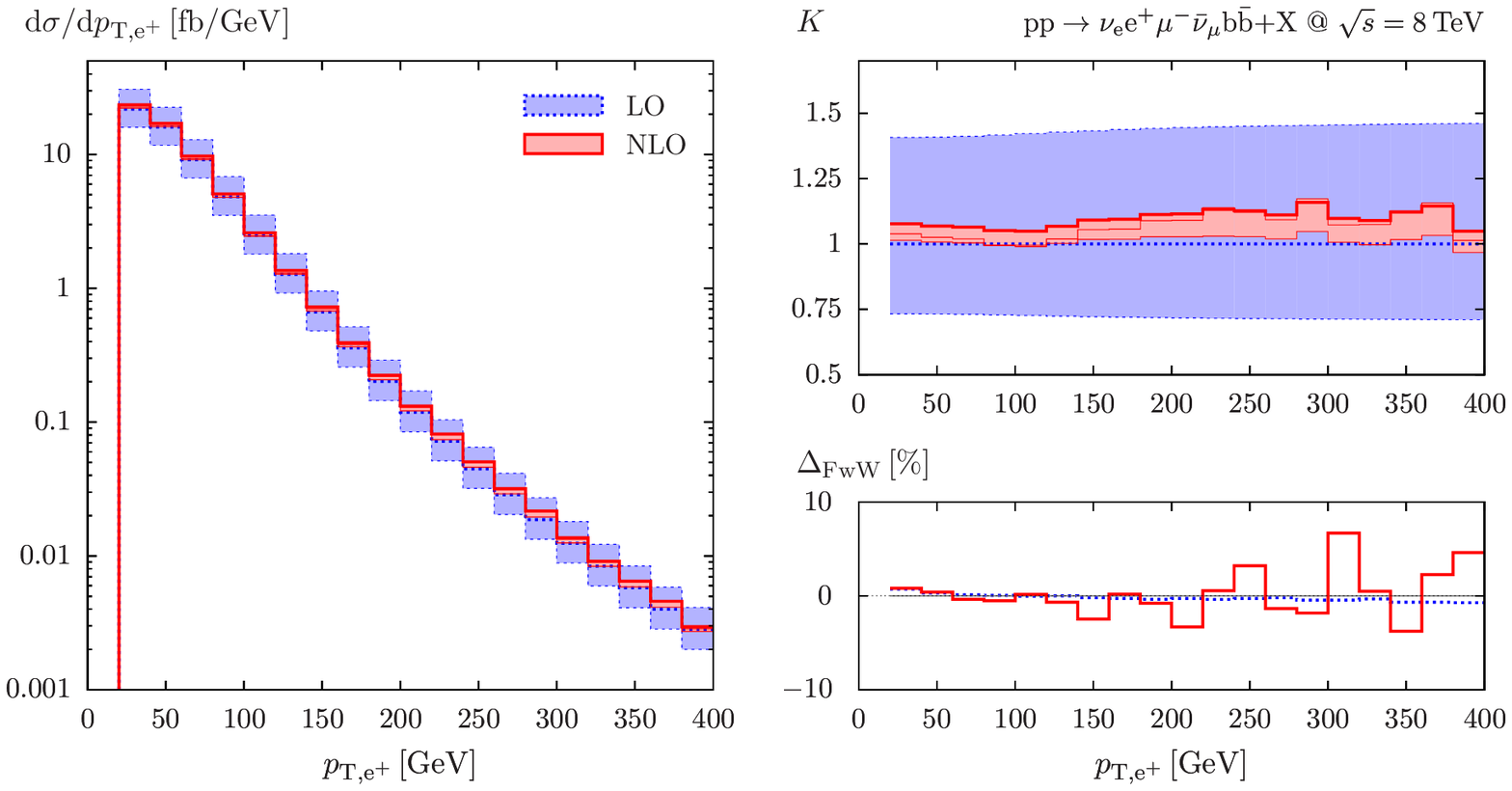}
\caption{Transverse-momentum distribution of the positron with standard cuts for the LHC at $\sqrt{s}=8\TeV$ for dynamical scale
  $\mu_0=\dynscale/2$.}
\label{fi:ptep_lhc8_vs}
\vspace*{4ex}
\includegraphics[width=\textwidth]{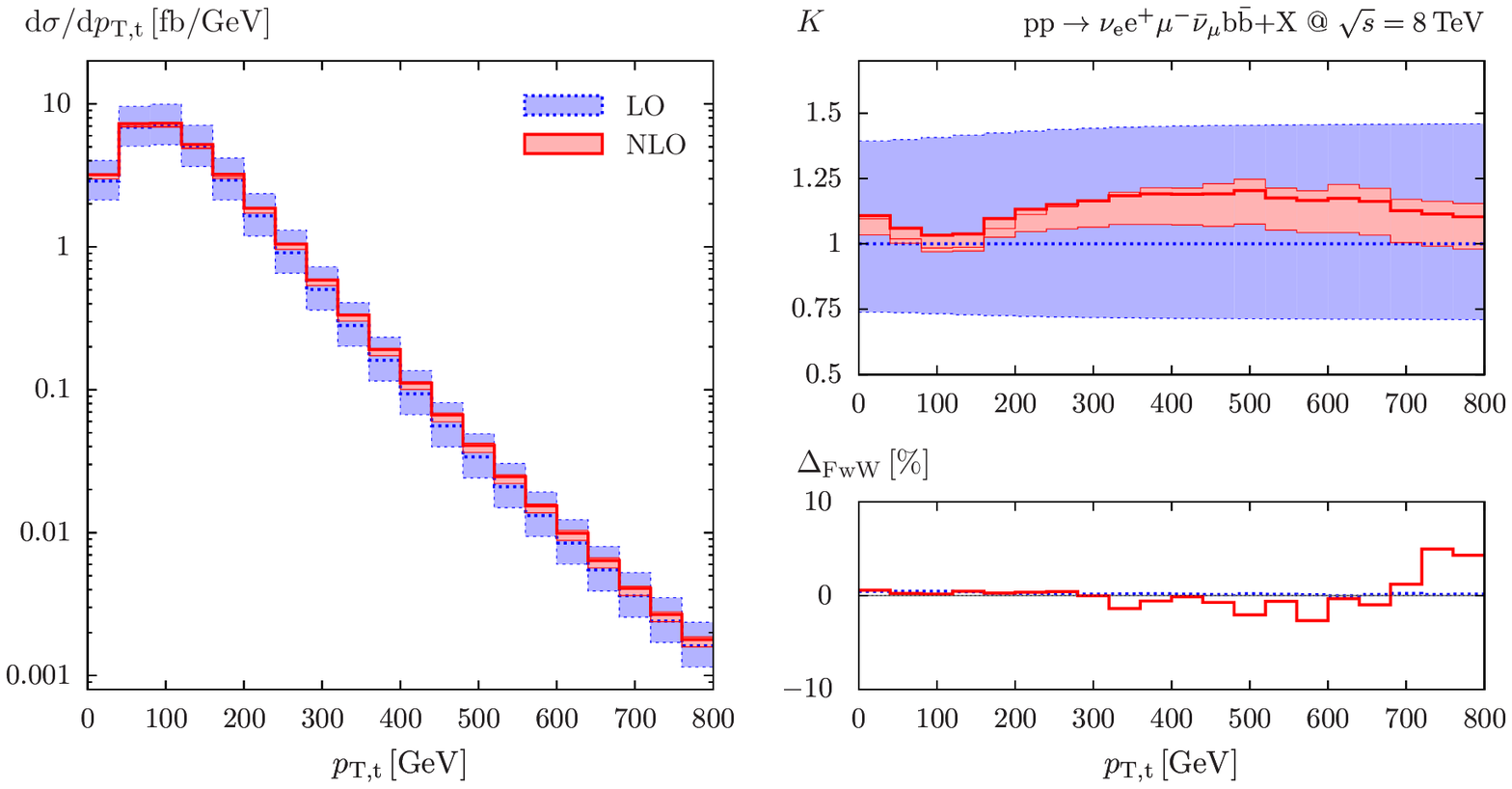}
\caption{Transverse-momentum distribution of the top quark with standard cuts for the LHC at $\sqrt{s}=8\TeV$ for dynamical scale
  $\mu_0=\dynscale/2$.}
\label{fi:ptt_lhc8_vs}
\end{figure}

The $p_{\rT,\Pep}$ and $p_{\rT,\Pt}$ distributions,
displayed in \reffis{fi:ptep_lhc8_vs} and \ref{fi:ptt_lhc8_vs},
show that the dynamical scale choice leads to a drastic improvement of the perturbative stability.
Now the $K$ factor changes only slightly with $\pt$ (within
$20\%$), and the NLO band lies within the LO band.  

\begin{figure}
\includegraphics[width=\textwidth]{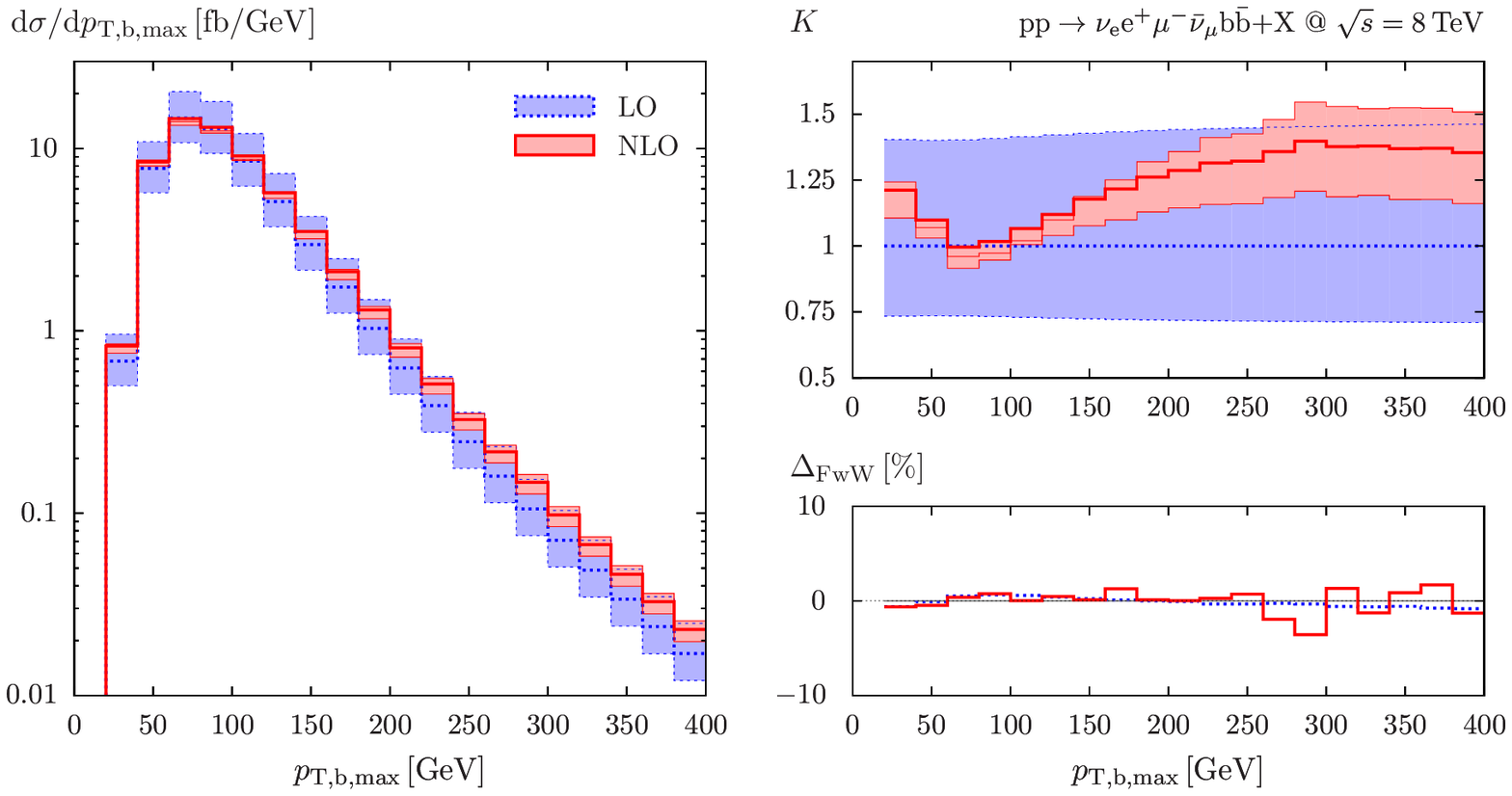}
\caption{Transverse-momentum distribution of the harder $\Pb$ jet with standard cuts for the LHC at $\sqrt{s}=8\TeV$ for dynamical scale
  $\mu_0=\dynscale/2$.}
\label{fi:ptbmax_lhc8_vs}
\vspace*{4ex}
\includegraphics[width=\textwidth]{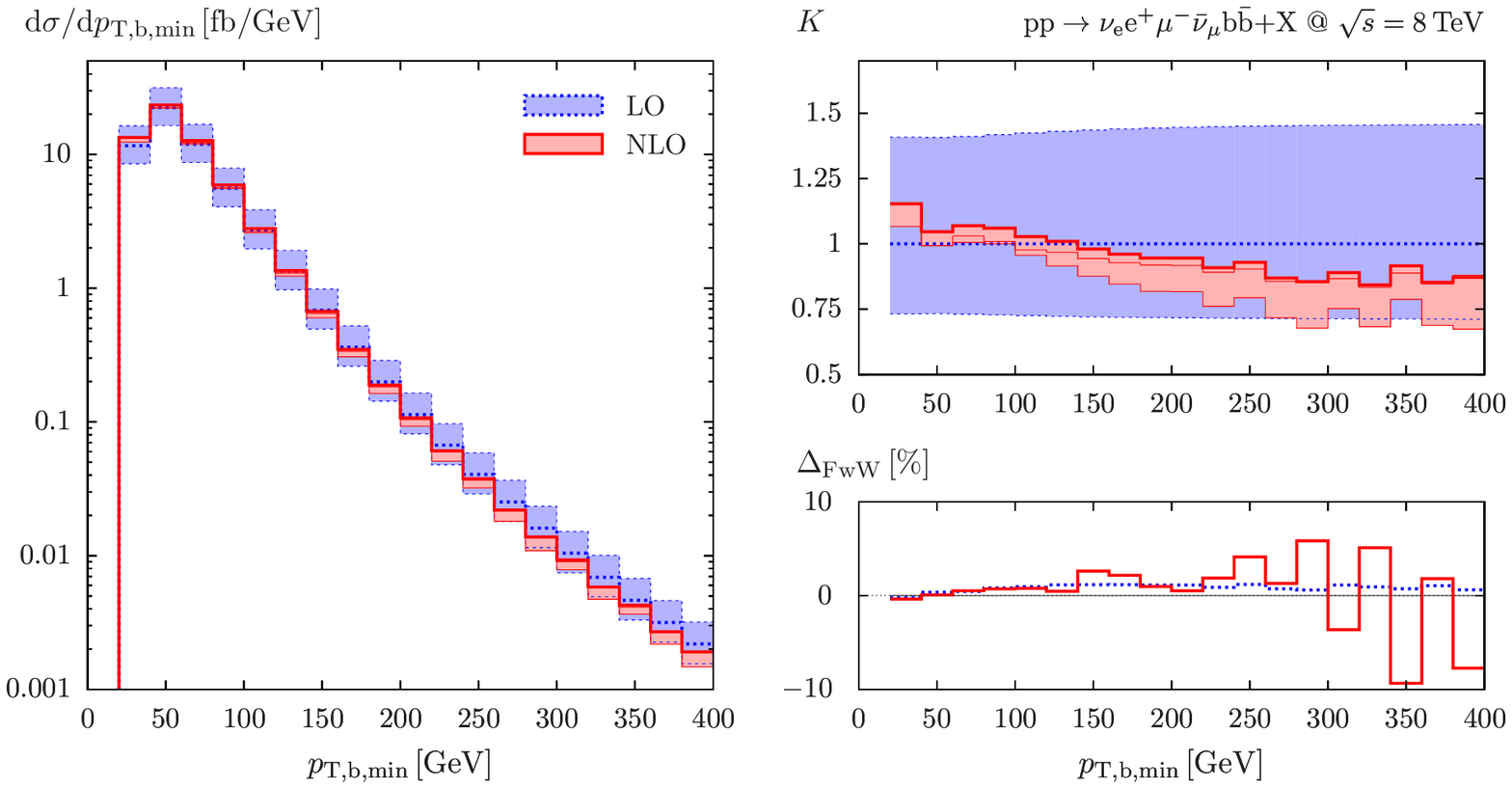}
\caption{Transverse-momentum distribution of the softer $\Pb$ jet with standard cuts for the LHC at $\sqrt{s}=8\TeV$ for dynamical scale
  $\mu_0=\dynscale/2$.}
\label{fi:ptbmin_lhc8_vs}
\end{figure}

Next we consider the distributions in the transverse momenta of the
harder b~jet (the one with larger $\pt$) and the softer b~jet
in \reffis{fi:ptbmax_lhc8_vs} and \ref{fi:ptbmin_lhc8_vs}.
The corrections to the transverse-momentum distribution of the harder
$\Pb$ jet are small near the maximum of the distribution and increase
slowly for higher $\pt$ by about $40\%$, while those for the
distribution of the softer $\Pb$ jet decrease by $20\%$.  
Using the fixed scale 
$\mu_0=\Mt/2$ (not shown), 
the $K$ factor to the $p_{\rT,\Pb,\max}$ distribution
decreases by $40\%$ and the one for the $p_{\rT,\Pb,\min}$ distribution
even by $90\%$.  FwW effects are at the level of
$1\%$ or below.

\begin{figure}
\includegraphics[width=\textwidth]{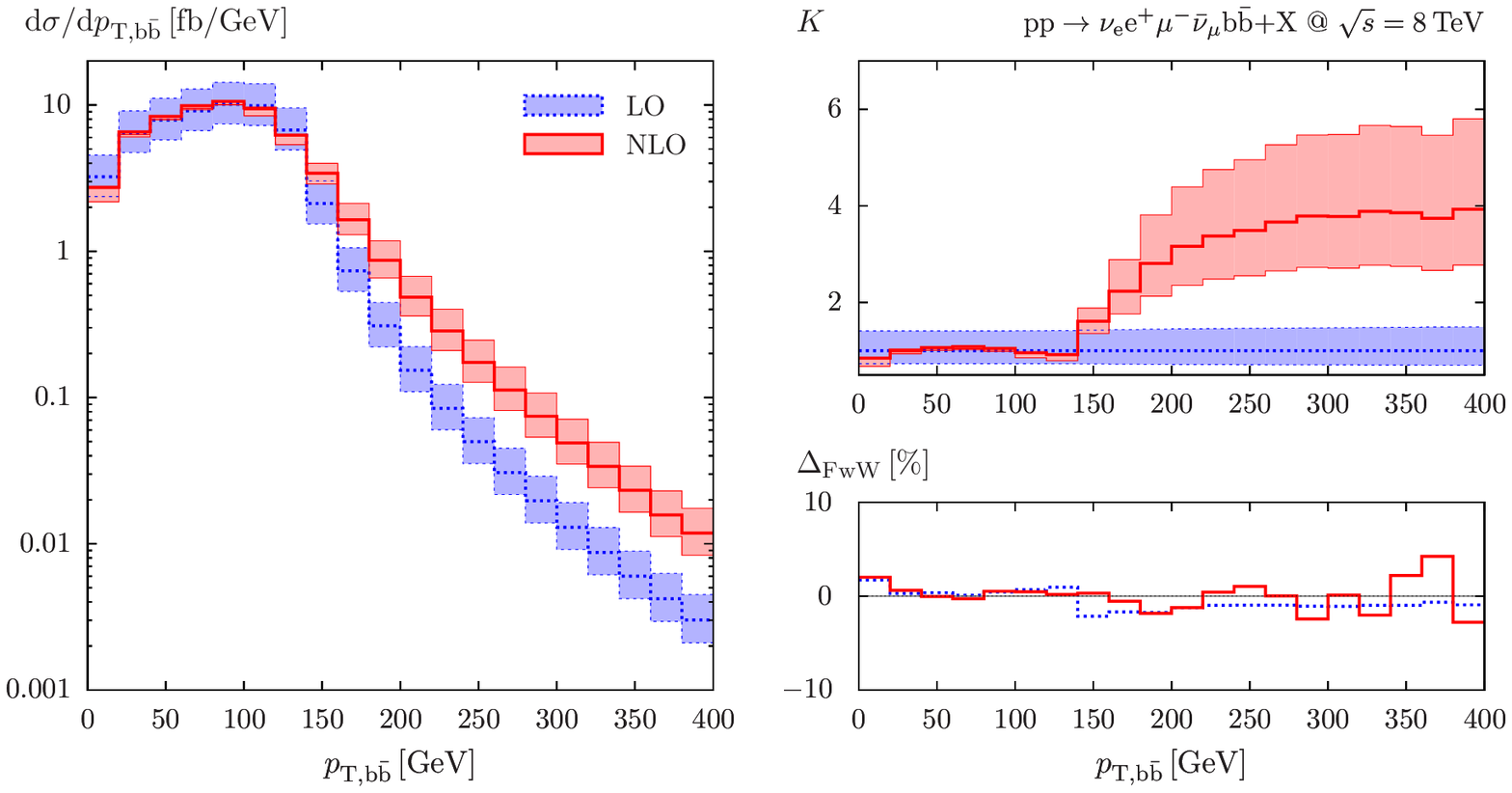}
\caption{Transverse-momentum distribution of the  $\Pb\Pbbar$ pair with standard cuts for the LHC at $\sqrt{s}=8\TeV$ for dynamical scale
  $\mu_0=\dynscale/2$.}
\label{fi:ptbB_lhc8_vs}
\vspace*{4ex}
\includegraphics[width=\textwidth]{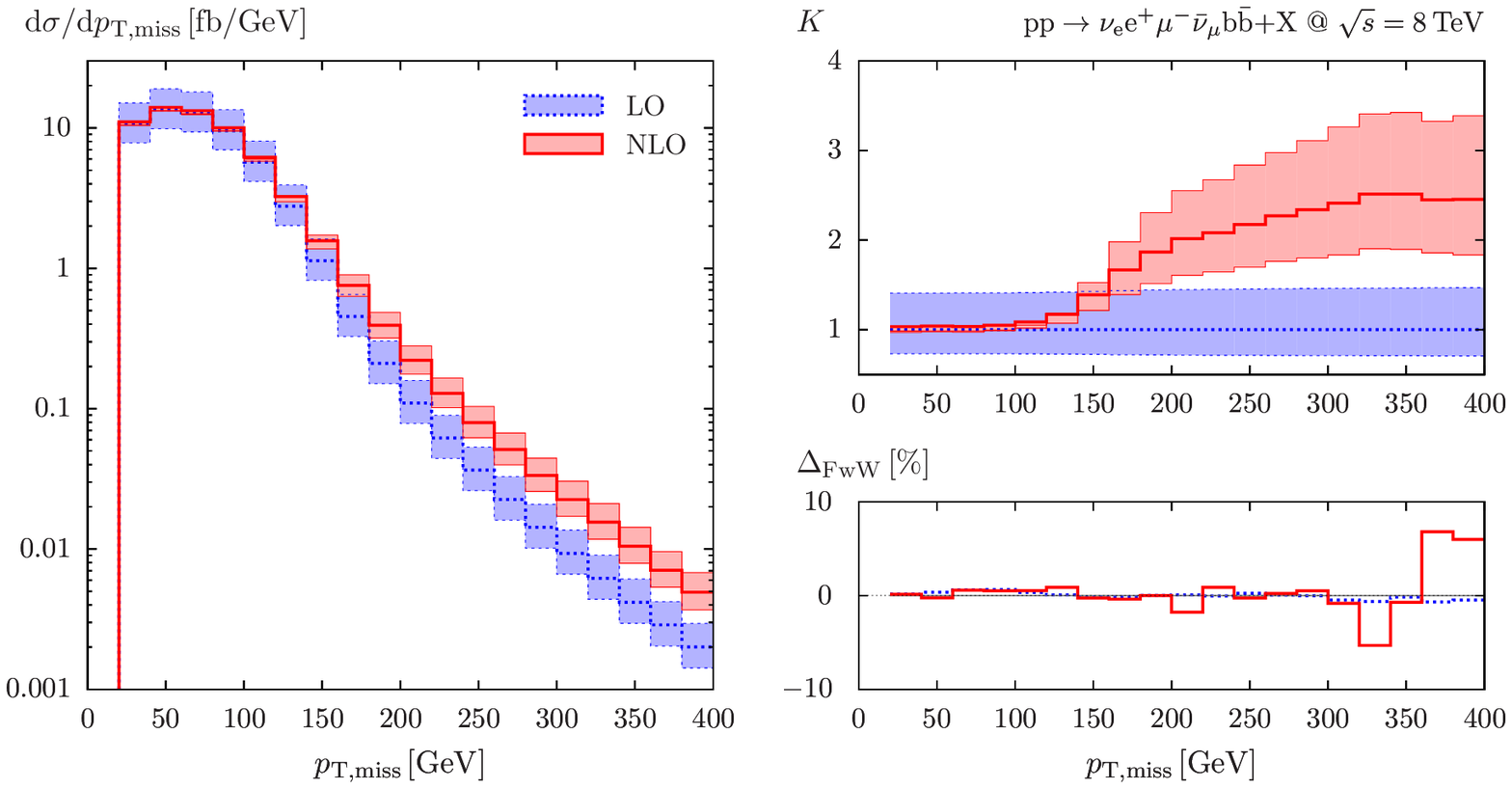}
\caption{Missing-transverse-momentum distribution with standard cuts for the LHC at $\sqrt{s}=8\TeV$ for dynamical scale
  $\mu_0=\dynscale/2$.}
\label{fi:ptmiss_lhc8_vs}
\end{figure}

The distribution in the transverse momentum of the $\Pb\Pbbar$ pair
is presented in  \reffi{fi:ptbB_lhc8_vs}.
This variable plays an important role in boosted-Higgs searches with a
large $\ttb$ background.  In particular, the strategy proposed in
\citeres{Butterworth:2008iy,ATL-2009-088} to extract a $\Pp\Pp\to
\PH(\to\Pb\bar\Pb)\PW$ signal at the LHC is based on the selection of
boosted $\PH\to \Pb\bar\Pb$ candidates with
$p_{\rT,\Pb\bar\Pb}>200\,\GeV$, which permits to reduce $\ttb$
contamination (and other backgrounds) in a very efficient way.  As can
be seen from \reffi{fi:ptbB_lhc8_vs}, the suppression of $\ttb$
production is indeed particularly strong at $p_{\rT,\Pb\bar\Pb}\gsim
150\,\GeV$.  This is due to the fact that, in NtWA, $\Pb$ quarks need
to be boosted via the $\pt$ of their parent (anti)top quarks in order
to acquire $p_{\rT,\Pb}> (\Mt^2-\MW^2)/(2\Mt)\simeq 65\,\GeV$, and a
$\Pb\bar\Pb$ system with high $\pt$ is kinematically strongly
disfavoured at LO, since top and antitop quarks have opposite
transverse momenta.  The NLO corrections undergo less stringent
kinematic restrictions, resulting in a significant enhancement of
$\wwbb$ events at large $p_{\rT,\Pb\bar\Pb}$.  This is clearly
reflected in the differences between the LO and NLO curves in the left
plot of Fig.~\ref{fi:ptbB_lhc8_vs}.  At NLO the $\ttb$ system can
acquire large transverse momentum by recoiling against extra jet
radiation. As indicated by the upper--right plot, the NLO correction
represents $50{-}80\%$ of the cross section at high $\pt$,
corresponding to a huge $K$-factor of $2{-}4$.  FwW effects
(lower--right plot) stay at the level of $2\%$. On the other hand, FtW
effects (not shown here explicitly) 
become as large as 10--30\% for
$p_{\rT,\Pb\bar\Pb}>200\,\GeV$~\cite{Denner:lh2011}. This is most
likely due
to non-resonant topologies with direct
${\Pb\bar\Pb}$ production from a high-$\pt$ gluon that recoils against a
$\PW^+\PW^-$ system and splits into a $\Pb\bar\Pb$ pair.

\reffi{fi:ptmiss_lhc8_vs} displays the distribution in the missing
transverse momentum,
$p_{\rT,\mathrm{miss}}=|\vec{p}_{\rT,\nu_\Pe}+\vec{p}_{\rT,\bar\nu_{\mu}}|$.
The tail of this distribution, which is relevant for new-physics
searches based on missing transverse energy plus jets and leptons,
features a qualitatively similar behaviour as in the case of
$p_{\rT,\Pb\bar\Pb}$, owing to analogous kinematic constraints.
However, in the case of $p_\mathrm{T,miss}$ the corrections are less
pronounced: the $K$ factor only rises to $2.5$. 
FwW contributions are not significant.

\begin{figure}
\includegraphics[width=\textwidth]{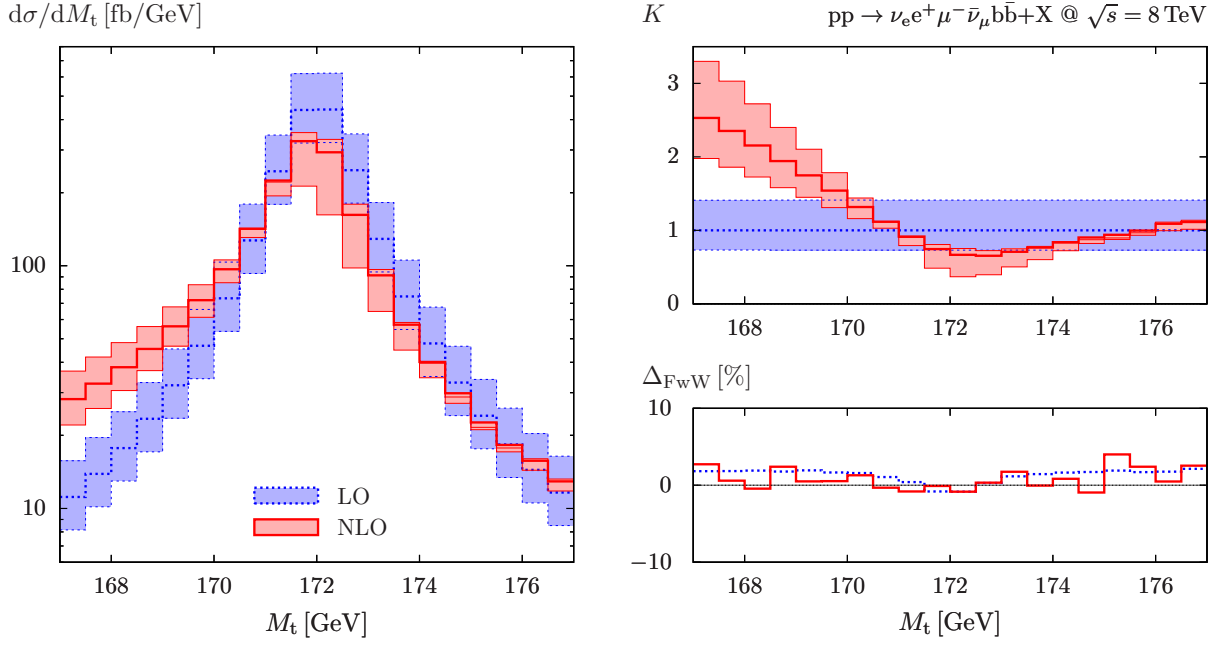}
\caption{Invariant-mass distribution of the top quark, $M_\Pt=M_{\nu_\Pe\Pep\Pb}$, with standard cuts for the LHC at $\sqrt{s}=8\TeV$ for dynamical scale
  $\mu_0=\dynscale/2$.}
\label{fi:mt_lhc8_vs}
\end{figure}

The distribution in the invariant mass of the top quark, $M_\Pt=M_{\nu_\Pe\Pep\Pb}$, in the
vicinity of its resonance is shown in \reffi{fi:mt_lhc8_vs}.
Below the resonance the NLO corrections develop a radiative tail of
positive corrections due to final-state gluon radiation that is not
recombined with the top-quark decay products. 
The NLO corrections induce a small shift in the peak of the
distribution, whose magnitude depends on the jet recombination.  
Note that the major part of this shift would also be
accounted for in a pure parton-shower approach to describe
higher-order QCD effects.  Thus, a thorough assessment of the impact
of the full NLO QCD calculation with off-shell top quarks on
observables that are used in the top mass measurement requires a
careful comparison with predictions based on parton showers.  Note
also that FwW corrections increase in the side bands of the resonance
to the level of a few per cent.

\begin{figure}
\includegraphics[width=\textwidth]{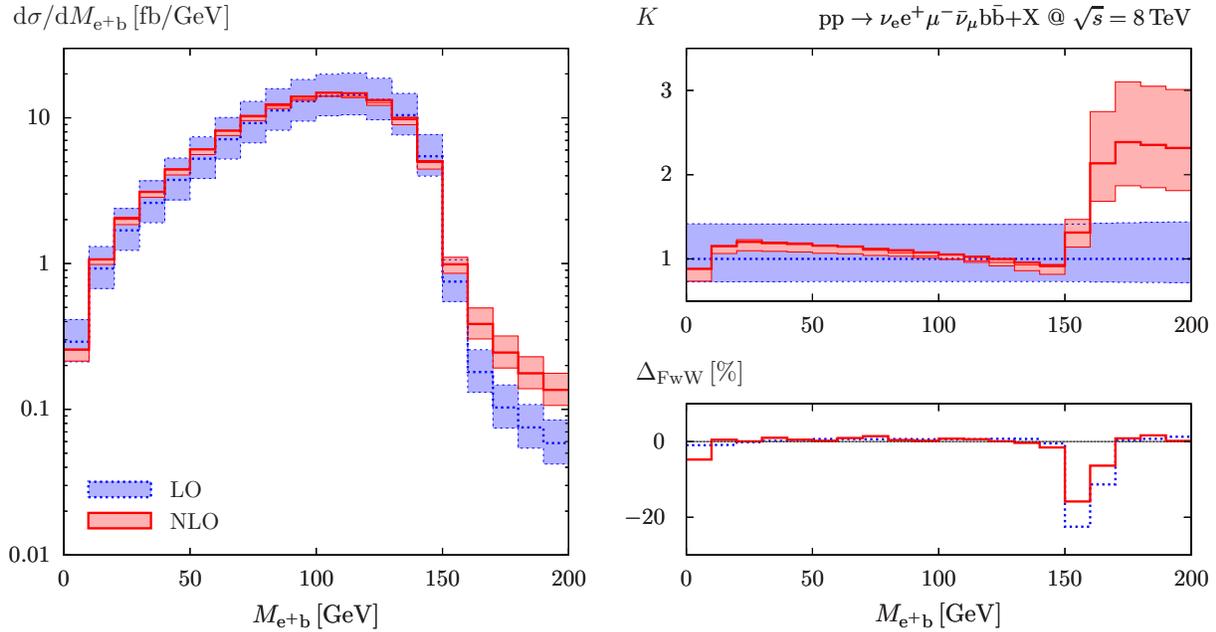}
\caption{Invariant-mass distribution of positron--\Pb-jet system with standard cuts for the LHC at $\sqrt{s}=8\TeV$ for dynamical scale
  $\mu_0=\dynscale/2$.}
\label{fi:mepb_lhc8_vs}
\end{figure}

Figure~\ref{fi:mepb_lhc8_vs}  displays 
the distribution in the invariant mass of the positron
and a $\Pb$~jet, \ie the visible products of a top-quark decay.
We here use the  Monte Carlo truth to select the negatively charged
b-quark jet.%
\footnote{In \citere{Melnikov:2009dn} the $\Pe^+\Pb$ pair is  
formed by selecting the $\Pb$~jet that yields the smallest invariant 
mass.}
In narrow-top-width and LO approximation this kinematic quantity is
characterized by a sharp upper bound, \eqintext{$M^2_{\Pe^+\Pb}<
  \Mt^2-\MW^2\simeq(152\,\GeV)^2$}, which renders it very sensitive to
the top-quark mass.  The value of $\Mt$ can be extracted with high
precision using, for instance, the invariant-mass distribution of a
positron and a $J/\psi$ from a B-meson
decay~\cite{Kharchilava:1999yj,Biswas:2010sa}, an observable that is
closely related to $M_{\Pe^+\Pb}$.  In the region below the kinematic
bound, the NLO corrections to $M_{\Pe^+\Pb}$ vary between $-10\%$ and
$20\%$, and the impact of the NLO shape distortion on a precision
$\Mt$ measurement is certainly significant.  For
$M_{\Pe^+\Pb}<150\,\GeV$, the NwWA is very good. Above the kinematic
bound, NLO corrections become clearly visible, giving rise to a tail
that extends above $M^2_{\Pe^+\Pb}=\Mt^2-\MW^2$, and also FwW
corrections become sizeable. In this kinematic region also the finite
top width causes effects at the level of 50\%~\cite{Denner:lh2011}.
While the contribution to the total cross section from the region
above $150\GeV$ is fairly small, the impact of these contributions on
the top-mass measurement might be non-negligible, given the high $\Mt$
sensitivity of the $M^2_{\Pe^+\Pb}\simeq \Mt^2-\MW^2$ region.  Again a
careful comparison between NLO off-shell calculation and parton-shower
approach would be required to quantify 
off-shell effects on the $\Mt$ measurement.

\begin{figure}
\includegraphics[width=\textwidth]{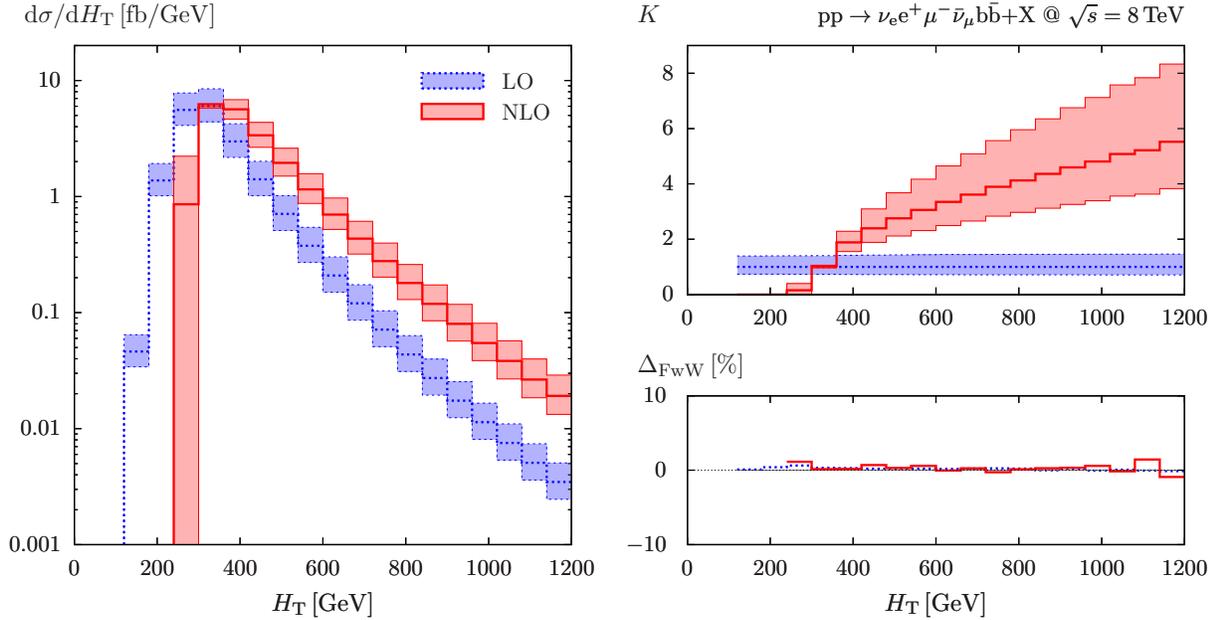}
\caption{Distribution in the  total
  transverse energy $H_\rT$ with standard cuts for the LHC at
  $\sqrt{s}=8\TeV$ for dynamical scale $\mu_0=\dynscale/2$.}
\label{fi:ht_lhc8_vs}
\end{figure}

In \reffi{fi:ht_lhc8_vs} we display the distribution in the total
transverse energy,
\beq\label{eq:ht}
H_{\rT}=p_{\rT,\Pep}+p_{\rT,\Pmum}+p_{\rT,\miss}+\sum_{j} p_{\rT,j},
\eeq
where the scalar
sum over the jet transverse momenta, $p_{\rT,j}$, includes the two b-quark jets 
as well as possible extra jets contributing to NLO real emission.
The $H_\rT$ distribution 
is relevant for new-physics searches.
The large correction above $H_{\rT}=400\GeV$ results from the
inclusion of the gluon jet, which shifts the
distribution to higher $H_{\rT}$. If this is not included, as e.g.\ in
\citere{Bevilacqua:2010qb}, the $K$ factor is flat and varies between
$0.8$ and $1.2$. 

\begin{figure}
\includegraphics[width=\textwidth]{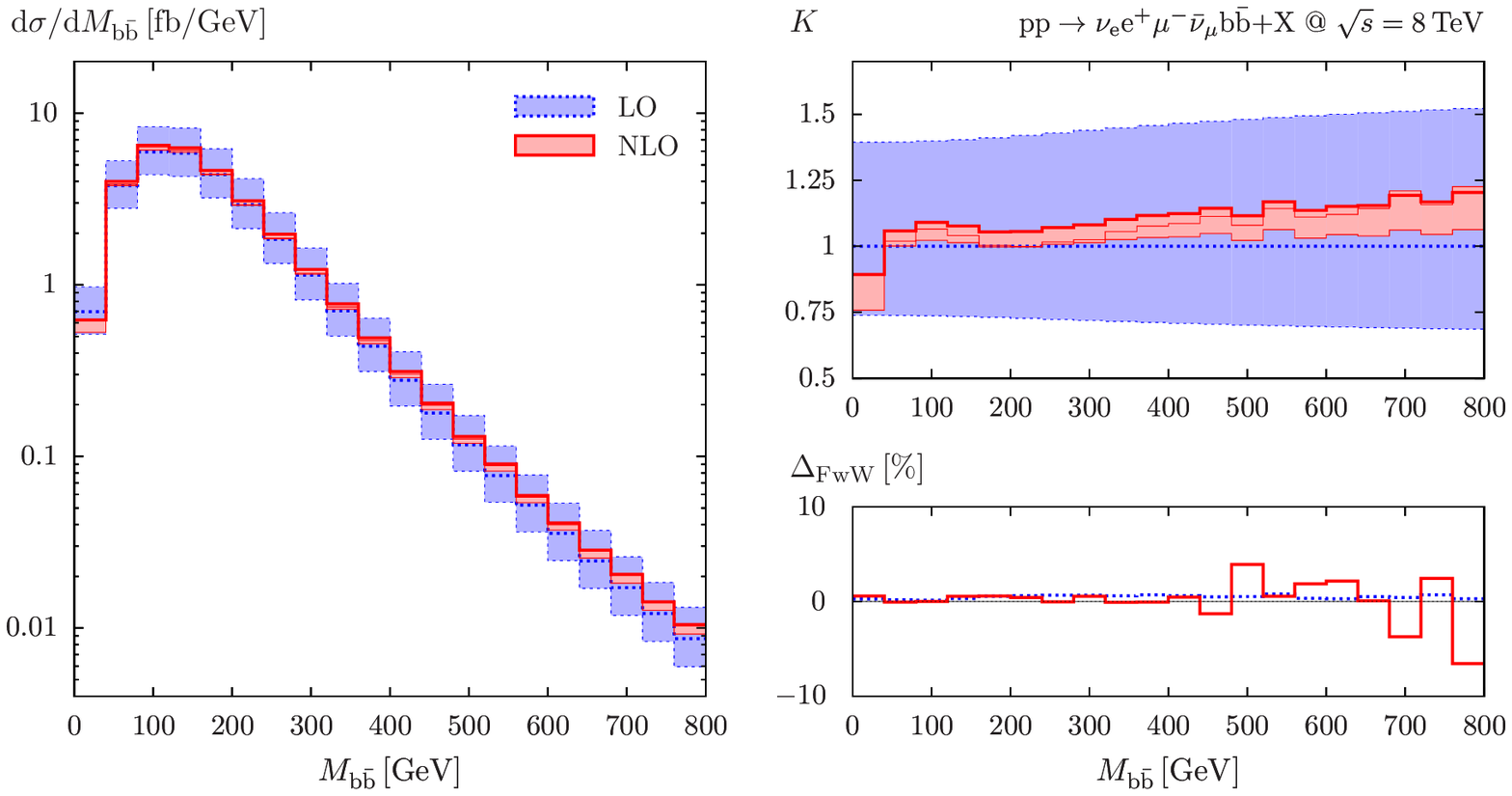}
\caption{Invariant-mass distribution of the  $\Pb\Pbbar$ pair with standard cuts for the LHC at $\sqrt{s}=8\TeV$ for dynamical scale
  $\mu_0=\dynscale/2$.}
\label{fi:mbB_lhc8_vs}
\vspace*{4ex}
\includegraphics[width=\textwidth]{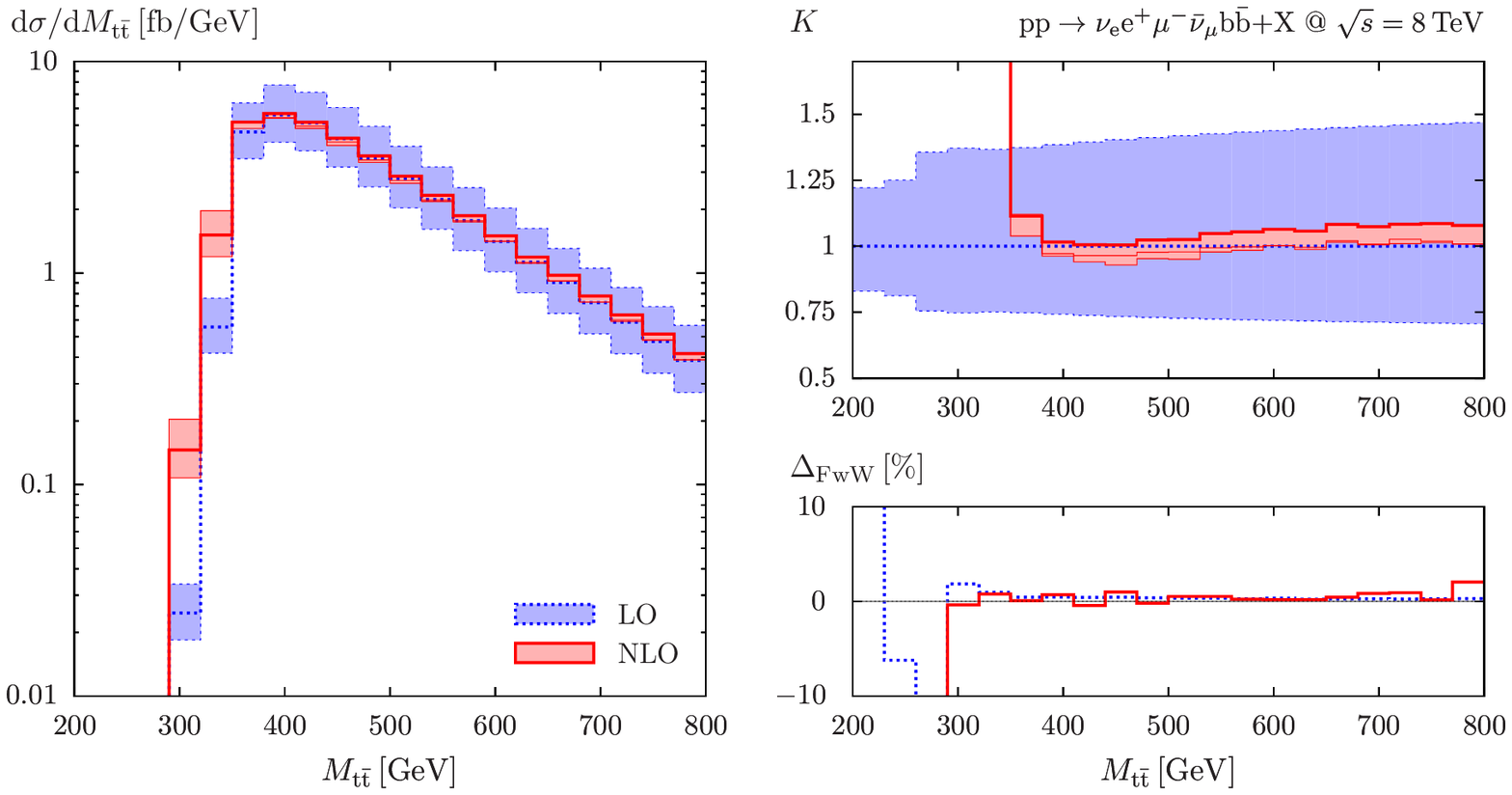}
\caption{Distribution in the invariant mass of the $\Pt\Ptbar$ pair
with standard cuts for the LHC at $\sqrt{s}=8\TeV$ for dynamical scale
$\mu_0=\dynscale/2$.}
\label{fi:mtT_lhc8_vs}
\end{figure}

\begin{figure}
\includegraphics[width=\textwidth]{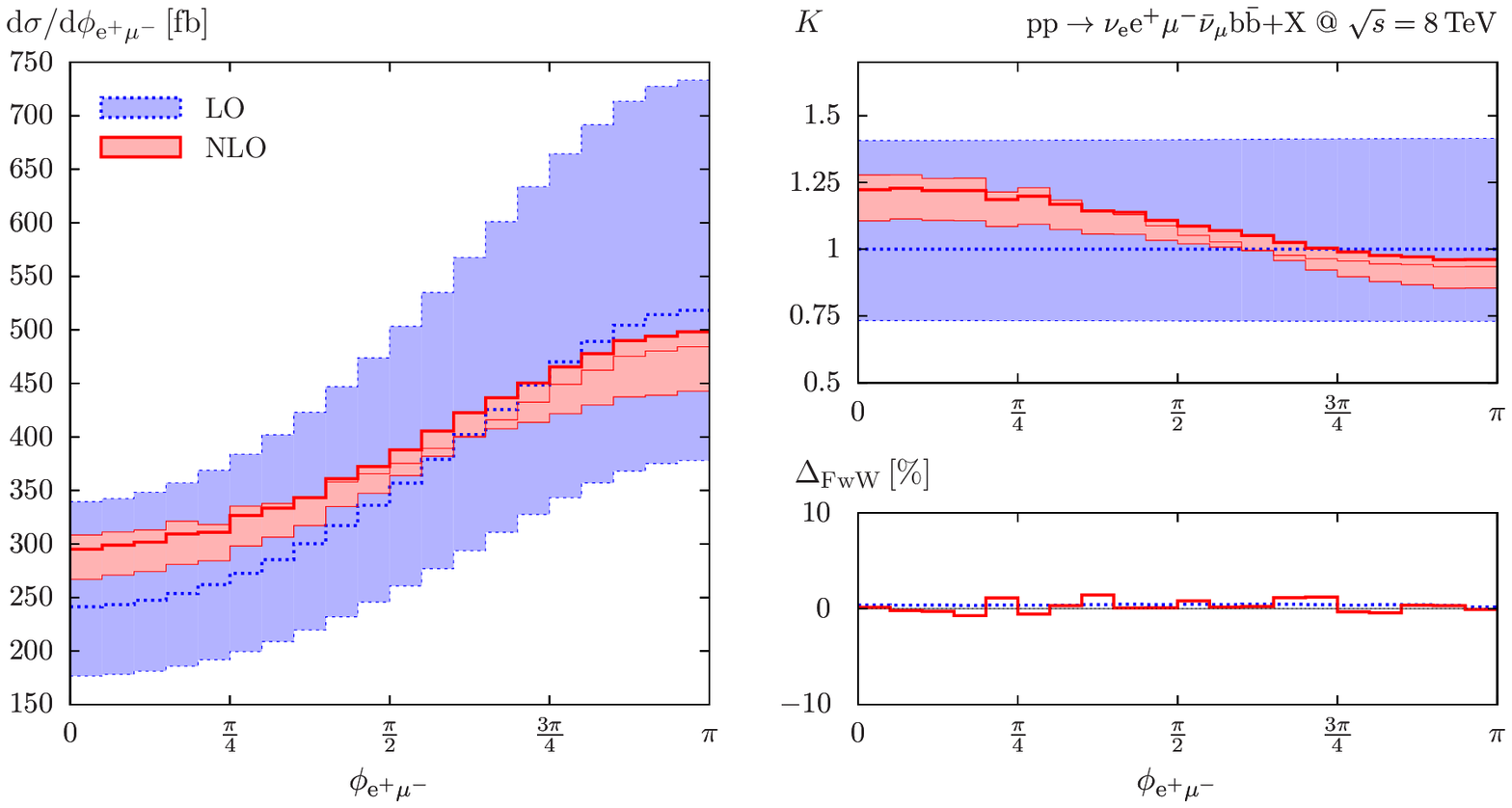}
\caption{Distribution in the azimuthal angle between the positron and
  the muon in the transverse plane, $\phi_{\Pep\Pmum}$,
with standard cuts for the LHC at $\sqrt{s}=8\TeV$ for dynamical scale
$\mu_0=\dynscale/2$.}
\label{fi:phiepmum_lhc8_vs}
\vspace*{4ex}
\includegraphics[width=\textwidth]{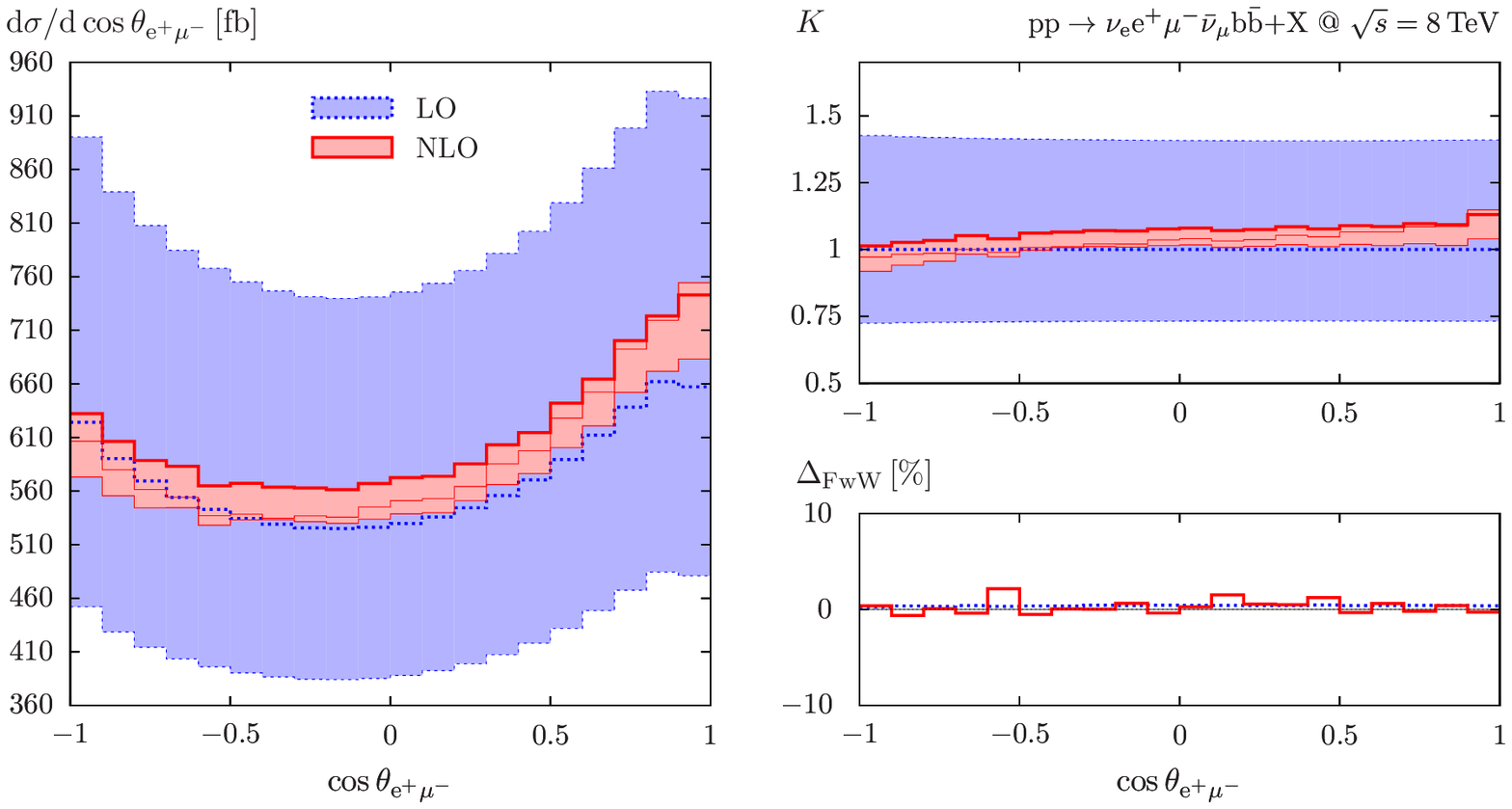}
\caption{Distribution in the cosine of the angle between the positron and
  the muon, $\cos\theta_{\Pep\Pmum}$,
with standard cuts for the LHC at $\sqrt{s}=8\TeV$ for dynamical scale
$\mu_0=\dynscale/2$.}
\label{fi:cthepmum_lhc8_vs}
\end{figure}

Finally, we show the distributions in the invariant masses of the 
$\Pb\Pbbar$ and $\Pt\Ptbar$ pairs, in \reffis{fi:mbB_lhc8_vs} and
\ref{fi:mtT_lhc8_vs}, and the distributions in the azimuthal angle
between the two charged leptons 
in the transverse plane and the cosine of the angle between
them, in  \reffis{fi:phiepmum_lhc8_vs} and
\ref{fi:cthepmum_lhc8_vs}.
In general these distributions receive moderate shape corrections, and 
FwW effects are tiny.  Sizable NLO and FwW effects show up 
only in the $M_{\Pt\Ptbar}$ distribution
below the $\Pt\Ptbar$ threshold.

\subsubsection{Differential distributions for the LHC at $14\TeV$}
\begin{figure}
\includegraphics[width=\textwidth]{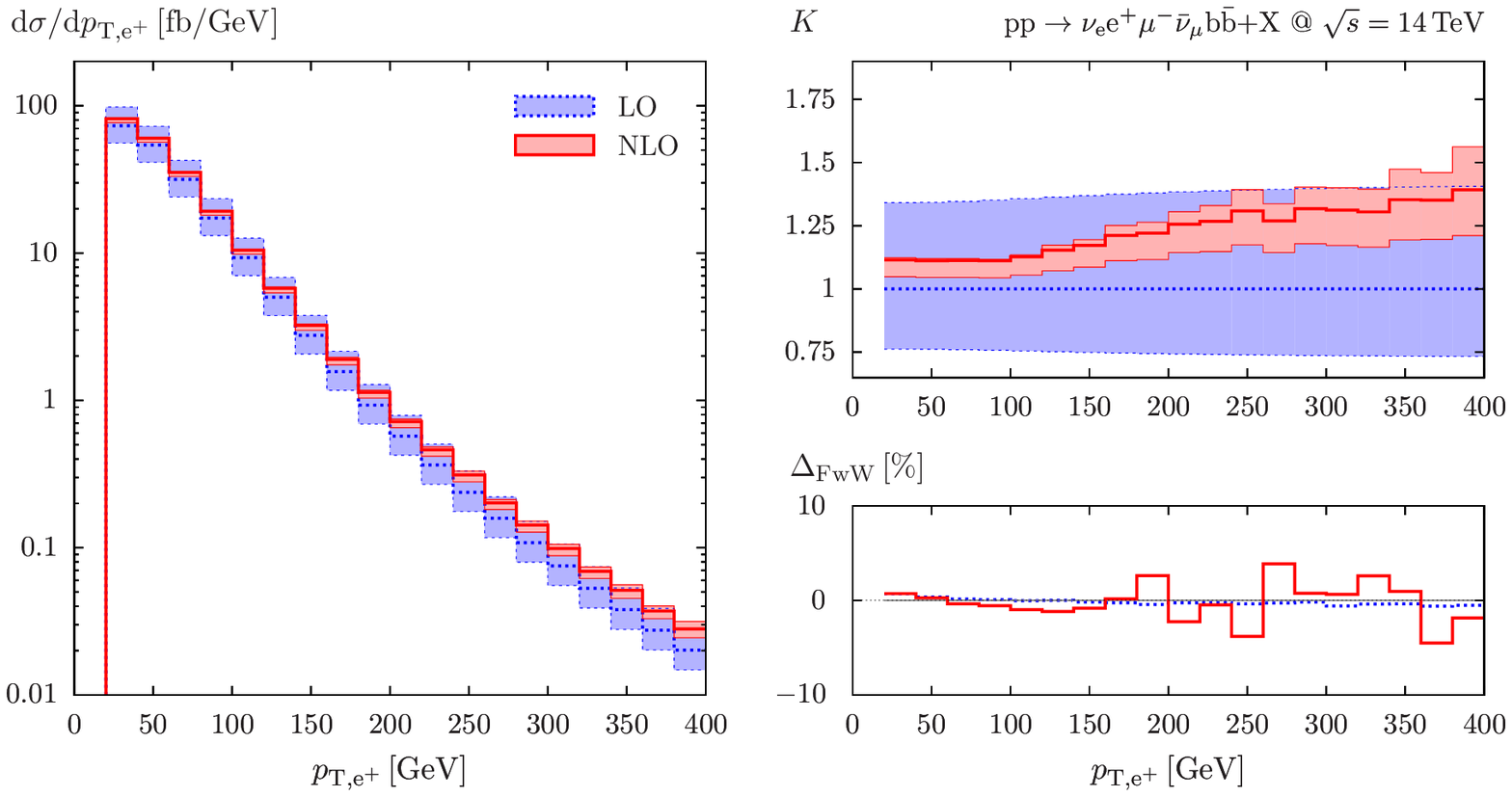}
\caption{Transverse-momentum distribution of the positron with standard cuts for the LHC at $\sqrt{s}=14\TeV$ for dynamical scale
$\mu_0=\dynscale/2$.}
\label{fi:ptep_lhc14_vs}
\vspace*{4ex}
\includegraphics[width=\textwidth]{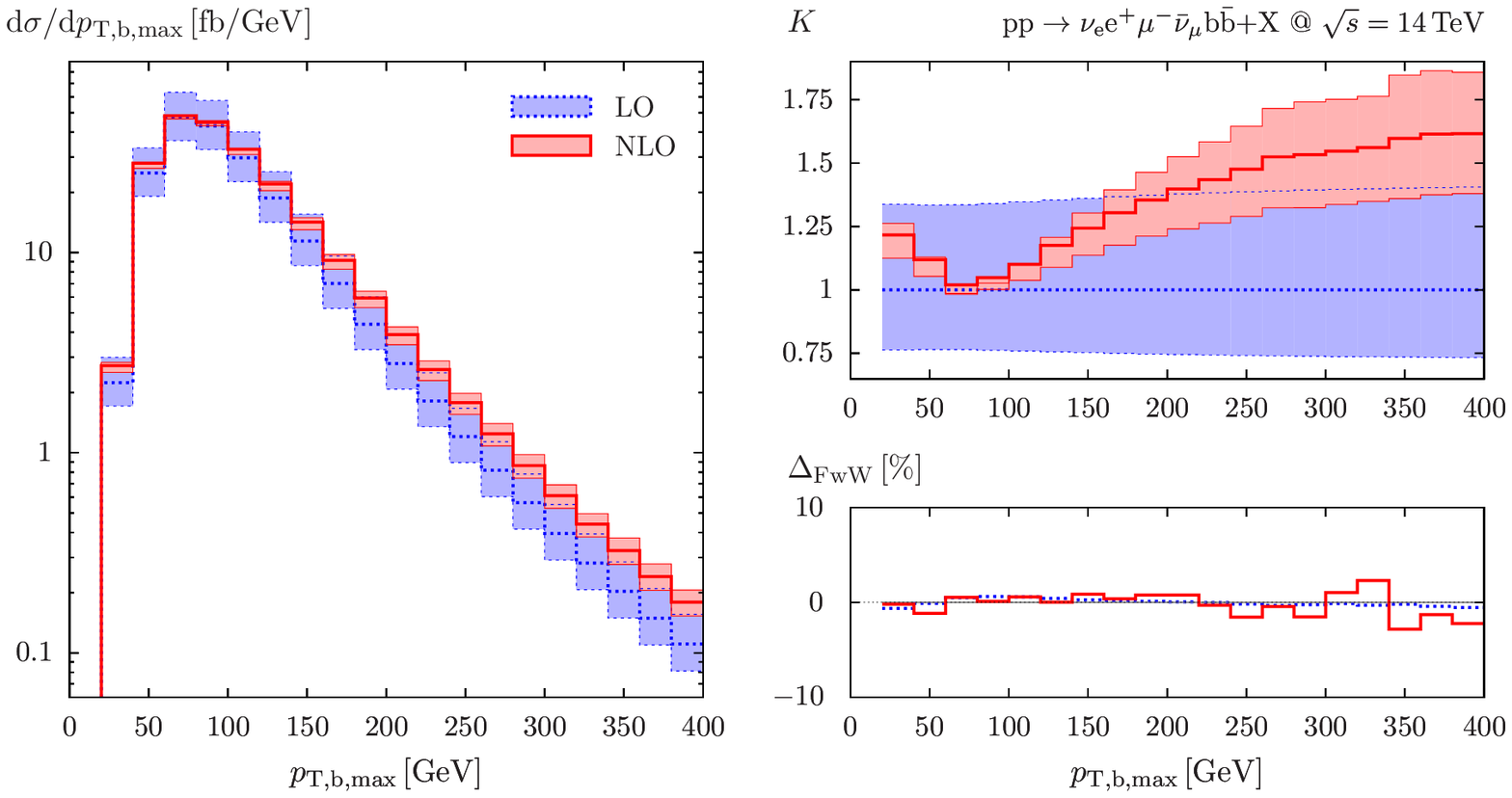}
\caption{Transverse-momentum distribution of the harder $\Pb$ jet with standard cuts for the LHC at $\sqrt{s}=14\TeV$ for dynamical scale
$\mu_0=\dynscale/2$.}
\label{fi:ptbmax_lhc14_vs}
\end{figure}

At the LHC with $14\TeV$, using the dynamical scale $\mu_0=\dynscale/2$,
we observe that
the shapes of the distributions and the NLO
corrections look qualitatively similar as for
$8\TeV$. However, the $K$ factors for transverse-momentum
distributions grow faster with $\pt$ than at $8\TeV$. In
\reffis{fi:ptep_lhc14_vs} and \ref{fi:ptbmax_lhc14_vs} we show
results for the distributions in the transverse momenta of the
positron and of the harder b~jet. 
\renewcommand{\distrdir}{LHC14.vs2.distributionCV.next.cp05KfactorCV.cp03Delta}
Above $\pt=100\GeV$, the $K$ factor grows by about $30\%$ for the
$p_{\rT,\Pep}$ distribution and by about $60\%$ for the $p_{\rT,\Pb,\max}$
distribution.  Using the fixed scale $\mu=\Mt/2$  instead (not shown), the $K$
factor decreases by about $40\%$ for the $p_{\rT,\Pep}$ distribution
and by about $20\%$ for the $p_{\rT,\Pb,\max}$ distribution.
In general, FwW corrections remain similarly suppressed as at $8\TeV$.

\subsubsection{Differential distributions for Tevatron}
\begin{figure}
\includegraphics[width=\textwidth]{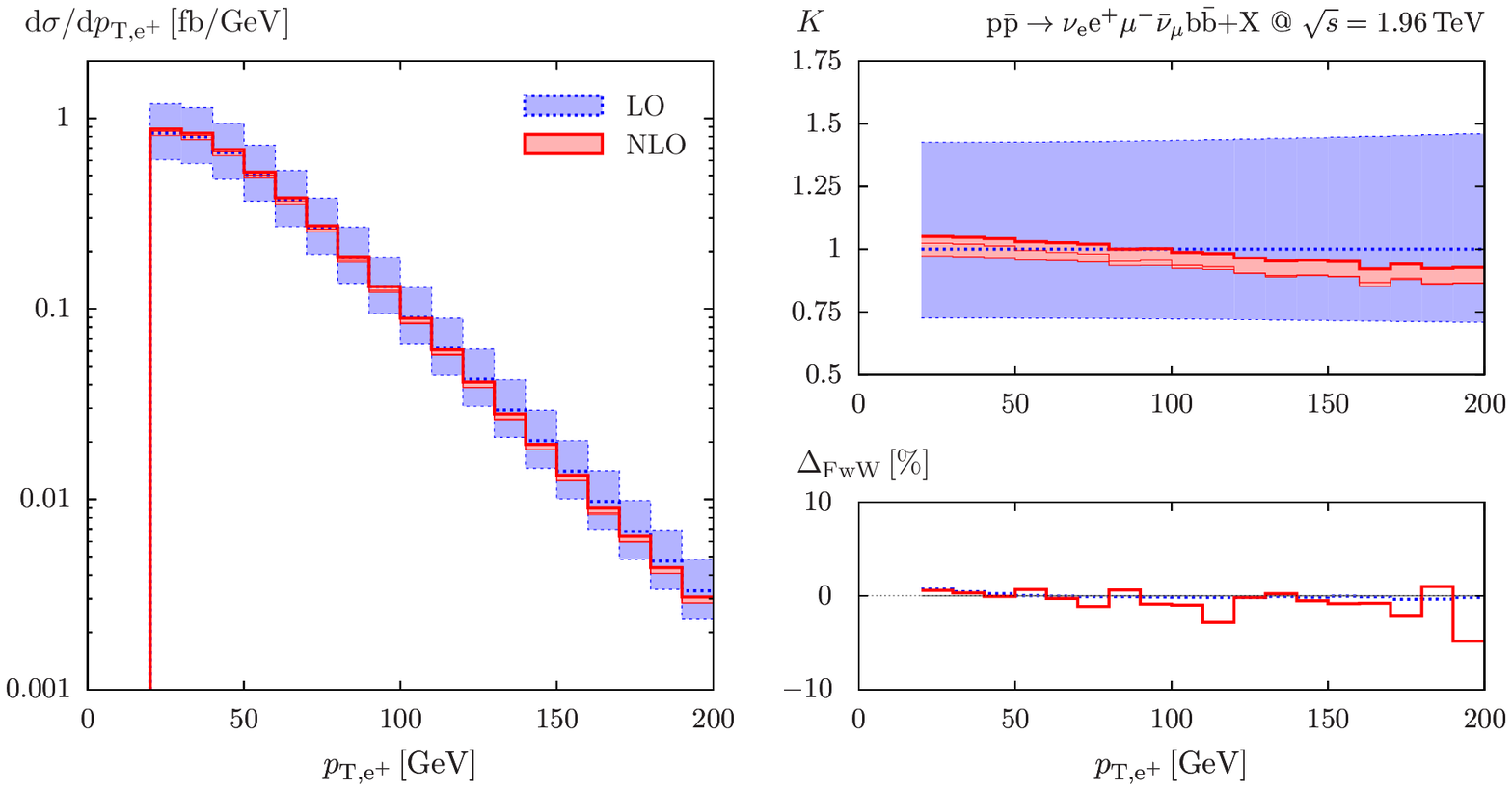}
\caption{Transverse-momentum distribution of the positron with standard cuts for Tevatron for dynamical scale
  $\mu_0=\dynscale$.}
\label{fi:ptep_tev_vs}
\end{figure}

\renewcommand{\distrdir}{Tevatron196.vs1.distributionCV.next.cp05KfactorCV.cp03Delta}
Next we show a few distributions for Tevatron, using the
dynamical scale $\mu_0=\dynscale$.
The transverse-momentum distribution of the positron 
is presented in
\reffi{fi:ptep_tev_vs}.
The $K$ factor decreases slightly with increasing $\pt$. 
FwW effects are completely negligible.

\begin{figure}
\includegraphics[width=\textwidth]{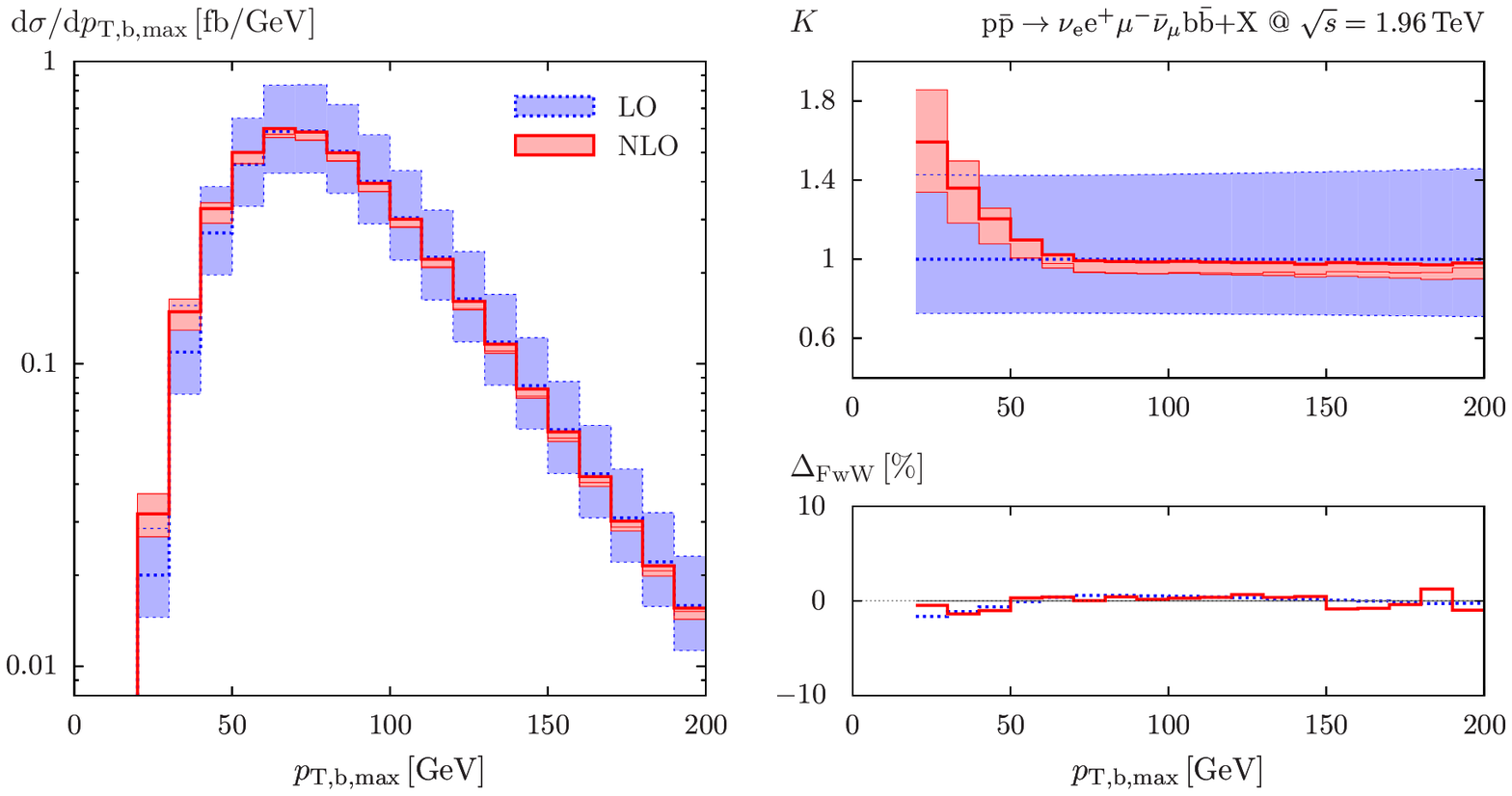}
\caption{Transverse-momentum distribution of the harder $\Pb$ jet with standard cuts for Tevatron for dynamical scale
  $\mu_0=\dynscale$.}
\label{fi:ptbmax_tev_vs}
\vspace*{4ex}
\includegraphics[width=\textwidth]{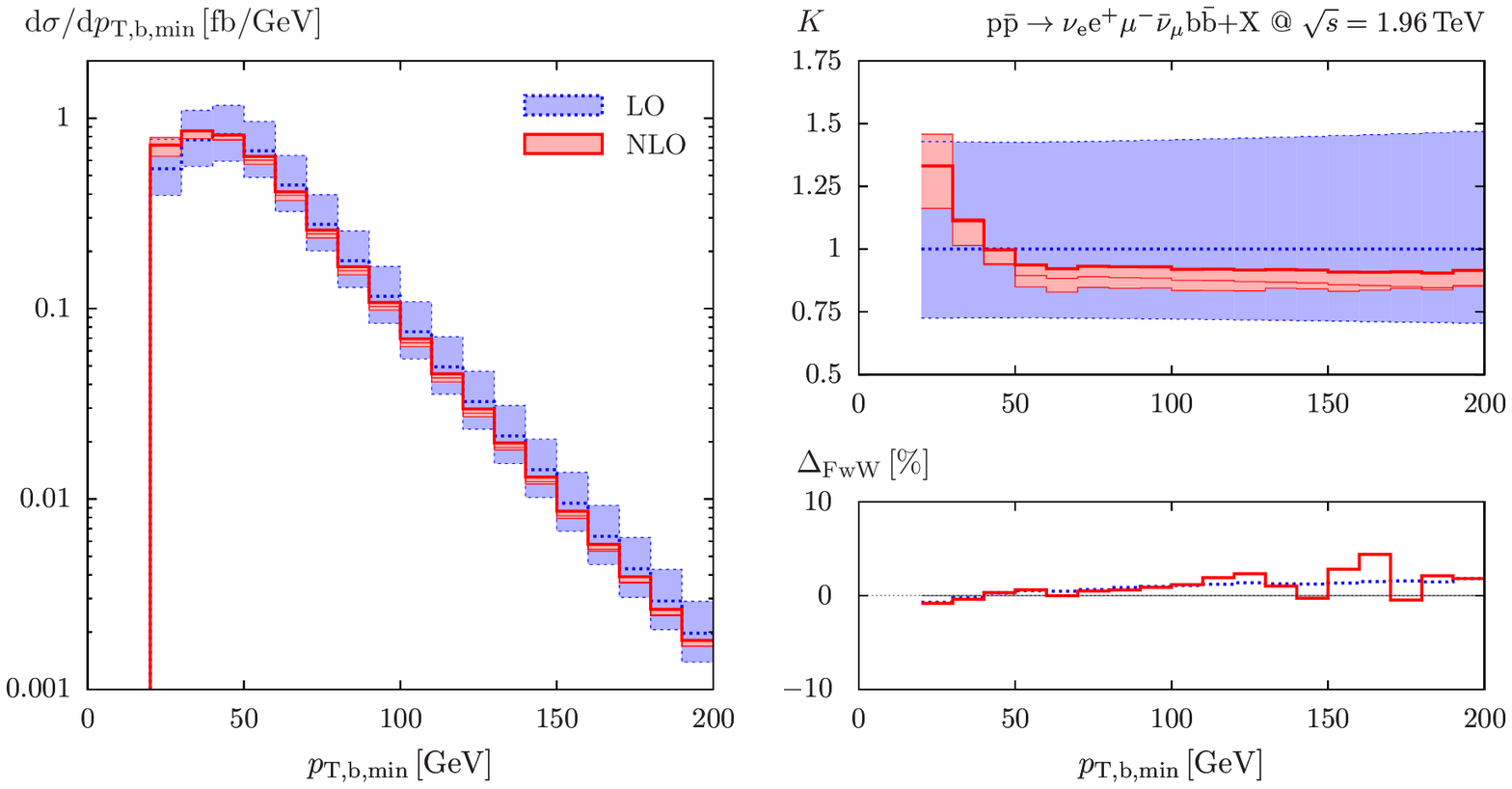}
\caption{Transverse-momentum distribution of the softer $\Pb$ jet with standard cuts for Tevatron for dynamical scale
  $\mu_0=\dynscale$.}
\label{fi:ptbmin_tev_vs}
\end{figure}

We turn to the distributions in the transverse momenta of the
harder b~jet and the softer b~jet
in \reffis{fi:ptbmax_tev_vs} and \ref{fi:ptbmin_tev_vs}.
Below the maxima of the distributions we find a positive $K$ factor at
the level of $40\%$, but above $p_{\rT,\Pb}=70\GeV$ the $K$ factor is pretty
flat and close to one. For large $p_{\rT,\Pb,\min}$ FwW 
effects at the level of $2\%$ show up.

\begin{figure}
\includegraphics[width=\textwidth]{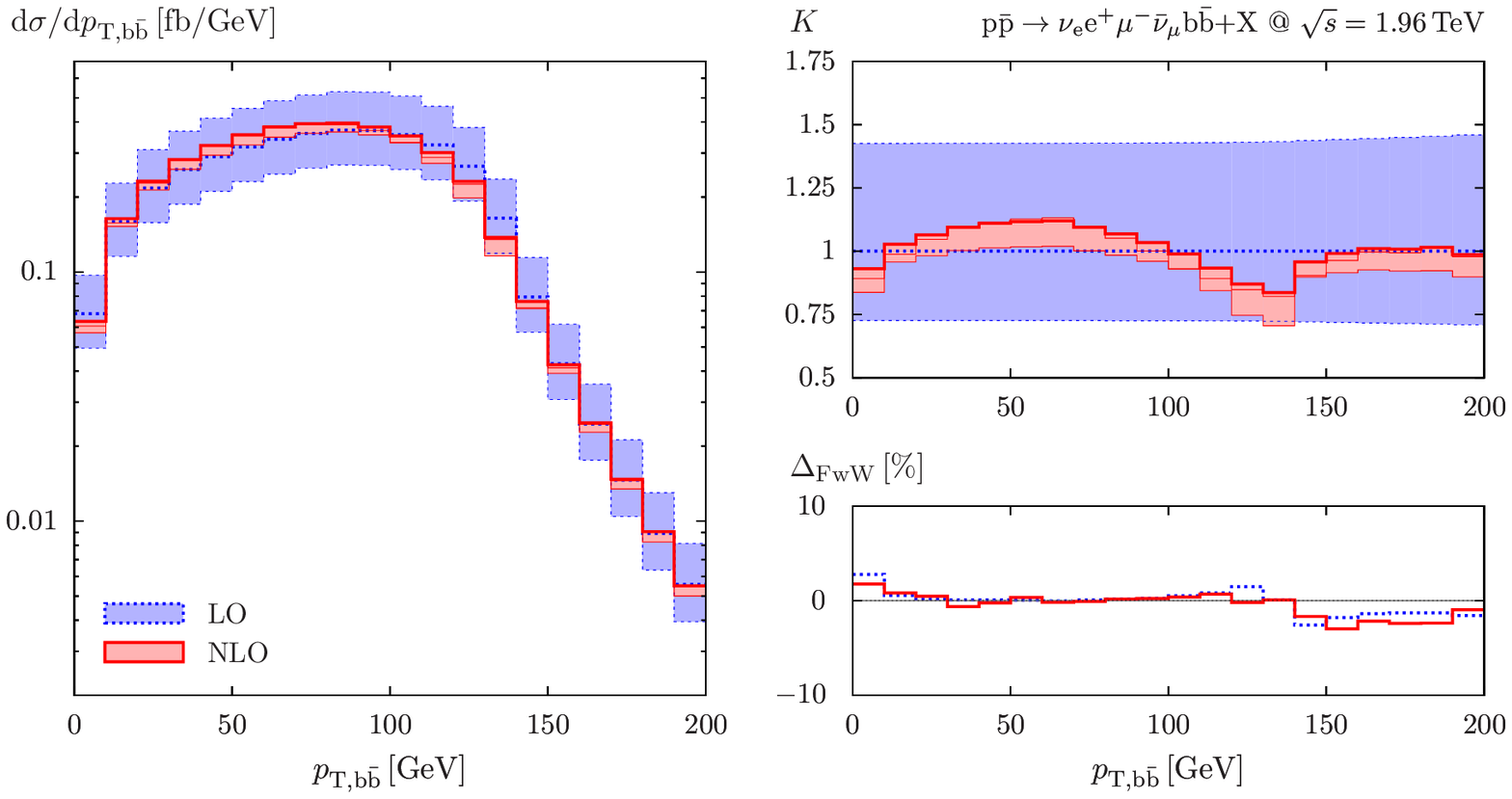}
\caption{Transverse-momentum distribution of the  $\Pb\Pbbar$ pair with standard cuts for Tevatron for dynamical scale
  $\mu_0=\dynscale$.}
\label{fi:ptbB_tev_vs}
\end{figure}

The distribution in the transverse momentum of the $\Pb\Pbbar$ pair
is presented in  \reffi{fi:ptbB_tev_vs}. 
In contrast to the LHC there is no enhancement of $\wwbb$ events at
large $p_{\rT,\Pb\bar\Pb}$, and the $K$ factor stays near one in the
complete considered $\pt$ range. A similar behaviour can be observed
for the missing-transverse-momentum distribution
(not shown).

\begin{figure}
\includegraphics[width=\textwidth]{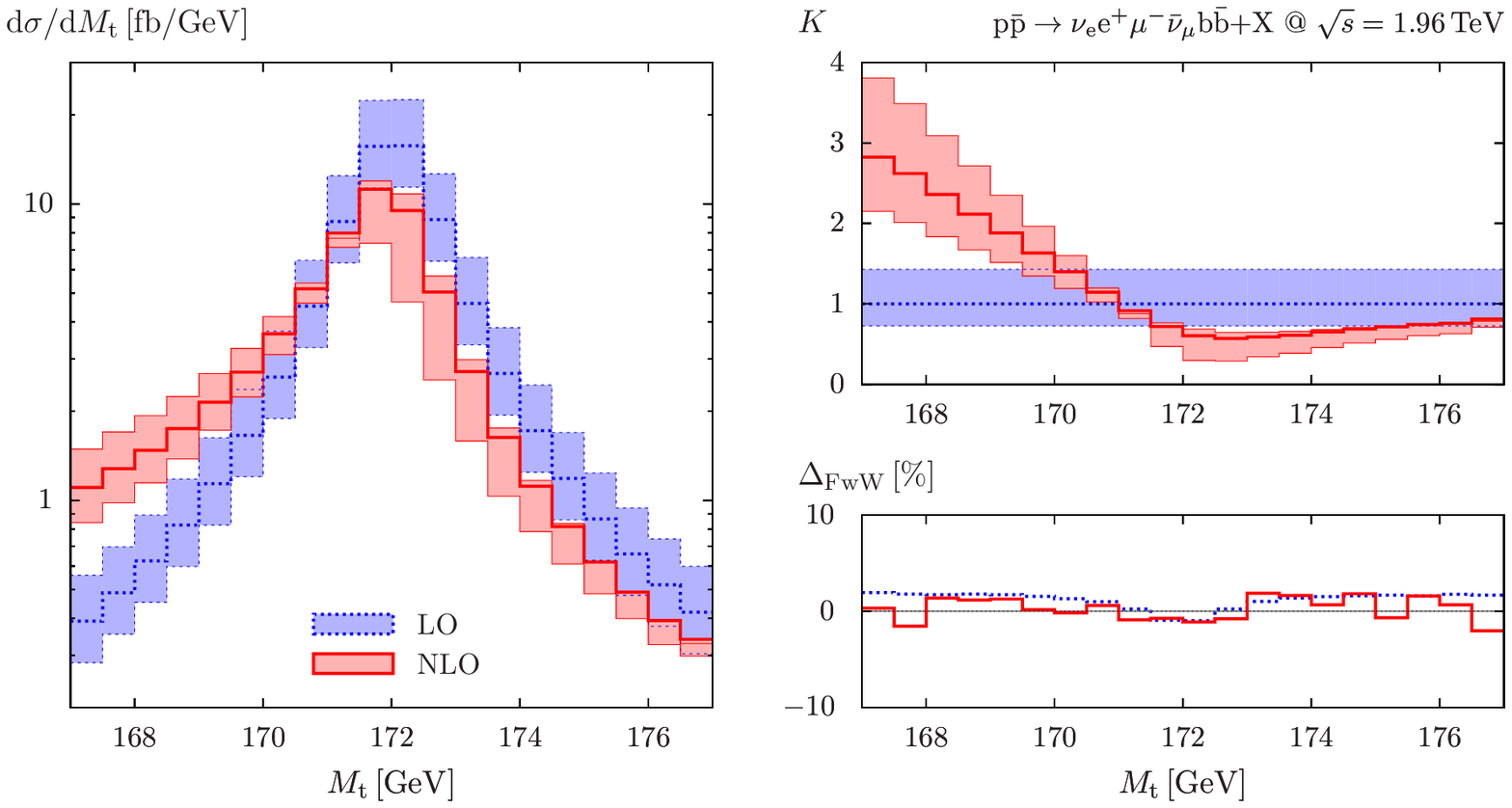}
\caption{Invariant-mass distribution of the top quark, $M_\Pt=M_{\nu_\Pe\Pep\Pb}$, with standard cuts for Tevatron for dynamical scale
  $\mu_0=\dynscale$.}
\label{fi:mt_tev_vs}
\end{figure}

The distribution in the invariant mass of the top quark in the
vicinity of its resonance is shown in \reffi{fi:mt_tev_vs}.
While the enhancement of the distribution by NLO corrections below the
resonance is similar as for the LHC, the rise of the $K$ factor above
the resonance is weaker. In the side bands of the resonance FwW
corrections are at the level of a few per cent as at the LHC.

\begin{figure}
\includegraphics[width=\textwidth]{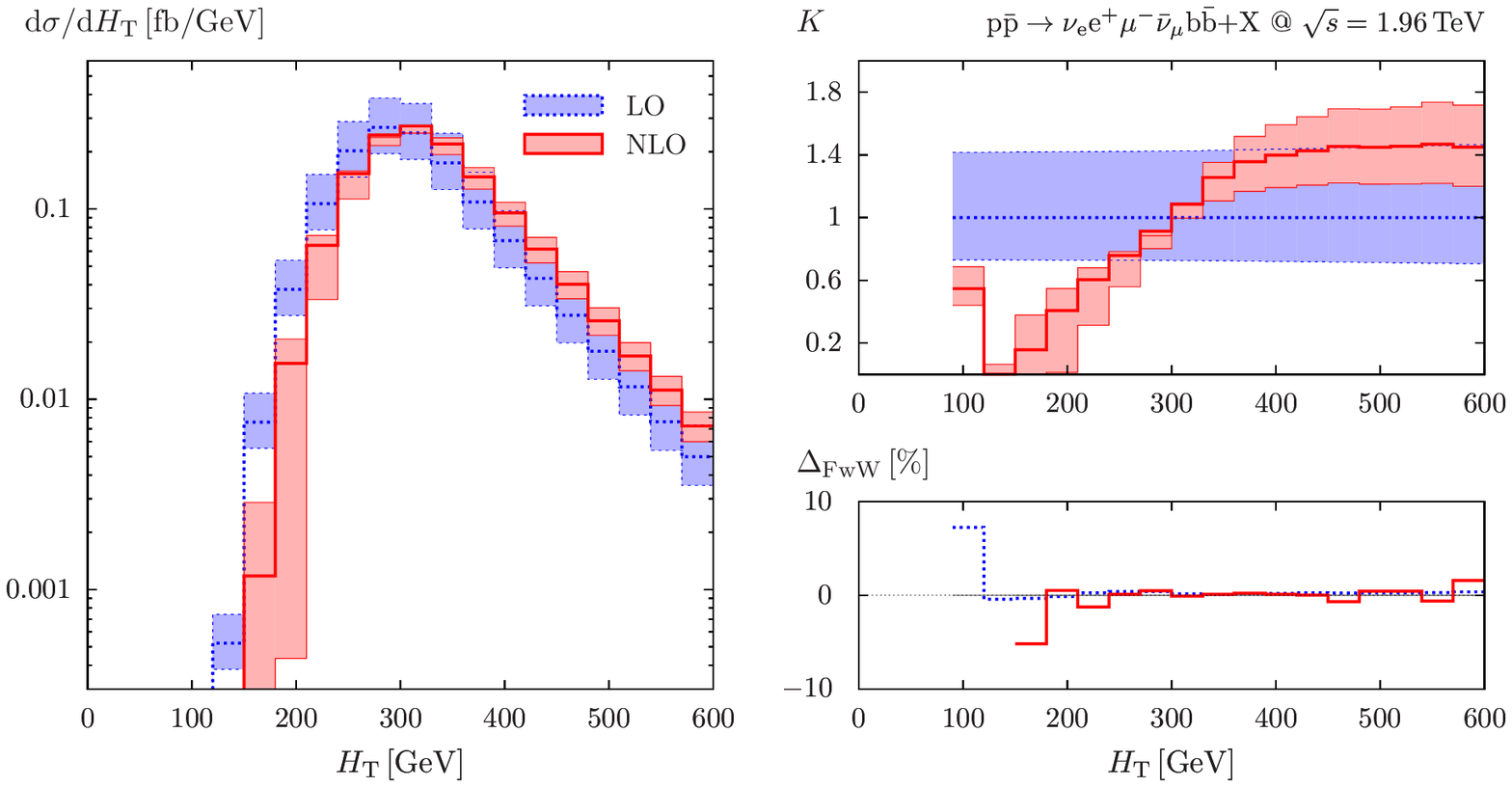}
\caption{Distribution in the total transverse energy with standard cuts for Tevatron for dynamical scale
  $\mu_0=\dynscale$.}
\label{fi:ht_tev_vs}
\end{figure}

The distribution in the total transverse energy as defined in
\refeq{eq:ht} is presented in \reffi{fi:ht_tev_vs}.
The effect of the NLO corrections is qualitatively similar as for the
LHC, but the enhancement of the distribution for high $H_\rT$ is only
$40\%$. As for the LHC, the $K$ factor would become much flatter if the
gluon $\pt$ was not included in $H_\rT$.

\subsection{Limit of on-shell top quarks}
\label{se:FtWeffects}
\begin{figure}
\vspace*{0.5em}
\begin{center}
{\includegraphics[width=.47\textwidth]{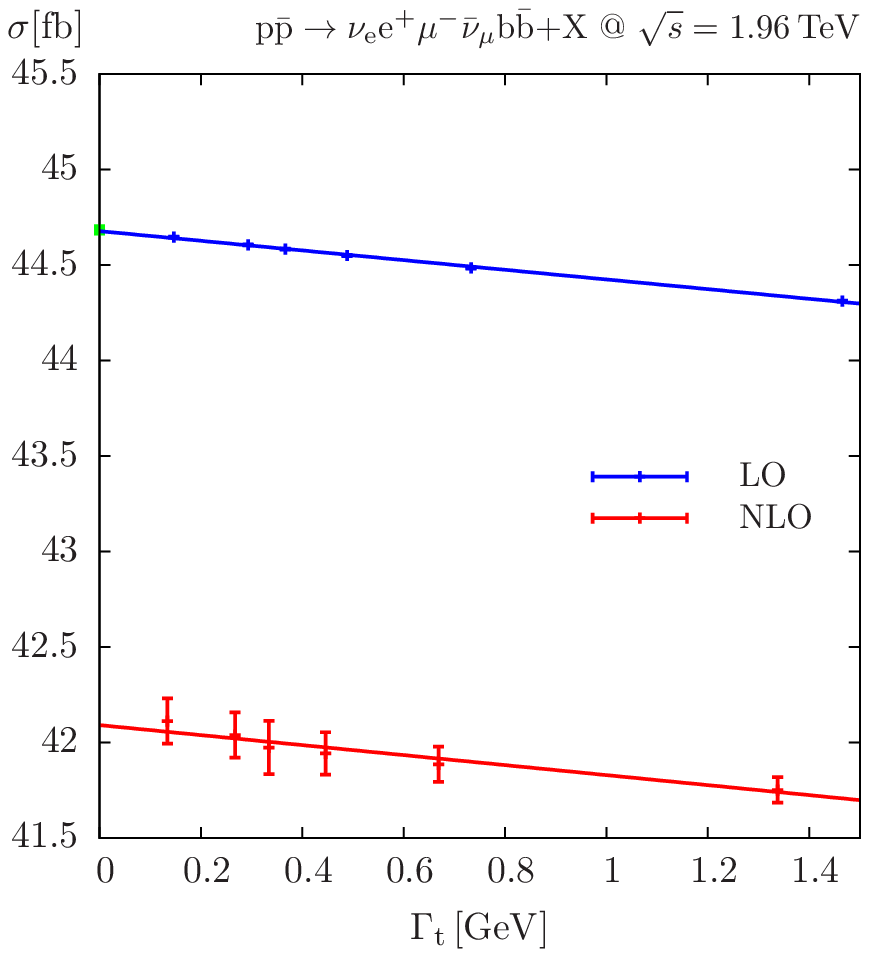}}
\hspace{5mm}
{\includegraphics[width=.47\textwidth]{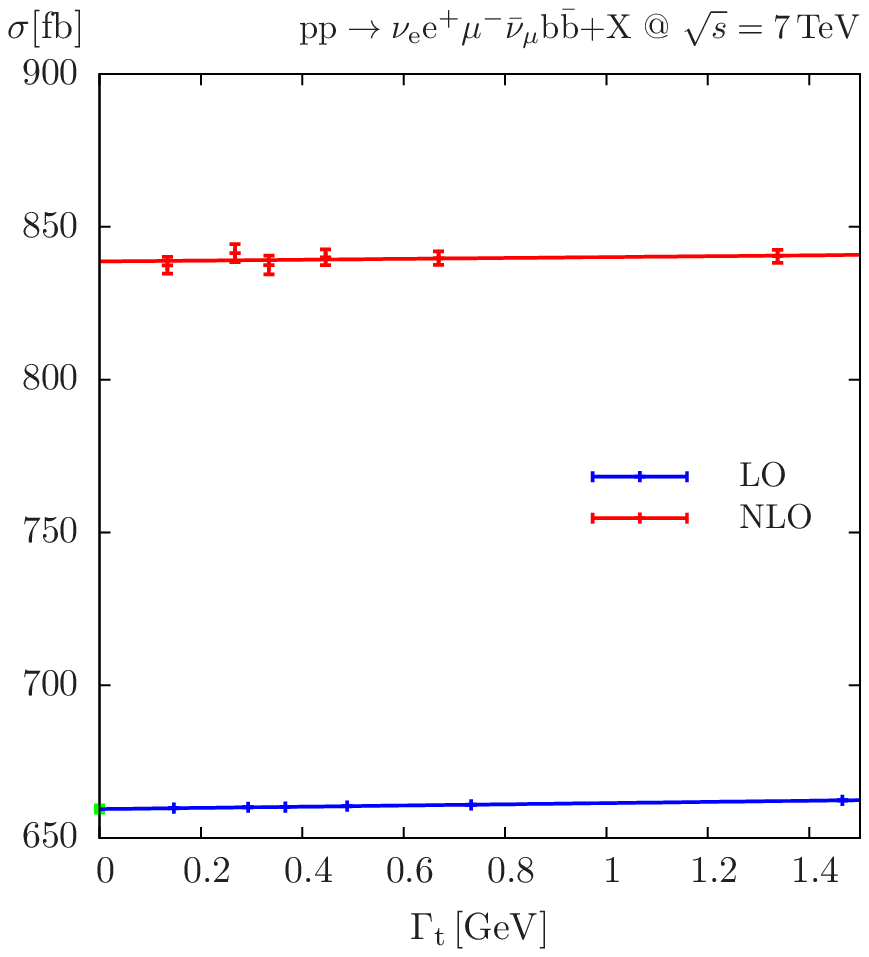}}
\end{center}
\vspace*{-1em}
\caption{Numerical narrow-top-width extrapolation of the LO and NLO \eqintext{$\leptfs$} 
 cross section at Tevatron and the LHC at $\sqrt{s}=7\TeV$. 
This study does not include finite-W-width effects and is based on the same fixed-scale choice, $\mu_0=\Mt$,
for Tevatron and LHC.
}
\label{fig:nwest}
\end{figure}
To quantify non-resonant and off-shell-top contributions to the
\eqintext{$\leptfs$} integrated cross section, we have
investigated its narrow-top-width limit, $\Gamma_\Pt\to 0$, by means of a
numerical extrapolation.  This is shown in~\reffi{fig:nwest}, where we plot 
\beqar
\label{extrapolation}
\bar\sigma(\Gamma_\Pt)&=&
\sigma(\Gamma_\Pt)\left(
\frac{\Gamma_\Pt}{\Gamma_\Pt^{\mathrm{phys}}}
\right)^2
\eeqar
in the range $0<\Gamma_\Pt\le \Gamma_\Pt^{\mathrm{phys}}$,
where $\Gamma_\Pt^{\mathrm{phys}}$ is the physical top-quark width.
The factor $(\Gamma_\Pt/\Gamma_\Pt^{\mathrm{phys}})^2$ 
compensates deviations of $\Gamma_\Pt$ from 
$\Gamma_\Pt^{\mathrm{phys}}$ 
in such a way that the effective 
top-decay branching fractions
remain constant.
The consistency of the $\Gamma_\Pt\to 0$ extrapolation at LO 
is demonstrated by the green squares in  \reffi{fig:nwest}, 
which were obtained with an explicit Born calculation in NtWA.
At NLO, 
when \eqintext{$\Gamma_\Pt\to 0$} 
the virtual and
real parts of NLO corrections are individually enhanced by soft-gluon
logarithmic singularities \mbox{$\propto\alphas\ln\Gamma_\Pt$},
which are not reshuffled via Catani--Seymour
dipoles.  The precise numerical cancellation of such singularities in
the sum of virtual and real corrections manifests itself in the
quality of the linear convergence of the $\Gamma_\Pt\to 0$
extrapolation. This provides a non-trivial confirmation of the
consistency and numerical stability of the calculation. Finite-top-width
effects are extracted by comparing results at
$\Gamma_\Pt=\Gamma_\Pt^{\mathrm{phys}}$ and $\Gamma_\Pt\to 0$.  At the
Tevatron, 
we find that FtW effects shift the LO(NLO) cross section by
$-0.8\%(-0.9\%)$.  This is fairly close to the numerical value of
$\Gamma_\Pt/\Mt$, which represents the expected order of magnitude.
At the LHC, 
with $\sqrt{s}=7\TeV$,
FtW effects turn out to be even smaller: $+0.4\%$
at LO and $+0.2\%$ (comparable to the Monte Carlo statistical error)
at NLO.  
Most likely this 
difference to the result of $\sim\!1\%$ at Tevatron
is due to the influence of singly-top-resonant diagrams corresponding to 
associated $\Pt\PW$ production, 
whose positive contribution even overcompensates the negative effects from off-shell $\ttb$ production.

The results of this extrapolation have been confirmed 
in a dedicated comparison \cite{Denner:lh2011}
between our calculation and the one of 
\citere{Melnikov:2009dn} for on-shell top quarks.
A similar $\Gamma_\Pt\to 0$ extrapolation has been
presented in \citere{Bevilacqua:2010qb}.

\subsection{Comparison with HELAC-NLO}
\label{se:helacnlo}
Our predictions for the hadronic $\leptfs$ cross section have been
successfully reproduced by the {\sc HELAC-NLO} collaboration and vice
versa.  On the one hand, our results for the Tevatron in NwWA
\cite{Denner:2010jp} have been checked by the {\sc HELAC-NLO} group
finding agreement within integration errors \cite{worek:private}.  On
the other hand, we compared the FwW variant of our calculation against
the {\sc HELAC-NLO} predictions of \citere{Bevilacqua:2010qb}.  The
corresponding results obtained with the anti-$\kt$ jet algorithm and
the input parameters, cuts, and PDFs of \citere{Bevilacqua:2010qb} are
listed in \refta{tab:comp_opp}.
\begin{table}
$$\arraycolsep10pt
\begin{array}{c|c|c|c|c|c}
\mathrm{collider} & \mathrm{energy} & 
\si_{\LO}[\mathrm{fb}] &
 \si_{\LO}[\mathrm{fb}]&
\si_{\NLO}[\mathrm{fb}] &
 \si_{\NLO}[\mathrm{fb}] \\
\hline
& & 
\mbox{\citere{Bevilacqua:2010qb}} &
\mbox{this work} &
\mbox{\citere{Bevilacqua:2010qb}} &
\mbox{this work} \\
\hline
\mathrm{Tevatron} &1.96\TeV & 34.922(14) & 34.923(4) & 35.705(47) & 35.777(24)\\    
\mathrm{LHC} &7\TeV         & 550.54(18) & 550.30(6) & 808.46(98) & 807.56(52)\\
\mathrm{LHC} &10\TeV        & 1394.72(75) & 1394.51(14) & 1993.3(2.5) & 1992.1(1.4)\\ 

\end{array}
$$
\caption{Cross sections for off-shell top-pair production for various
  colliders in the setup of \citere{Bevilacqua:2010qb} with anti-$\kt$
  algorithm.}
\label{tab:comp_opp}
\end{table}
At LO and NLO both calculations coincide within 
$(1{-}2)\si$, which is for NLO at the level of $0.2\%$. We note,
however, that the numerical values of the input parameters
$\Gamma^\LO_\Pt$ and $\Gamma^\NLO_\Pt$ used in
\citere{Bevilacqua:2010qb}, and adopted for the technical comparison
in \refta{tab:comp_opp}, do not include FwW corrections. As discussed
in \refses{se:wdec} and \ref{se:xsec}, this induces a fake FwW shift
of roughly $-3\%$ in the integrated cross section.

\section{Conclusions}
\label{se:conclusion}
The precise understanding of top--antitop production is important for
precision tests of the Standard Model and in searches for 
the Higgs boson and new physics.
In this paper, we have presented 
NLO QCD predictions for the
process $\Pp\Pp/\Pp\bar\Pp\to\leptfs+X$,
which corresponds to $\ttb$ production and decay in the 
di-lepton channel.
The calculation includes all finite-top-width effects, \ie 
off-shell top contributions,  diagrams involving 
less than two top resonances, and all factorizable and non-factorizable NLO corrections.
The leptonic W-boson decays have been described alternatively within a narrow-W-width
approximation or including finite-W-width effects.

The employed techniques, based on Feynman diagrams and tensor integrals,
provide high numerical stability and CPU efficiency.
Each Feynman diagram is algebraically reduced to a canonical 
form and automatically processed to {\sc Fortran} output, while
tensor loop integrals are reduced to scalar integrals
using numerical algorithms that avoid
instabilities due to inverse Gram determinants 
and other spurious singularities.
The real corrections are integrated using the dipole subtraction method.  

We have presented a detailed study of integrated and differential
cross sections and of asymmetries at Tevatron and the LHC at $\sqrt{s}=7\TeV$, 
$8\TeV$, and $14\TeV$. 
We pointed out that a fixed QCD scale leads to perturbative instabilities in
the tails of transverse-momentum distributions, while using
a dynamical scale related to the transverse energy of the top quarks
yields stable NLO predictions.
For the LHC, where top--antitop production mainly proceeds via
gluon--gluon annihilation, we decided to reduce the QCD scale by a factor two
as compared to the Tevatron, where the quark--antiquark channel dominates. 
This can be motivated by resummation of higher-order corrections, and leads
to smaller $K$ factors and a somewhat reduced residual scale dependence.

We also discussed the consequences of truncating the perturbative
expansion at NLO in presence of unstable intermediate particles.
Although the NLO corrections to the production and decay of top quarks
are separately well known, a fixed-order NLO description of the entire
process, \ie production and decay, does not include the product of
these two corrections.  Such contributions are of NNLO kind but can
largely exceed one per cent.  To take them into account, we introduced
an approach based on the idea of matching the inclusive $\leptfs$
cross section, in the $\Gt\to 0$ limit, to the on-shell $\ttb$ cross
section. Numerically, the corresponding correction amounts to a few
per cent.

Effects associated to the finite top-quark width are formally
suppressed by $\Gt/\Mt$ and turn out to be below 1\% for the total
cross sections at Tevatron and the LHC at $\sqrt{s}=7\TeV$. 
In contrast, finite-top-width effects can reach tens of per
cent in phase space regions where top--antitop production is suppressed
as a signal, but remains important as a background to Higgs and
new-physics searches, as discussed in \citere{Denner:lh2011}
where our results are compared to the ones of the
(spin-correlated) narrow-width approximation of \citere{Melnikov:2009dn}.

In spite of the larger numerical value of $\Gamma_\PW/\MW$ as compared
to $\Gamma_\Pt/\Mt$, finite-W-width effects turn out to be less
important than finite-top-width effects.  For observables that are
dominated by on-shell top quarks they remain below 0.5\%, and for all
other observables considered in the present paper they never exceed a
few per cent. The only exception is the $M_{\Pe^+\Pb}$ distribution
in the vicinity of $M_{\Pe^+\Pb}^2=\Mt^2-\MW^2$, an observable which is, 
however, important for the precision measurement of the top-quark mass.
The small size of finite-W-width contributions (to
inclusive observables) is due to a double-suppression mechanism, which
involves subtle cancellations between finite-W-width corrections to
$\leptfs$ matrix elements and to the $\Gamma_\Pt$ input parameter.
Besides these observations, one should keep in mind that observables
where top-quark and/or W-boson resonances are kinematically excluded
can receive much larger finite-width contributions.

The step from on-shell top quarks to the inclusion of the top-quark decays
with its full off-shell effects and the differential information on the
decay products, including NLO QCD corrections, is an important step
towards pushing the simulation of $\Pt\bar\Pt$ production at hadron colliders
to a higher level. 
A natural next development, apart from extending the calculation to hadronic
W~decays as well, consists in a proper matching of the new NLO QCD
corrections to off-shell $\Pt\bar\Pt$ production with QCD parton showers.
Further future improvements of the calculation should also aim at
including NLO electroweak corrections, a task that actually leads to a
horrible NLO calculation for a $2\to6$ particle process. However, our setup
of using a double-pole approximation for the W~resonances effectively
factorizes the full process again into a $2\to4$ ($\PW\PW\Pb\bar\Pb$) production and 
the $1\to2$ W-boson decays, rendering this calculation possible, though still 
very demanding.

\section*{Acknowledgements}
S.P.~would like to acknowledge financial support from the SNSF.

\appendix

\section{Benchmark numbers for the virtual corrections}
\label{app:benchmark}
In order to facilitate a comparison to our calculation, 
we provide explicit numbers for the 
squared tree amplitude and the corresponding virtual correction for 
the partonic processes  $\Pu\bar\Pu/\Pg\Pg\to \leptfs$ 
evaluated at the phase-space point specified in \refta{tab:point}.
\begin{table}
$$\arraycolsep10pt
\begin{array}{c|cccc}
\mathrm{particle} &  p^0\,[\GeV] &  p^1\,[\GeV] &  p^2\,[\GeV] &  p^3\,[\GeV]\\ 
\hline
a               & \scriptstyle  \pz250                  & \scriptstyle  \pz0                  & \scriptstyle  \pz0                  & \scriptstyle  \pz250              \\
b               & \scriptstyle  \pz250                  & \scriptstyle  \pz0                  & \scriptstyle  \pz0                  & \scriptstyle    -250              \\
\nu_\Pe         & \scriptstyle  \pz48.56688853967434    & \scriptstyle    -5.434957739109776  & \scriptstyle    -19.57544900121795  & \scriptstyle  \pz44.11355452919156        \\
\Pep            & \scriptstyle  \pz44.98131949428399    & \scriptstyle  \pz27.80765763034380  & \scriptstyle    -19.96763010585797  & \scriptstyle    -29.17785167753225        \\
\mu^-           & \scriptstyle  \pz81.94400052131301    & \scriptstyle  \pz35.50973810316283  & \scriptstyle  \pz57.00882135433334  & \scriptstyle    -46.94541520821219        \\
\bar\nu_\mu     & \scriptstyle  \pz75.44248660279438    & \scriptstyle    -13.21633653678938  & \scriptstyle    -1.177064626981959  & \scriptstyle    -74.26649144952955        \\
\Pb             & \scriptstyle  \pz156.4430715983779    & \scriptstyle    -88.78355499018868  & \scriptstyle    -95.32442077834540  & \scriptstyle  \pz86.63238319307789        \\
\bar\Pb         & \scriptstyle  \pz92.62223324355639    & \scriptstyle  \pz44.11745353258120  & \scriptstyle  \pz79.03574315806995  & \scriptstyle  \pz19.64382061300453        \\ 
\hline
\end{array}
$$
\caption{Phase-space point for the partonic process $ab\to \leptfs$. All scattering particles are massless, 
intermediate top quarks are (slightly) off shell, and intermediate 
W~bosons are exactly on shell, \ie $(p_{\nu_\Pe}+p_\Pep)^2=(p_{\mu^-}+p_{\bar\nu_\mu})^2=\MW^2$.}
\label{tab:point}
\end{table}
The amplitudes are computed using the input parameters of
\refse{se:setup} and the complex top-quark mass \refeq{e:tcms}.  For
the top-quark width we use the NLO value in \refeq{eq:GtfwW}, which
includes FwW corrections, both in the tree and one-loop matrix
elements.  Since we compute virtual matrix elements handling W-boson
resonances in double-pole approximation, we consider a phase-space
point with on-shell intermediate W~bosons.  Moreover, as discussed in
\refse{se:wdec}, we treat the W- and Z-boson masses as real
parameters, we set $\Gamma_\PZ=0$, and we insert a non-zero W-boson
width only in the W-boson propagator, which contributes a factor
$1/(\Gamma_\PW\MW)$ for each W resonance at $p_\PW^2=\MW^2$.  For the
renormalization scale and the scale of dimensional regularization we
choose $\mu=\mu_{\mathrm{R}}=\Mt$.  The corresponding values of the
strong coupling constant in the renormalization scheme described in
\refse{se:numres} are
\beq
\alpha_{\mathrm{s}}(\Mt)|_{\mathrm{LO}} =
0.1258086856923967, \qquad
\alpha_{\mathrm{s}}(\Mt)|_{\mathrm{NLO}}=
0.1095186120831399.
\eeq
In the following we use the NLO strong coupling everywhere, \ie also in
tree matrix elements.

For the spin- and colour-averaged squared LO amplitudes in DPA we obtain
\beqar
|\M_{\Pu\bar\Pu,\DPA}^{\LO}|^2  &=&   1.568863069202805\cdot10^{-5}\GeV^{-8}, \nl
|\M_{\Pg\Pg,\DPA}^{\LO}|^2      &=&   4.554053154627972\cdot10^{-5}\GeV^{-8},
\eeqar
while including doubly- and singly-W-resonant diagrams in the 
complex-mass scheme \refeq{e:WZcms}
with input parameters \refeq{eq:GammaWZ} yields
\beqar
|\M_{\Pu\bar\Pu,\full}^{\LO}|^2 &=&   1.567938932324559\cdot10^{-5}\GeV^{-8}, \nl
|\M_{\Pg\Pg,\full}^{\LO}|^2     &=&   4.551376909529072\cdot10^{-5}\GeV^{-8}.
\eeqar

We express virtual NLO contributions in the $2\to 6$ phase space
as Laurent series in $\epsilon=(4-D)/2$,
\beqar
|\M_\DPA|^2&=&
\left(1+c_\Gamma\sum_{k=0}^2 \de_\NLO^{(k)}\epsilon^{-k}\right)
|\M_\DPA^{\LO}|^2,
\eeqar
where we factor out the LO term in DPA and the normalization factor
\beqar
c_\Gamma&=&
\frac{(4\pi)^\epsilon \Gamma(1+\epsilon)\Gamma^2(1-\epsilon)}
{\Gamma(1-2\epsilon)}
=\frac{(4\pi)^\epsilon}
{\Gamma(1-\epsilon)}+\mathcal{O}(\epsilon^3)
\nn\\
&=&(4\pi)^\epsilon
\Gamma(1+\epsilon)
-\frac{\pi^2}{6}\epsilon^2
+\mathcal{O}(\epsilon^3).
\eeqar
We split the result into the two parts,
\newcommand{\loops}{\mathrm{loops}}
\beqar
\de_\NLO^{(k)}&=&
\de_\loops^{(k)}+
\de_\rI^{(k)},
\eeqar
which correspond to the contributions of renormalized loop diagrams
(loops) and the $I$ operator of the dipole subtraction function as
defined in \citere{Catani:2002hc}.  Here we work with purely massless
light fermions, i.e.\ in pure dimensional regularization.  In
$\de_\loops^{(k)}$ we include counterterm contributions from field,
coupling, and mass renormalization.  The numbers in
\refta{Tab:benchmark} have been obtained in the 't~Hooft--Feynman
gauge using the 't Hooft--Veltman variant of dimensional
regularization (four-dimensional external partons).

The agreement between our two independent versions of the virtual
corrections, as can be seen from \refta{Tab:benchmark}, is typically about 
11~digits at non-exceptional phase-space points.
\begin{table}
$$
\begin{array}{|c|c|c|c|c|}
\hline
\multicolumn{2}{|c|}{\Pu\bar\Pu\; \mathrm{channel}} & \de^{(2)} & \de^{(1)} & \de^{(0)} \\
\hline
\hline
\loops  & \mathrm{v1}   & -0.0929622851927013 & -0.103079416107610 & \pz0.346530980271734 \\
        & \mathrm{v2}   & -0.0929622851925008 & -0.103079416107340 & \pz0.346530980272341 \\
\hline
\rI     & \mathrm{v1}   & \pz0.0929622851925095 & \pz0.103079416107488  & -0.041671620626660 \\
        & \mathrm{v2}   & \pz0.0929622851925092 & \pz0.103079416107487  & -0.041671620626660  \\
\hline\hline
\NLO    & \mathrm{v1}   &   -0.0000000000001918 &   -0.000000000000122  & \pz0.304859359645074 \\
        & \mathrm{v2}   & \pz0.0000000000000084 & \pz0.000000000000146  & \pz0.304859359645681 \\
\hline
\hline
\multicolumn{2}{|c|}{\Pg\Pg\; \mathrm{channel}} & \de^{(2)} & \de^{(1)} & \de^{(0)} \\
\hline
\hline
\loops  & \mathrm{v1}   & -0.1510637134378864 & -0.032125892699063 & \pz0.571739679133372 \\
        & \mathrm{v2}   & -0.1510637134378313 & -0.032125892699187 & \pz0.571739679132260  \\
\hline
\rI     & \mathrm{v1}   & \pz0.1510637134378276 & \pz0.032125892699080 & -0.155205928257365 \\
        & \mathrm{v2}   & \pz0.1510637134378276 & \pz0.032125892699079 & -0.155205928257365  \\
\hline\hline
\NLO    & \mathrm{v1}   & -0.0000000000000588 & \pz0.000000000000017  & \pz0.416533750876007 \\
        & \mathrm{v2}   & -0.0000000000000037 &   -0.000000000000107  & \pz0.416533750874894 \\
\hline
\end{array}
$$
\caption{Various contributions to the virtual NLO corrections
to $\Pu\bar\Pu/\Pg\Pg \to \leptfs$ at the phase-space point specified in 
\refta{Tab:benchmark}. For each contribution we provide the results of the 
two independent calculations (v1 and v2).
}
\label{Tab:benchmark}
\end{table}

\bibliography{ppwwbb}

\providecommand{\href}[2]{#2}\begingroup\raggedright\begin{thebibliography}{10%
0}

\bibitem{Nason:1989zy}
P.~Nason et~al., {\it {The One Particle Inclusive Differential Cross-Section
  for Heavy Quark Production in Hadronic Collisions}},  {\em Nucl. Phys.} {\bf
  B327} (1989) 49--92.

\bibitem{Beenakker:1990maa}
W.~Beenakker, W.~van Neerven, R.~Meng, G.~Schuler, and J.~Smith, {\it {QCD
  corrections to heavy quark production in hadron hadron collisions}},  {\em
  Nucl. Phys.} {\bf B351} (1991) 507--560.

\bibitem{Mangano:1991jk}
M.~L. Mangano, P.~Nason, and G.~Ridolfi, {\it {Heavy quark correlations in
  hadron collisions at next-to-leading order}},  {\em Nucl. Phys.} {\bf B373}
  (1992) 295--345.

\bibitem{Frixione:1995fj}
S.~Frixione, M.~L. Mangano, P.~Nason, and G.~Ridolfi, {\it {Top quark
  distributions in hadronic collisions}},  {\em Phys. Lett.} {\bf B351} (1995)
  555--561, [\href{http://xxx.lanl.gov/abs/hep-ph/9503213}{{\tt
  hep-ph/9503213}}].

\bibitem{Beenakker:1993yr}
W.~Beenakker, A.~Denner, W.~Hollik, R.~Mertig, T.~Sack, et~al., {\it
  {Electroweak one loop contributions to top pair production in hadron
  colliders}},  {\em Nucl. Phys.} {\bf B411} (1994) 343--380.

\bibitem{Moretti:2006nf}
S.~Moretti, M.~R. Nolten, and D.~A. Ross, {\it {Weak corrections to
  gluon-induced top-antitop hadro-production}},  {\em Phys. Lett.} {\bf B639}
  (2006) 513--519, [\href{http://xxx.lanl.gov/abs/hep-ph/0603083}{{\tt
  hep-ph/0603083}}].

\bibitem{Kuhn:2006vh}
J.~H. {K\"uhn}, A.~Scharf, and P.~Uwer, {\it {Electroweak effects in top-quark
  pair production at hadron colliders}},  {\em Eur. Phys. J.} {\bf C51} (2007)
  37--53, [\href{http://xxx.lanl.gov/abs/hep-ph/0610335}{{\tt
  hep-ph/0610335}}].

\bibitem{Bernreuther:2008aw}
W.~Bernreuther, M.~{F\"ucker}, and Z.-G. Si, {\it {Electroweak corrections to
  $t\bar{t}$ production at hadron colliders}},  {\em Nuovo Cim.} {\bf B123}
  (2008) 1036--1044, [\href{http://xxx.lanl.gov/abs/0808.1142}{{\tt
  arXiv:0808.1142}}].

\bibitem{Hollik:2007sw}
W.~Hollik and M.~Kollar, {\it {NLO QED contributions to top-pair production at
  hadron collider}},  {\em Phys. Rev.} {\bf D77} (2008) 014008.

\bibitem{Kuhn:2011ri}
J.~H. {K\"uhn} and G.~Rodrigo, {\it {Charge asymmetries of top quarks at hadron
  colliders revisited}},  {\em JHEP} {\bf 1201} (2012) 063,
  [\href{http://xxx.lanl.gov/abs/1109.6830}{{\tt arXiv:1109.6830}}].

\bibitem{Beneke:2009rj}
M.~Beneke, P.~Falgari, and C.~Schwinn, {\it {Soft radiation in heavy-particle
  pair production: All-order colour structure and two-loop anomalous
  dimension}},  {\em Nucl. Phys.} {\bf B828} (2010) 69--101,
  [\href{http://xxx.lanl.gov/abs/0907.1443}{{\tt arXiv:0907.1443}}].

\bibitem{Czakon:2009zw}
M.~Czakon, A.~Mitov, and G.~F. Sterman, {\it {Threshold Resummation for
  Top-Pair Hadroproduction to Next-to-Next-to-Leading Log}},  {\em Phys. Rev.}
  {\bf D80} (2009) 074017, [\href{http://xxx.lanl.gov/abs/0907.1790}{{\tt
  arXiv:0907.1790}}].

\bibitem{Ahrens:2010zv}
V.~Ahrens, A.~Ferroglia, M.~Neubert, B.~D. Pecjak, and L.~L. Yang, {\it
  {Renormalization-Group Improved Predictions for Top-Quark Pair Production at
  Hadron Colliders}},  {\em JHEP} {\bf 1009} (2010) 097,
  [\href{http://xxx.lanl.gov/abs/1003.5827}{{\tt arXiv:1003.5827}}].

\bibitem{Kidonakis:2010dk}
N.~Kidonakis, {\it {Next-to-next-to-leading soft-gluon corrections for the top
  quark cross section and transverse momentum distribution}},  {\em Phys. Rev.}
  {\bf D82} (2010) 114030, [\href{http://xxx.lanl.gov/abs/1009.4935}{{\tt
  arXiv:1009.4935}}].

\bibitem{Dittmaier:2007wz}
S.~Dittmaier, P.~Uwer, and S.~Weinzierl, {\it {NLO QCD corrections to t anti-t
  + jet production at hadron colliders}},  {\em Phys. Rev. Lett.} {\bf 98}
  (2007) 262002, [\href{http://xxx.lanl.gov/abs/hep-ph/0703120}{{\tt
  hep-ph/0703120}}].

\bibitem{Kniehl:2008fd}
B.~Kniehl, Z.~Merebashvili, J.~{K\"orner}, and M.~Rogal, {\it {Heavy quark pair
  production in gluon fusion at next-to- next-to-leading $O(\alpha_s^{4})$
  order: One-loop}},  {\em Phys. Rev.} {\bf D78} (2008) 094013,
  [\href{http://xxx.lanl.gov/abs/0809.3980}{{\tt arXiv:0809.3980}}].

\bibitem{Anastasiou:2008vd}
C.~Anastasiou and S.~M. Aybat, {\it {The one-loop gluon amplitude for
  heavy-quark production at NNLO}},  {\em Phys. Rev.} {\bf D78} (2008) 114006,
  [\href{http://xxx.lanl.gov/abs/0809.1355}{{\tt arXiv:0809.1355}}].

\bibitem{Czakon:2007ej}
M.~Czakon, A.~Mitov, and S.~Moch, {\it {Heavy-quark production in massless
  quark scattering at two loops in QCD}},  {\em Phys. Lett.} {\bf B651} (2007)
  147--159, [\href{http://xxx.lanl.gov/abs/0705.1975}{{\tt arXiv:0705.1975}}].

\bibitem{Czakon:2007wk}
M.~Czakon, A.~Mitov, and S.~Moch, {\it {Heavy-quark production in gluon fusion
  at two loops in QCD}},  {\em Nucl. Phys.} {\bf B798} (2008) 210,
  [\href{http://xxx.lanl.gov/abs/0707.4139}{{\tt arXiv:0707.4139}}].

\bibitem{Czakon:2008zk}
M.~Czakon, {\it {Tops from Light Quarks: Full Mass Dependence at Two-Loops in
  QCD}},  {\em Phys. Lett.} {\bf B664} (2008) 307--314,
  [\href{http://xxx.lanl.gov/abs/0803.1400}{{\tt arXiv:0803.1400}}].

\bibitem{Bonciani:2008az}
R.~Bonciani, A.~Ferroglia, T.~Gehrmann, D.~Maitre, and C.~Studerus, {\it
  {Two-Loop Fermionic Corrections to Heavy-Quark Pair Production: The
  Quark-Antiquark Channel}},  {\em JHEP} {\bf 07} (2008) 129,
  [\href{http://xxx.lanl.gov/abs/0806.2301}{{\tt arXiv:0806.2301}}].

\bibitem{Bonciani:2009nb}
R.~Bonciani et~al., {\it {Two-Loop Planar Corrections to Heavy-Quark Pair
  Production in the Quark-Antiquark Channel}},  {\em JHEP} {\bf 0908} (2009)
  067, [\href{http://xxx.lanl.gov/abs/0906.3671}{{\tt arXiv:0906.3671}}].

\bibitem{Bonciani:2010mn}
R.~Bonciani, A.~Ferroglia, T.~Gehrmann, A.~Manteuffel, and C.~Studerus, {\it
  {Two-Loop Leading Color Corrections to Heavy-Quark Pair Production in the
  Gluon Fusion Channel}},  {\em JHEP} {\bf 1101} (2011) 102,
  [\href{http://xxx.lanl.gov/abs/1011.6661}{{\tt arXiv:1011.6661}}].

\bibitem{GehrmannDeRidder:2009fz}
A.~Gehrmann-De~Ridder and M.~Ritzmann, {\it {NLO Antenna Subtraction with
  Massive Fermions}},  {\em JHEP} {\bf 0907} (2009) 041,
  [\href{http://xxx.lanl.gov/abs/0904.3297}{{\tt arXiv:0904.3297}}].

\bibitem{Czakon:2010td}
M.~Czakon, {\it {A novel subtraction scheme for double-real radiation at
  NNLO}},  {\em Phys. Lett.} {\bf B693} (2010) 259--268,
  [\href{http://xxx.lanl.gov/abs/1005.0274}{{\tt arXiv:1005.0274}}].

\bibitem{Baernreuther:2012ws}
P.~Baernreuther, M.~Czakon, and A.~Mitov, {\it {Percent level precision physics
  at the Tevatron: first genuine NNLO QCD corrections to $q\bar q\to t \bar t+
  X$}},  \href{http://xxx.lanl.gov/abs/1204.5201}{{\tt arXiv:1204.5201}}.

\bibitem{Kauer:2001sp}
N.~Kauer and D.~Zeppenfeld, {\it {Finite width effects in top quark production
  at hadron colliders}},  {\em Phys.Rev.} {\bf D65} (2002) 014021,
  [\href{http://xxx.lanl.gov/abs/hep-ph/0107181}{{\tt hep-ph/0107181}}].

\bibitem{Bernreuther:2004jv}
W.~Bernreuther, A.~Brandenburg, Z.~Si, and P.~Uwer, {\it {Top quark pair
  production and decay at hadron colliders}},  {\em Nucl. Phys.} {\bf B690}
  (2004) 81--137, [\href{http://xxx.lanl.gov/abs/hep-ph/0403035}{{\tt
  hep-ph/0403035}}].

\bibitem{Melnikov:2009dn}
K.~Melnikov and M.~Schulze, {\it {NLO QCD corrections to top quark pair
  production and decay at hadron colliders}},  {\em JHEP} {\bf 08} (2009) 049,
  [\href{http://xxx.lanl.gov/abs/0907.3090}{{\tt arXiv:0907.3090}}].

\bibitem{Bernreuther:2010ny}
W.~Bernreuther and Z.-G. Si, {\it {Distributions and correlations for top quark
  pair production and decay at the Tevatron and LHC}},  {\em Nucl. Phys.} {\bf
  B837} (2010) 90, [\href{http://xxx.lanl.gov/abs/1003.3926}{{\tt
  arXiv:1003.3926}}].

\bibitem{Campbell:2012uf}
J.~M. Campbell and R.~K. Ellis, {\it {Top-quark processes at NLO in production
  and decay}},  \href{http://xxx.lanl.gov/abs/1204.1513}{{\tt
  arXiv:1204.1513}}.

\bibitem{Denner:2010jp}
A.~Denner, S.~Dittmaier, S.~Kallweit, and S.~Pozzorini, {\it {NLO QCD
  corrections to WWbb production at hadron colliders}},  {\em Phys. Rev. Lett.}
  {\bf 106} (2011) 052001, [\href{http://xxx.lanl.gov/abs/1012.3975}{{\tt
  arXiv:1012.3975}}].

\bibitem{Bevilacqua:2010qb}
G.~Bevilacqua et~al., {\it {Complete off-shell effects in top quark pair
  hadroproduction with leptonic decay at next-to-leading order}},  {\em JHEP}
  {\bf 1102} (2011) 083, [\href{http://xxx.lanl.gov/abs/1012.4230}{{\tt
  arXiv:1012.4230}}].

\bibitem{Binoth:2010ra}
{\bf {SM and NLO Multileg Working Group}} Collaboration, J.~R. Andersen et~al.,
  {\it {The SM and NLO multileg working group: Summary report}},
  \href{http://xxx.lanl.gov/abs/1003.1241}{{\tt arXiv:1003.1241}}.

\bibitem{Bredenstein:2009aj}
A.~Bredenstein, A.~Denner, S.~Dittmaier, and S.~Pozzorini, {\it {NLO QCD
  corrections to pp $\to$ t anti-t b anti-b + X at the LHC}},  {\em
  Phys.Rev.Lett.} {\bf 103} (2009) 012002,
  [\href{http://xxx.lanl.gov/abs/0905.0110}{{\tt arXiv:0905.0110}}].

\bibitem{Bevilacqua:2009zn}
G.~Bevilacqua, M.~Czakon, C.~Papadopoulos, R.~Pittau, and M.~Worek, {\it
  {Assault on the NLO Wishlist: pp $\to$ t anti-t b anti-b}},  {\em JHEP} {\bf
  0909} (2009) 109, [\href{http://xxx.lanl.gov/abs/0907.4723}{{\tt
  arXiv:0907.4723}}].

\bibitem{Bevilacqua:2011aa}
G.~Bevilacqua, M.~Czakon, C.~Papadopoulos, and M.~Worek, {\it {Hadronic
  top-quark pair production in association with two jets at Next-to-Leading
  Order QCD}},  {\em Phys.Rev.} {\bf D84} (2011) 114017,
  [\href{http://xxx.lanl.gov/abs/1108.2851}{{\tt arXiv:1108.2851}}].

\bibitem{Ellis:2009zw}
R.~Ellis, K.~Melnikov, and G.~Zanderighi, {\it {Generalized unitarity at work:
  first NLO QCD results for hadronic $W$ + 3jet production}},  {\em JHEP} {\bf
  0904} (2009) 077, [\href{http://xxx.lanl.gov/abs/0901.4101}{{\tt
  arXiv:0901.4101}}].

\bibitem{Melia:2011dw}
T.~Melia, K.~Melnikov, R.~Rontsch, and G.~Zanderighi, {\it {NLO QCD corrections
  for $W^+W^-$ pair production in association with two jets at hadron
  colliders}},  {\em Phys.Rev.} {\bf D83} (2011) 114043,
  [\href{http://xxx.lanl.gov/abs/1104.2327}{{\tt arXiv:1104.2327}}].

\bibitem{Berger:2009zg}
C.~Berger, Z.~Bern, L.~J. Dixon, F.~Febres~Cordero, D.~Forde, et~al., {\it
  {Precise Predictions for $W$ + 3 Jet Production at Hadron Colliders}},  {\em
  Phys.Rev.Lett.} {\bf 102} (2009) 222001,
  [\href{http://xxx.lanl.gov/abs/0902.2760}{{\tt arXiv:0902.2760}}].

\bibitem{Berger:2010vm}
C.~Berger, Z.~Bern, L.~J. Dixon, F.~Febres~Cordero, D.~Forde, et~al., {\it
  {Next-to-Leading Order QCD Predictions for $Z,\gamma^*$+3-Jet Distributions
  at the Tevatron}},  {\em Phys.Rev.} {\bf D82} (2010) 074002,
  [\href{http://xxx.lanl.gov/abs/1004.1659}{{\tt arXiv:1004.1659}}].

\bibitem{Berger:2010zx}
C.~Berger, Z.~Bern, L.~J. Dixon, F.~Febres~Cordero, D.~Forde, et~al., {\it
  {Precise Predictions for W + 4 Jet Production at the Large Hadron Collider}},
   {\em Phys.Rev.Lett.} {\bf 106} (2011) 092001,
  [\href{http://xxx.lanl.gov/abs/1009.2338}{{\tt arXiv:1009.2338}}].

\bibitem{Ita:2011wn}
H.~Ita, Z.~Bern, L.~Dixon, F.~Febres~Cordero, D.~Kosower, et~al., {\it {Precise
  Predictions for Z + 4 Jets at Hadron Colliders}},  {\em Phys.Rev.} {\bf D85}
  (2012) 031501, [\href{http://xxx.lanl.gov/abs/1108.2229}{{\tt
  arXiv:1108.2229}}].

\bibitem{Bern:2011ep}
Z.~Bern, G.~Diana, L.~Dixon, F.~Febres~Cordero, S.~Hoeche, et~al., {\it
  {Four-Jet Production at the Large Hadron Collider at Next-to-Leading Order in
  QCD}},  \href{http://xxx.lanl.gov/abs/1112.3940}{{\tt arXiv:1112.3940}}.

\bibitem{Campanario:2011ud}
F.~Campanario, C.~Englert, M.~Rauch, and D.~Zeppenfeld, {\it {Precise
  predictions for $W \gamma \gamma+$jet production at hadron colliders}},  {\em
  Phys.Lett.} {\bf B704} (2011) 515--519,
  [\href{http://xxx.lanl.gov/abs/1106.4009}{{\tt arXiv:1106.4009}}].

\bibitem{Greiner:2011mp}
N.~Greiner, A.~Guffanti, T.~Reiter, and J.~Reuter, {\it {NLO QCD corrections to
  the production of two bottom-antibottom pairs at the LHC}},  {\em
  Phys.Rev.Lett.} {\bf 107} (2011) 102002,
  [\href{http://xxx.lanl.gov/abs/1105.3624}{{\tt arXiv:1105.3624}}].

\bibitem{Greiner:2012im}
N.~Greiner, G.~Heinrich, P.~Mastrolia, G.~Ossola, T.~Reiter, et~al., {\it {NLO
  QCD corrections to the production of W+ W- plus two jets at the LHC}},  {\em
  Phys.Lett.} {\bf B713} (2012) 277--283,
  [\href{http://xxx.lanl.gov/abs/1202.6004}{{\tt arXiv:1202.6004}}].

\bibitem{Becker:2011vg}
S.~Becker, D.~Goetz, C.~Reuschle, C.~Schwan, and S.~Weinzierl, {\it {NLO
  results for five, six and seven jets in electron--positron annihilation}},
  {\em Phys.Rev.Lett.} {\bf 108} (2012) 032005,
  [\href{http://xxx.lanl.gov/abs/1111.1733}{{\tt arXiv:1111.1733}}].

\bibitem{Bevilacqua:2012em}
G.~Bevilacqua and M.~Worek, {\it {Constraining BSM Physics at the LHC: Four top
  final states with NLO accuracy in perturbative QCD}},
  \href{http://xxx.lanl.gov/abs/1206.3064}{{\tt arXiv:1206.3064}}.

\bibitem{Denner:2005es}
A.~Denner, S.~Dittmaier, M.~Roth, and L.~H. Wieders, {\it {Complete electroweak
  $\mathcal{O}(\alpha)$ corrections to charged-current $e^+ e^- \to$ 4 fermion
  processes}},  {\em Phys. Lett.} {\bf B612} (2005) 223,
  [\href{http://xxx.lanl.gov/abs/hep-ph/0502063}{{\tt hep-ph/0502063}}].
  Erratum-ibid.~{\bf B704} (2011) 667-668.

\bibitem{Denner:2005fg}
A.~Denner, S.~Dittmaier, M.~Roth, and L.~Wieders, {\it {Electroweak corrections
  to charged-current $\Pep\Pem\to$ 4 fermion processes: Technical details and
  further results}},  {\em Nucl.Phys.} {\bf B724} (2005) 247--294,
  [\href{http://xxx.lanl.gov/abs/hep-ph/0505042}{{\tt hep-ph/0505042}}].
  Erratum-ibid.~{\bf B854} (2012) 504--507.

\bibitem{Denner:2002ii}
A.~Denner and S.~Dittmaier, {\it {Reduction of one-loop tensor 5-point
  integrals}},  {\em Nucl. Phys.} {\bf B658} (2003) 175--202.

\bibitem{Denner:2005nn}
A.~Denner and S.~Dittmaier, {\it {Reduction schemes for one-loop tensor
  integrals}},  {\em Nucl. Phys.} {\bf B734} (2006) 62--115.

\bibitem{'tHooft:1978xw}
G.~'t~Hooft and M.~J.~G. Veltman, {\it {Scalar One Loop Integrals}},  {\em
  Nucl. Phys.} {\bf B153} (1979) 365--401.

\bibitem{Denner:2010tr}
A.~Denner and S.~Dittmaier, {\it {Scalar one-loop 4-point integrals}},  {\em
  Nucl. Phys.} {\bf B844} (2011) 199--242,
  [\href{http://xxx.lanl.gov/abs/1005.2076}{{\tt arXiv:1005.2076}}].

\bibitem{Catani:1996vz}
S.~Catani and M.~H. Seymour, {\it {A general algorithm for calculating jet
  cross sections in NLO QCD}},  {\em Nucl. Phys.} {\bf B485} (1997) 291--419.

\bibitem{Dittmaier:1999mb}
S.~Dittmaier, {\it {A General approach to photon radiation off fermions}},
  {\em Nucl.Phys.} {\bf B565} (2000) 69--122,
  [\href{http://xxx.lanl.gov/abs/hep-ph/9904440}{{\tt hep-ph/9904440}}].

\bibitem{Phaf:2001gc}
L.~Phaf and S.~Weinzierl, {\it {Dipole formalism with heavy fermions}},  {\em
  JHEP} {\bf 0104} (2001) 006,
  [\href{http://xxx.lanl.gov/abs/hep-ph/0102207}{{\tt hep-ph/0102207}}].

\bibitem{Catani:2002hc}
S.~Catani, S.~Dittmaier, M.~H. Seymour, and Z.~Trocsanyi, {\it {The Dipole
  formalism for next-to-leading order QCD calculations with massive partons}},
  {\em Nucl.Phys.} {\bf B627} (2002) 189--265,
  [\href{http://xxx.lanl.gov/abs/hep-ph/0201036}{{\tt hep-ph/0201036}}].

\bibitem{Berends:1994pv}
F.~A. Berends, R.~Pittau, and R.~Kleiss, {\it {All electroweak four fermion
  processes in electron--positron collisions}},  {\em Nucl. Phys.} {\bf B424}
  (1994) 308--342.

\bibitem{Denner:1999gp}
A.~Denner, S.~Dittmaier, M.~Roth, and D.~Wackeroth, {\it {Predictions for all
  processes $e^+e^-\to$ 4 fermions + $\gamma$}},  {\em Nucl. Phys.} {\bf B560}
  (1999) 33--65, [\href{http://xxx.lanl.gov/abs/hep-ph/9904472}{{\tt
  hep-ph/9904472}}].

\bibitem{Dittmaier:2002ap}
S.~Dittmaier and M.~Roth, {\it {LUSIFER: A LUcid approach to SIx FERmion
  production}},  {\em Nucl. Phys.} {\bf B642} (2002) 307--343,
  [\href{http://xxx.lanl.gov/abs/hep-ph/0206070}{{\tt hep-ph/0206070}}].

\bibitem{Denner:lh2011}
A.~Denner, S.~Dittmaier, S.~Kallweit, S.~Pozzorini, and M.~Schulze, {\it
  {Finite-width effects in top-quark pair production and decay at the LHC}},
  in {\em {The SM and NLO Multileg and SM MC Working Groups: Summary Report}}
  (J.~A. Maestre, S.~Alioli, J.~Andersen, R.~Ball, A.~Buckley, et~al., eds.),
  pp.~55--63, 2012.
\newblock \href{http://xxx.lanl.gov/abs/1203.6803}{{\tt arXiv:1203.6803}}.

\bibitem{ftwpaper}
{A. Denner, S. Dittmaier, S. Kallweit, S. Pozzorini and M. Schulze,
  \textit{Off-shell effects in top--antitop production at hadron colliders}, in
  preparation.}

\bibitem{Bevilacqua:2011xh}
G.~Bevilacqua, M.~Czakon, M.~Garzelli, A.~van Hameren, A.~Kardos, et~al., {\it
  {HELAC-NLO}},  \href{http://xxx.lanl.gov/abs/1110.1499}{{\tt
  arXiv:1110.1499}}.

\bibitem{Fadin:1993dz}
V.~S. Fadin, V.~A. Khoze, and A.~D. Martin, {\it {Interference radiative
  phenomena in the production of heavy unstable particles}},  {\em Phys.Rev.}
  {\bf D49} (1994) 2247--2256.

\bibitem{Fadin:1993kt}
V.~S. Fadin, V.~A. Khoze, and A.~D. Martin, {\it {How suppressed are the
  radiative interference effects in heavy instable particle production?}},
  {\em Phys.Lett.} {\bf B320} (1994) 141--144,
  [\href{http://xxx.lanl.gov/abs/hep-ph/9309234}{{\tt hep-ph/9309234}}].

\bibitem{Melnikov:1993np}
K.~Melnikov and O.~I. Yakovlev, {\it {Top near threshold: All $\alpha_S$
  corrections are trivial}},  {\em Phys.Lett.} {\bf B324} (1994) 217--223,
  [\href{http://xxx.lanl.gov/abs/hep-ph/9302311}{{\tt hep-ph/9302311}}].

\bibitem{Kublbeck:1990xc}
J.~{K\"ublbeck}, M.~{B\"ohm}, and A.~Denner, {\it {FEYNARTS: computer algebraic
  generation of Feynman graphs and amplitudes}},  {\em Comput. Phys. Commun.}
  {\bf 60} (1990) 165--180.

\bibitem{Hahn:2000kx}
T.~Hahn, {\it {Generating Feynman diagrams and amplitudes with FeynArts 3}},
  {\em Comput. Phys. Commun.} {\bf 140} (2001) 418--431.

\bibitem{Bredenstein:2008zb}
A.~Bredenstein, A.~Denner, S.~Dittmaier, and S.~Pozzorini, {\it {NLO QCD
  corrections to top anti-top bottom anti-bottom production at the LHC: 1.
  quark-antiquark annihilation}},  {\em JHEP} {\bf 08} (2008) 108.

\bibitem{Bredenstein:2010rs}
A.~Bredenstein, A.~Denner, S.~Dittmaier, and S.~Pozzorini, {\it {NLO QCD
  corrections to top anti-top bottom anti-bottom production at the LHC: 2. full
  hadronic results}},  {\em JHEP} {\bf 1003} (2010) 021.

\bibitem{Hahn:1998yk}
T.~Hahn and M.~Perez-Victoria, {\it {Automatized one-loop calculations in four
  and D dimensions}},  {\em Comput. Phys. Commun.} {\bf 118} (1999) 153--165.

\bibitem{Dittmaier:1998nn}
S.~Dittmaier, {\it {Weyl-van-der-Waerden formalism for helicity amplitudes of
  massive particles}},  {\em Phys. Rev.} {\bf D59} (1999) 016007.

\bibitem{Binoth:2005ff}
T.~Binoth, J.~Guillet, G.~Heinrich, E.~Pilon, and C.~Schubert, {\it {An
  Algebraic/numerical formalism for one-loop multi-leg amplitudes}},  {\em
  JHEP} {\bf 0510} (2005) 015,
  [\href{http://xxx.lanl.gov/abs/hep-ph/0504267}{{\tt hep-ph/0504267}}].

\bibitem{Fleischer:2010sq}
J.~Fleischer and T.~Riemann, {\it {A Complete algebraic reduction of one-loop
  tensor Feynman integrals}},  {\em Phys.Rev.} {\bf D83} (2011) 073004,
  [\href{http://xxx.lanl.gov/abs/1009.4436}{{\tt arXiv:1009.4436}}].

\bibitem{Melrose:1965kb}
D.~Melrose, {\it {Reduction of Feynman diagrams}},  {\em Nuovo Cim.} {\bf 40}
  (1965) 181--213.

\bibitem{Passarino:1978jh}
G.~Passarino and M.~J.~G. Veltman, {\it {One Loop Corrections for $e^+ e^-$
  Annihilation Into $\mu^+ \mu^-$ in the Weinberg Model}},  {\em Nucl. Phys.}
  {\bf B160} (1979) 151.

\bibitem{Ferroglia:2002mz}
A.~Ferroglia, M.~Passera, G.~Passarino, and S.~Uccirati, {\it {All purpose
  numerical evaluation of one loop multileg Feynman diagrams}},  {\em
  Nucl.Phys.} {\bf B650} (2003) 162--228,
  [\href{http://xxx.lanl.gov/abs/hep-ph/0209219}{{\tt hep-ph/0209219}}].

\bibitem{Giele:2004ub}
W.~Giele, E.~Glover, and G.~Zanderighi, {\it {Numerical evaluation of one-loop
  diagrams near exceptional momentum configurations}},  {\em
  Nucl.Phys.Proc.Suppl.} {\bf 135} (2004) 275--279,
  [\href{http://xxx.lanl.gov/abs/hep-ph/0407016}{{\tt hep-ph/0407016}}].

\bibitem{Beenakker:1988jr}
W.~Beenakker and A.~Denner, {\it {Infrared divergent scalar box integrals with
  applications in the electroweak standard model}},  {\em Nucl. Phys.} {\bf
  B338} (1990) 349--370.

\bibitem{Dittmaier:2003bc}
S.~Dittmaier, {\it {Separation of soft and collinear singularities from
  one-loop N-point integrals}},  {\em Nucl. Phys.} {\bf B675} (2003) 447--466,
  [\href{http://xxx.lanl.gov/abs/hep-ph/0308246}{{\tt hep-ph/0308246}}].

\bibitem{Stuart:1991xk}
R.~G. Stuart, {\it {Gauge invariance, analyticity and physical observables at
  the $Z_0$ resonance}},  {\em Phys.Lett.} {\bf B262} (1991) 113--119.

\bibitem{Aeppli:1993rs}
A.~Aeppli, G.~J. van Oldenborgh, and D.~Wyler, {\it {Unstable particles in one
  loop calculations}},  {\em Nucl.Phys.} {\bf B428} (1994) 126--146,
  [\href{http://xxx.lanl.gov/abs/hep-ph/9312212}{{\tt hep-ph/9312212}}].

\bibitem{Grunewald:2000ju}
M.~W. {Gr\"unewald}, G.~Passarino, E.~Accomando, A.~Ballestrero, P.~Bambade,
  et~al., {\it {Reports of the Working Groups on Precision Calculations for
  LEP2 Physics: Proceedings. Four fermion production in electron--positron
  collisions}},  \href{http://xxx.lanl.gov/abs/hep-ph/0005309}{{\tt
  hep-ph/0005309}}.

\bibitem{Denner:2000bj}
A.~Denner, S.~Dittmaier, M.~Roth, and D.~Wackeroth, {\it {Electroweak radiative
  corrections to $e^+ e^-\to W W \to$ 4 fermions in double pole approximation:
  The RACOONWW approach}},  {\em Nucl.Phys.} {\bf B587} (2000) 67--117,
  [\href{http://xxx.lanl.gov/abs/hep-ph/0006307}{{\tt hep-ph/0006307}}].

\bibitem{Beenakker:1998gr}
W.~Beenakker, F.~A. Berends, and A.~Chapovsky, {\it {Radiative corrections to
  pair production of unstable particles: results for $e^+ e^-\to$ four
  fermions}},  {\em Nucl.Phys.} {\bf B548} (1999) 3--59,
  [\href{http://xxx.lanl.gov/abs/hep-ph/9811481}{{\tt hep-ph/9811481}}].

\bibitem{Alwall:2007st}
J.~Alwall et~al., {\it {MadGraph/MadEvent v4: The New Web Generation}},  {\em
  JHEP} {\bf 09} (2007) 028.

\bibitem{Cascioli:2011va}
F.~Cascioli, P.~{Maierh\"ofer}, and S.~Pozzorini, {\it {Scattering Amplitudes
  with Open Loops}},  {\em Phys.Rev.Lett.} {\bf 108} (2012) 111601,
  [\href{http://xxx.lanl.gov/abs/1111.5206}{{\tt arXiv:1111.5206}}].

\bibitem{Dittmaier:2009un}
S.~Dittmaier, S.~Kallweit, and P.~Uwer, {\it {NLO QCD corrections to $p p/p
  \bar{p}$ $\to$ WW+jet+X including leptonic W-boson decays}},  {\em Nucl.
  Phys.} {\bf B826} (2010) 18--70,
  [\href{http://xxx.lanl.gov/abs/0908.4124}{{\tt arXiv:0908.4124}}].

\bibitem{Campanario:2010hp}
F.~Campanario, C.~Englert, S.~Kallweit, M.~Spannowsky, and D.~Zeppenfeld, {\it
  {NLO QCD corrections to WZ+jet production with leptonic decays}},  {\em JHEP}
  {\bf 07} (2010) 076, [\href{http://xxx.lanl.gov/abs/1006.0390}{{\tt
  arXiv:1006.0390}}].

\bibitem{Hasegawa:2009tx}
K.~Hasegawa, S.~Moch, and P.~Uwer, {\it {AutoDipole: Automated generation of
  dipole subtraction terms}},  {\em Comput. Phys. Commun.} {\bf 181} (2010)
  1802--1817, [\href{http://xxx.lanl.gov/abs/0911.4371}{{\tt
  arXiv:0911.4371}}].

\bibitem{Kleiss:1994qy}
R.~Kleiss and R.~Pittau, {\it {Weight optimization in multichannel Monte
  Carlo}},  {\em Comput. Phys. Commun.} {\bf 83} (1994) 141--146.

\bibitem{Bredenstein:2005zk}
A.~Bredenstein, S.~Dittmaier, and M.~Roth, {\it {Four-fermion production at
  gamma gamma colliders. 2. Radiative corrections in double-pole
  approximation}},  {\em Eur.Phys.J.} {\bf C44} (2005) 27--49,
  [\href{http://xxx.lanl.gov/abs/hep-ph/0506005}{{\tt hep-ph/0506005}}].

\bibitem{Ciccolini:2007jr}
M.~Ciccolini, A.~Denner, and S.~Dittmaier, {\it {Strong and electroweak
  corrections to the production of Higgs + 2jets via weak interactions at the
  LHC}},  {\em Phys. Rev. Lett.} {\bf 99} (2007) 161803,
  [\href{http://xxx.lanl.gov/abs/0707.0381}{{\tt arXiv:0707.0381}}].

\bibitem{Martin:2009iq}
A.~Martin, W.~Stirling, R.~Thorne, and G.~Watt, {\it {Parton distributions for
  the LHC}},  {\em Eur. Phys. J.} {\bf C63} (2009) 189--285,
  [\href{http://xxx.lanl.gov/abs/0901.0002}{{\tt arXiv:0901.0002}}].

\bibitem{Jezabek:1988iv}
M.~Jezabek and J.~H. {K\"uhn}, {\it {QCD Corrections to Semileptonic Decays of
  Heavy Quarks}},  {\em Nucl. Phys.} {\bf B314} (1989) 1.

\bibitem{Cacciari:2008gp}
M.~Cacciari, G.~P. Salam, and G.~Soyez, {\it {The anti-kt jet clustering
  algorithm}},  {\em JHEP} {\bf 04} (2008) 063,
  [\href{http://xxx.lanl.gov/abs/0802.1189}{{\tt arXiv:0802.1189}}].

\bibitem{Bonciani:1998vc}
R.~Bonciani, S.~Catani, M.~L. Mangano, and P.~Nason, {\it {NLL resummation of
  the heavy quark hadroproduction cross-section}},  {\em Nucl.Phys.} {\bf B529}
  (1998) 424--450, [\href{http://xxx.lanl.gov/abs/hep-ph/9801375}{{\tt
  hep-ph/9801375}}].

\bibitem{Aaltonen:2011kc}
{\bf CDF} Collaboration, T.~Aaltonen et~al., {\it {Evidence for a Mass
  Dependent Forward-Backward Asymmetry in Top Quark Pair Production}},  {\em
  Phys.Rev.} {\bf D83} (2011) 112003,
  [\href{http://xxx.lanl.gov/abs/1101.0034}{{\tt arXiv:1101.0034}}].

\bibitem{Abazov:2011rq}
{\bf D0} Collaboration, V.~M. Abazov et~al., {\it {Forward-backward asymmetry
  in top quark-antiquark production}},  {\em Phys.Rev.} {\bf D84} (2011)
  112005, [\href{http://xxx.lanl.gov/abs/1107.4995}{{\tt arXiv:1107.4995}}].

\bibitem{Aaltonen:2012}
{\bf CDF} Collaboration, T.~Aaltonen et~al., {\it {Study of the Top Quark
  Production Asymmetry and Its Mass and Rapidity Dependence in the Full Run II
  Tevatron Dataset}},  \href{http://xxx.lanl.gov/abs/{CDF-NOTE-10807}}{{\tt
  {CDF-NOTE-10807}}}.

\bibitem{Hollik:2011ps}
W.~Hollik and D.~Pagani, {\it {The electroweak contribution to the top quark
  forward-backward asymmetry at the Tevatron}},  {\em Phys.Rev.} {\bf D84}
  (2011) 093003, [\href{http://xxx.lanl.gov/abs/1107.2606}{{\tt
  arXiv:1107.2606}}].

\bibitem{Bernreuther:2012sx}
W.~Bernreuther and Z.-G. Si, {\it {Top quark and leptonic charge asymmetries
  for the Tevatron and LHC}},  \href{http://xxx.lanl.gov/abs/1205.6580}{{\tt
  arXiv:1205.6580}}.

\bibitem{Halzen:1987xd}
F.~Halzen, P.~Hoyer, and C.~Kim, {\it {Forward--backward asymmetry of
  hadroproduced heavy quarks in QCD}},  {\em Phys.Lett.} {\bf B195} (1987) 74.

\bibitem{Kuhn:1998kw}
J.~H. {K\"uhn} and G.~Rodrigo, {\it {Charge asymmetry of heavy quarks at hadron
  colliders}},  {\em Phys.Rev.} {\bf D59} (1999) 054017,
  [\href{http://xxx.lanl.gov/abs/hep-ph/9807420}{{\tt hep-ph/9807420}}].

\bibitem{Antunano:2007da}
O.~Antunano, J.~H. {K\"uhn}, and G.~Rodrigo, {\it {Top quarks, axigluons and
  charge asymmetries at hadron colliders}},  {\em Phys.Rev.} {\bf D77} (2008)
  014003, [\href{http://xxx.lanl.gov/abs/0709.1652}{{\tt arXiv:0709.1652}}].

\bibitem{ATLAS-Atop:2011}
{\bf ATLAS} Collaboration, {\it {Measurement of the charge asymmetry in top
  quark pair production in pp collisions at sqrt{s}=7 TeV using the ATLAS
  detector}},  \href{http://xxx.lanl.gov/abs/{ATLAS CONF-2011-106}}{{\tt {ATLAS
  CONF-2011-106}}}.

\bibitem{CMS-Atop:2010}
{\bf CMS} Collaboration, {\it {Measurement of the charge asymmetry in top quark
  pair production with the CMS experiment}},
  \href{http://xxx.lanl.gov/abs/{CMS-PAS-TOP-10-010}}{{\tt
  {CMS-PAS-TOP-10-010}}}.

\bibitem{Butterworth:2008iy}
J.~M. Butterworth, A.~R. Davison, M.~Rubin, and G.~P. Salam, {\it {Jet
  substructure as a new Higgs search channel at the LHC}},  {\em Phys. Rev.
  Lett.} {\bf 100} (2008) 242001,
  [\href{http://xxx.lanl.gov/abs/0802.2470}{{\tt arXiv:0802.2470}}].

\bibitem{ATL-2009-088}
J.~M. Butterworth, A.~R. Davison, K.~Jakobs, V.~E. Oezcan, G.~Piacquadio, and
  C.~Weiser, {\it {ATLAS sensitivity to the standard model Higgs in the HW and
  HZ channels at high transverse momenta}},
  \href{http://xxx.lanl.gov/abs/{ATL-PHYS-PUB-2009-088}}{{\tt
  {ATL-PHYS-PUB-2009-088}}}.

\bibitem{Kharchilava:1999yj}
A.~Kharchilava, {\it {Top mass determination in leptonic final states with
  $J/\Psi$}},  {\em Phys. Lett.} {\bf B476} (2000) 73--78.

\bibitem{Biswas:2010sa}
S.~Biswas, K.~Melnikov, and M.~Schulze, {\it {Next-to-leading order QCD effects
  and the top quark mass measurements at the LHC}},  {\em JHEP} {\bf 1008}
  (2010) 048, [\href{http://xxx.lanl.gov/abs/1006.0910}{{\tt
  arXiv:1006.0910}}].

\bibitem{worek:private}
{M. Worek, private communication.}

\end{thebibliography}\endgroup

\end{document}